\documentclass[11pt]{amsart}
\usepackage{latexsym,amssymb,calc,amsmath,amsthm,mathrsfs,amssymb,color,graphicx,graphics,psfrag}
\usepackage[bf]{caption2}
\usepackage{fancyhdr,wrapfig}
\usepackage[windows]{umlaute}
\usepackage{ulem}

\usepackage{color}

\definecolor{colorRed}{rgb}{0.,0.,1.}
\definecolor{colorBlue}{rgb}{1.,0.,0.}

\newcommand{\ds}{\displaystyle}
\newcommand{\ts}{\textstyle}
\newcommand{\pt}{\partial_t}
\newcommand{\stress}{\mathbf{S}}

\newcommand{\V}{\mathbf{v}}

\newcommand{\esssup}[1]{\mathop{\rm ess\ sup}}
\newcommand{\essinf}[1]{\mathop{\rm ess\ inf}}
\newcommand{\N}{{\rm I\kern - 2.5pt N}}
\newcommand{\Z}{{\rm Z\kern - 5.5pt Z}}
\newcommand{\Q}{{\rm I\kern - 5.25pt Q}}
\newcommand{\C}{{\rm I\kern - 6.25pt C}}
\newcommand{\R}{{\rm I\kern - 2.5pt R}}
\newcommand{\D}{{\rm {\small -}\kern - 6pt D}}

\newcommand{\Lp}{L_{\lower0.05cm\hbox{\scriptsize X}}^{\lower-0.05cm\hbox{\scriptsize 1}}}
\newcommand{\Linf}{L_{\lower0.05cm\hbox{\scriptsize X}}^{\lower-0.05cm\hbox{\scriptsize $\infty$}}}
\newcommand{\CX}{C_{\lower0.05cm\hbox{\scriptsize X}}}
\newcommand{\ea}[1]{\lower 1mm \hbox{$|_{#1}$}}
\newcommand{\lims}[1]{\lower0.2cm\hbox{${\overline{\ds{\lim}_{}}\atop {#1}}$}}
\newcommand{\limi}[1]{\lower0.2cm\hbox{${\underline{\ds{\lim}_{}}\atop {#1}}$}}

\renewcommand{\div}{{\rm div \,}}
\newcommand{\tr}{{\rm tr \,}}

\begin{document}
\title[Reactive fluid mixtures]
{Continuum thermodynamics
of\\ chemically reacting fluid mixtures
%and the Maxwell-Stefan equations of multicomponent mass transport
}
\author[Dieter Bothe]{Dieter Bothe}
\address{Dieter Bothe, Mathematische Modellierung und Analysis\\
     Center of Smart Interfaces und Fachbereich Mathematik\\
         Technische Universit\"at Darmstadt\\
         Alarich-Weiss-Str.~10\\
         64287 Darmstadt, Germany}
\email{bothe@csi.tu-darmstadt.de}

\author[Wolfgang Dreyer]{Wolfgang Dreyer}
\address{Wolfgang Dreyer, Thermodynamische Modellierung und Analyse von Phasen\"uber\-g\"an\-gen\\
        Weierstrass Institut f\"ur Angewandte Analysis und Stochastik\\
         Mohrenstraße~39\\
         10117 Berlin, Germany}
\email{wolfgang.dreyer@wias-berlin.de}
\maketitle
%\tableofcontents
\vspace{-0.4in}
\begin{abstract}
We consider viscous, heat conducting mixtures of molecularly
miscible chemical species forming a fluid in which the
constituents can undergo chemical reactions. Assuming a common
temperature for all components, we derive a closed system of partial mass and partial momentum balances plus a mixture balance of internal energy. This is achieved
by careful exploitation of the entropy principle and requires appropriate definitions of absolute
temperature and chemical potentials, based on an adequate
definition of thermal energy excluding diffusive contributions.
The resulting interaction forces split into a thermo-mechanical and a chemical part, where the former turns out to be symmetric in case of binary interactions.
For chemically reacting systems and as a new result, the chemical interaction force is a contribution being non-symmetric outside of chemical equilibrium.
The theory also provides a rigorous derivation of the so-called
generalized thermodynamic driving forces, avoiding the use of
approximate solutions to the Boltzmann equations. Moreover,
using an appropriately extended version of the entropy principle
and introducing cross-effects already before closure as entropy
invariant couplings between principal dissipative mechanisms, the
Onsager symmetry relations become a strict consequence. With a
classification of the factors in the binary products of the
entropy production according to their parity--instead of the
classical partition into so-called fluxes and driving forces--the apparent
anti-symmetry of certain couplings is thereby also revealed.

If the diffusion velocities are small compared to the speed of
sound, the Maxwell-Stefan equations follow in the case without chemistry,
thereby neglecting wave phenomena in the diffusive motion. This results in a reduced model
with only mass being balanced individually. In the
reactive case, this approximation via a scale separation argument is no longer possible.
We introduce the new concept of entropy invariant model reduction, leaving the entropy production unchanged under the reduction from partial momentum balances to a
single mixture momentum balance. This result in an extension of the Maxwell-Stefan equations to
chemically active mixtures with an additional contribution to the transport coefficients
due to the chemical interactions.
%Within the considered model class for reactive fluid mixtures the new results are achieved for arbitrary free energy functions.
% It also shows that Truesdell's metaprinciple is only applicable within a
%model class, but does not govern the model reduction.
%\sout{The present derivations are aligned with the framework of
%Rational Thermodynamics, but care is taken to establish
%connections to the Engineering Literature.}

\end{abstract}
%\maketitle
%
{\small \noindent

\section{Introduction}
We consider fluid systems being composed of $N$ constituents, i.e.\ multicomponent
mixtures of gases or liquids, typically containing additional dissolved chemical components.
Main emphasis is put on chemically reacting mixtures, since the existing literature on fluid mixtures
does not account for chemical processes in depth and, moreover, a thermodynamically
consistent coupling between chemistry and fluid mechanics is often missing. This is especially true
in case of highly concentrated mixtures for which cross-effects and non-idealities are to be included.

Our approach is based on continuum physics, where we strictly distinguish between universal laws
and constitutive relations. We employ the field equations which follow from conservation laws for the extensive (i.e., mass additive) quantities {\it mass, momentum and energy}.
This leads to unclosed sets of partial differential equations which need to be complemented by
material dependent closure relations, so-called {\it constitutive equations}.
Within {\it continuum thermodynamics}, the balance of {\it entropy} of the system is added
to this set of differential equations together with the {\it second law of thermodynamics}.
This severely restricts the closure relations admissible for the description of the constitutive
functions appearing in these equations.

Let us start with some information about the different available continuum physics approaches, since our theory combines ideas from all these.
The systematic theory of non-equilibrium thermodynamics for continuous matter was started in the year 1940 by C.\ Eckart who wrote the two basic papers \cite{Eckart1, Eckart2}, entitled \textit{The thermodynamics of irreversible processes I: The simple fluid} and \textit{The thermodynamics of irreversible processes II: Fluid mixtures}.
In the following years the theory was completed by J.\ Meixner. In the seminal paper \cite{Meixner43},
\textit{Zur Thermodynamik der irreversiblen Prozesse in Gasen mit chemisch reagierenden, dissoziierenden and anregbaren Komponenten},
Meixner studied the phenomenological coefficients in detail and equipped these with Onsager-Casimir reciprocal relations. Finally in 1963, S.R.\ de Groot and P.\ Mazur reached a certain endpoint by putting everything together in the fundamental monograph \textit{Non-equilibrium thermodynamics} (\cite{dGM}). This theory is still in broad use under the labeling \textit{Thermodynamics of irreversible processes}, TIP.

In 1962, W.\ Noll (cf.\ \cite{TN2003}) introduced new ideas to rationalize the mechanical theory of continua. In the following years the school around C.\ Truesdell, in particular B.D.\ Coleman and W.\ Noll applied these strategies to irreversible thermodynamics and established a variant of thermodynamics that has become very popular, at least for a certain time, under the name \textit{Rational Thermodynamics}, RT. The first fundamental contribution from this school to mixture theory has been given by I.\ M\"{u}ller in \cite{M68}, correcting several erroneous
contributions by other authors (see also \cite{CT69}). Later, in \cite{M75}, M\"{u}ller
extended the concept of Lagrange multipliers for the exploitation of the entropy inequality,
which had been developed in his group by I-Shih Liu in \cite{Liu72}, to the case of mixtures.
For a more complete presentation of rational thermodynamics of continuous media
with Lagrange multipliers see \cite{M85}, \cite{Liu}. Thus there are two versions of RT, namely RT with Lagrange multipliers (RT-LM, for short, introduced by I.\ M\"uller and I-Shih Liu) and RT without Lagrange multipliers (due to B.D.\ Coleman and W.\ Noll).
An excellent overview on the initial difficulties in formulating rational continuum mixture theories has been provided by R.M.\ Bowen \cite{Bowen}.
%
%
%WOLLEN WIR DIESES BUCH ZITIEREN: \cite{Samohyl} ?
%

Both theories, TIP and RT, are not very different from each other. In particular, both split the laws of thermodynamics into universal laws, relying on balance equations, and characterizations of the material at hand via constitutive equations. Another agreement between TIP and RT is the fact that the description
of diffusive transport processes leads to parabolic PDEs with infinite speed of propagation of disturbances. Since the latter apparently contradicts the fundamental physical principle that
nothing should be faster than the speed of light, several researchers started
a quest for a hyperbolic theory. This led to the invention of {\it Extended Thermodynamics}, ET,
see \cite{IM-Ruggeri}.
Within ET, fluxes of the mentioned extensive quantities, like the heat flux in case of the energy balance, are themselves considered as governed by balance equations.
%For an overview of the history of these thermodynamic theories, we refer to \cite{IM-history}.

The five versions TIP, RT, RT-LM, ET and the present theory, which we call Continuum Thermodynamics (CT for short), have a common basic strategy:
\begin{enumerate}
\item
Definition of a set of independent variables.
\item
Formulation of the corresponding equations of balance.
\item
Identification of the constitutive quantities, i.e.\ those quantities appearing in the
balance equations but not in the set of variables.
\item
Selection of classes of constitutive equations for the different constitutive quantities,
where all admissible constitutive functions obey the principle of material frame indifference.\footnote{\, Within TIP this comes down to Curie's principle}
\item
Formulation of the entropy principle as a tool to further restrict the classes of
constitutive equations so that the entropy inequality is satisfied for any
thermodynamic process occurring in the modeled class of materials.
\end{enumerate}

\noindent
The main differences between the different approaches are summarized in the following
Table~1.\\[1ex]
\noindent
\begin{tabular}{|l|l|l|l|l|l|} \hline
  & TIP & RT & RT-LM & ET & CT\\
\hline
 &   &   &
  & $\rho_i,\, {\bf v}_i,\, \rho e$,  &  \\
Choice of & $\rho_i,\, {\bf v},\, \rho e$ & $\rho_i,\, {\bf v}_i,\, \rho e$ &
$\rho_i,\, {\bf v}_i,\, \rho e$ & partial stress,  & $\rho_i,\, {\bf v}_i,\, \rho e$ \\
variables &   &   &  & heat flux, etc.\ &   \\
\hline
Selection & state space & equipresence &
equipresence & equipresence  & state space \\
of material  & only for $\rho s$ &  in constitutive  & in constitutive  & in constitutive  & only for $\rho s$ \\
functions &  &   equations &  equations &  equations &  \\
\hline
  & binary & binary & binary & binary  & binary \\
entropy  & structure  &  structure  & structure  & structure  & structure \\
production & postulated & not explicit &  not explicit &  calculated & postulated,\\
  &   &   &   &   & det.\ balance\\
\hline
  & determined &   &   &    & determined \\
entropy  & s.t.\ binary  &  postulated  & calculated  & calculated  & s.t.\ binary  \\
flux & structure &  &   &    & structure\\
  & results  &   &   &   & results\\
\hline
          &    & proved for       &  proved for dif-       &               & identifica- \\
Onsager-  &    &  diffusion       &  fusion mobilities   & usually       & tion of its  \\
Casimir   & postulated & mobilities  & and thermodif-  &  not          & origin for\\
relations &    &  under binary    &  fusion without  &  investigated & PDE-\\
          &    &  interactions    &  thermal drag  &               & systems \\
\hline
\end{tabular}\\[0.5ex]
{\small {\bf Table 1.} Comparison of different thermodynamic continuum theories}
$\mbox{ }$\\[2ex]
\noindent
Our approach to the continuum thermodynamics of reacting fluid mixtures
is a combination of elements of all three theories:
(i) We employ the partial balances of not only
mass, indispensable for any mixture theory, but also momentum. This corresponds to an extended
thermodynamics for the mass balances in which the time evolution of diffusion velocities is
also monitored.
Indeed, the contribution of chemical reactions to the diffusivities is similar
as the contribution to the heat conductivity obtained in relativistic ET; cf.\ \cite{IM-Ruggeri}.
Actually we start also with partial energy balances in order to extract some
structural information about the common energy balance on which the final modeling relies,
where we only consider a single, common temperature.

(ii) We use an axiomatic approach guided by RT, including the principle of material frame indifference
and also the strict separation into universal balance laws and material dependent constitutive relations.
But we do not impose the principle of equipresence which does not constitute a physical law.
Instead, we select the set of primitive thermodynamic variables when defining the
form of the constitutive equation for the entropy.
The dependence of all other quantities is
then a consequence of their relations to the latter.
Another important difference to RT lies in the entropy principle.
In fact, in our theory--in agreement with W.\ Alt and I.\ M\"{u}ller (see \cite{Al09, M85})--the form of the entropy flux depends on the material at hand and comes as a modeling result instead of being assumed via an axiom as in RT. Here we do not believe Gurtin's statement in \cite{Gurtin71} saying that the form of the entropy flux is ''a matter of taste''.

(iii) We consider the entropy principle by Alt and M\"{u}ller as the prototype.
However, we postulate an extended version by assuming further properties of the entropy production.
The most important extension concerns its structure: Relying on
classical TIP, we postulate it as a sum of binary products and exploit this structure to obtain
closure relations which satisfy the entropy inequality.
However, instead of the misleading
notion of ''fluxes'' and ''driving forces'' as used in TIP, we build on Meixner \cite{Meixner73}
in introducing
the parity of thermodynamic quantities and impose binary products of positive and negative parity factors.
Incorporating a priori knowledge on the physics, we also predefine which dissipative mechanisms are to be included in the modeling.
As compared to TIP, this allows for a strengthened version
of the principle of detailed balance: Every dissipative mechanism alone has to fulfil the
entropy inequality.
Finally, we observed several forms of non-uniqueness of both the entropy flux and the entropy production, arising by addition of new terms to these quantities, but which sum up to zero in total.
This way even new dissipative mechanisms can be introduced; cf.\ Remark~10 in Section~\ref{sec:fundamentals} for a non-trivial example.

All together, this enables a pragmatic evaluation of the entropy principle, at the same time
yielding a particularly strong framework for the constitutive theory.
Summarizing this work, we obtain the following main results:
\begin{enumerate}
\item
We provide various alternative forms of the energy balance and appropriately select the one which
enables to obtain information about the inter-species interactions from the entropy principle.
\item
We introduce a strengthened form of the entropy principle which includes a general version of the
principle of detailed balance.
\item
We employ the notion of parity to classify the co-factors in the binary products inside
the entropy production and show that the parity of a physical quantity is directly related
to its physical dimension.
\item
We introduce a way to couple between different dissipative mechanisms by zero additions
to the entropy production. This procedure of {\it entropy invariant mixing} describes cross-effects which automatically satisfy the Onsager symmetry.
%Hence the corresponding phenomenological cross-coefficients have arbitrary signs.
Combined with the concept
of parity, this way of introducing cross-effects also explains the appearance of the
Onsager-Casimir relations, i.e.\ the apparent anti-symmetry of some cross-coupling coefficients.
\item
We derive a consistent closed model for viscous, heat conducting and chemically
reacting fluid mixtures, comprising $N$ partial mass \underline{and} $N$
partial momentum balances. The model includes, in particular, a thermodynamically consistent
closure for the interaction forces in the partial momentum balances and for
the reaction rates as nonlinear functions of the affinities in the partial mass balances.
\item
We introduce the new method of {\it entropy invariant model reduction} to pass from
a model with partial momentum balances to one in which only a common momentum balance for the barycentric velocity appears.
This derivation relies on identification of corresponding quantities in the partial mass
and the mixture internal energy balance,
%since these are common to both model classes,
combined with the invariance of the entropy production under this reduction process.
%Without chemistry, the derivation is also possible (and essentially known) via a scale-limit; but the latter is not rational in the choice of the approximation.
\item
We extend the Maxwell-Stefan equations to the case of chemically reacting
constituents, leading to a non-symmetric addition. We also take into account partial viscosities, leading to a stress-driven diffusion contribution.
\item
We introduce a consistent general scheme to construct free energy functions for fluid mixtures.
These {\it elastic mixtures} are obtained by combining information about the chemical potentials
and the pressure-density relation at given composition.
\item
Incompressibility of mixtures can be defined in different ways. We show the passage to incompressible limits based on appropriate equations of state for the mixture pressure,
where we consider several alternatives concerning the dependence on the constituents.
We also revisit the apparent paradoxon concerning thermal expansion in an incompressible medium and explain how this is resolved.
\end{enumerate}

To conclude this introduction, let us mention that there are also kinetic
mixture theories available, the most famous one being the multi-species Boltzmann equations; cf.\ \cite{FK, Hirsch}.
While they provide a consistent derivation of a mixture theory, the admissible fluid
mixtures contain only rather simple materials, viz.\ so-called simple mixtures (see Section~\ref{sec:free-energies} below) of monatomic gases.
%\BA
For more details on corresponding mixture models without partial momentum balances we refer to \cite{FK},
kinetic models which account for partial momenta can be found in \cite{HecklM}.
%\EA
Let us note in passing that the Boltzmann equations can be motivated from first principles,
but their rigorous derivation from the
fundamental Liouville equation is still lacking; however, see \cite{Gallagher}.
%Let us also note that derived sets of closed PDEs for the first moments of the distribution function
%depend on the chosen method of approximation (Chapman/Enskog vs. Grad).
Nevertheless, the multi-species Boltzmann equations provide a valuable micro-theory and
we demand from our mixture theory that it contains the equations resulting from
kinetic theory as a special case.

In our opinion, the strict separation into balance equations being rigorously valid within the
continuum physics framework, and constitutive theory to model the material depend quantities,
is the crucial factor for the success of TIP, RT and ET.
There are other mixture continuum theories, relying on variational approaches, which
do not have this clear distinction. Moreover, these theories are developed for specific
classes of thermodynamic processes such as isentropic motions or isothermal systems.
We refer to \S 231 in \cite{Fluegge} for an assessment and
only mention the classical d'Alembert/Lagrange theory; see, e.g., \cite{Gouin-mixture}
for its extension to mixtures.
In recent years, a further variant of the variational approach, called GENERIC (\cite{GENERIC}), has emerged, where
another disadvantage is that, up to now, this theory can only describe adiabatically isolated, closed systems. An introduction to GENERIC, including a discussion of its shortcomings, may be found in \cite{Lebon}.
\section{Basic quantities and model classes}
We consider fluid systems being composed of $N$ constituents $A_1,\ldots ,A_N$, i.e.\ multicomponent
mixtures of gases or liquids, typically containing additional dissolved chemical components.
Special emphasis is put on chemically reacting mixtures, where we allow for
$N_R$ chemical reactions between the $A_i$ according to
\begin{equation}
\alpha_1^a \, A_1 + \ldots + \alpha_N^a \, A_N
\rightleftharpoons
\beta_1^a \, A_1 + \ldots + \beta_N^a \, A_N
\quad \mbox{ for } a=1,\ldots ,N_R
\end{equation}
with stoichiometric coefficients $\alpha_i^a, \beta_i^a \in \N_0$.

The mixture occupies a region $\Omega \subset \R^3$. At any time $t\geq 0$, the thermodynamic
state of the matter inside $\Omega$ is described by $N$ partial mass densities $\rho_1,
\ldots , \rho_N$, by $N$ partial velocities ${\bf v}_1, \ldots , {\bf v}_N$
of the constituents and by the
temperature $T$ of the mixture. These quantities will be functions of time $t\geq 0$ and
space ${\bf x}\in \R^3$. Partial mass densities and partial velocities are used to define the
total mass density $\rho$ and the barycentric velocity ${\bf v}$ of the mixture according to
\begin{equation}
\label{barycentric}
\rho :=  \sum_{i=1}^N \rho_i
\quad \mbox{ and } \quad
\rho\, {\bf v} := \sum_{i=1}^N \rho_i {\bf v}_i.
\end{equation}
The diffusion velocities ${\bf u}_i$ and the corresponding diffusional mass fluxes ${\bf j}_i$
are defined by
\begin{equation}
\label{diffusion-flux}
{\bf u}_i := {\bf v}_i - {\bf v}
\quad \mbox{ and } \quad {\bf j}_i :=  \rho_i {\bf u}_i
\quad \mbox{ with } \quad \sum_{i=1}^N {\bf j}_i =0.
\end{equation}
Finally, we introduce the molar concentrations and the molar fractions according to
\begin{equation}
\label{concentrations}
c_i := \frac{\rho_i}{M_i}
\quad \mbox{ with molar mass } M_i, \qquad
x_i:=\frac{c_i}{c}
\quad \mbox{ with } \quad c:=\sum_{i=1}^N c_i.
\end{equation}
Fluid mixtures can be modeled within three different levels of detail:
class-I, class-II and class-III. Class-I considers as primitive variables the mass densities $\rho_i$
of the constituents, the barycentric velocity ${\bf v}$ and the temperature $T$ of the mixture.
The basic variables of class-II are the mass densities $\rho_i$, the velocities ${\bf v}_i$ of
the constituents and the temperature $T$ of the mixture. Finally, in class-III one has the
partial mass densities $\rho_i$, the partial velocities ${\bf v}_i$ and individual temperatures
$T_i$ of the constituents. With this notion we follow \cite{Hutter-book}.

There are several motivations for the more complex models,
including in particular the following. For class-II model: (i) the
different components can experience different boundary conditions for the partial velocities;
(ii) the appearance of wave phenomena in the diffusive motion
which are often ignored on a macroscopic scale;
% -- however, current micro- and nano-technology applications may show such phenomena;
(iii) even for the derivation of class-I models, class-II models
are important because they offer more information concerning the
constitutive quantities. For class-III model: in plasma modeling,
the constituents experience individual temperatures on time scales
comparable to transport process times, hence separate energy
balances are necessary and, in general, the individual species
have different temperatures.

In the present study, we consider the model classes I and II.
In particular, we derive consistent closures for class-II models of chemically reacting, heat conducting viscous fluid mixtures. Based on these, we derive improved class-I models which contain information
from the class-II level within the constitutive relations. For example,
due to the explicit structure of the thermal energy balance
\eqref{eq:total-internal-energy-balance}, certain cross-effects with Onsager symmetry can already be derived.\\
%
%
%To describe the mixture as a whole, the partial balance equations above
%are added in order to exploit
%the conservation laws \eqref{conservation-laws}.
%Then appropriate further mixture quantities need to be defined, where the
%three meta-rules given by Truesdell (see \cite{CT69}) provide a guidance: {\small
%\begin{enumerate}
%\item[(I)]
% All properties of the mixture must be mathematical
% properties of the constituents.
%\item[(II)]
%So as to describe the motion of a constituent, we may in
%imagination isolate it from the rest of the mixture, provided
%we allow properly for the actions of the other constituents upon it.
%\item[(III)]
%The motion of the mixture is governed by the same equations as is a single body.
%\end{enumerate}}
%
%
%
%
%
\section{Balance equations for chemically reacting fluid mixtures}
\label{sec:balances}
{\bf Partial Balances.}
The starting point of our modeling are the {\it partial balances of mass, momentum and energy},
which will later be complemented by the {\it entropy balance}.
The local balances for the fluid components $A_i$ read as\\[1ex]
{\bf partial mass balance}
\begin{equation}
\label{partial-mass}
\pt \rho_i + \div (\rho_i {\bf v}_i)= r_i,
\end{equation}
{\bf partial momentum balance}
\begin{equation}
\label{partial-mom}
\pt (\rho_i {\bf v}_i) + \div (\rho_i {\bf v}_i \otimes {\bf v}_i - \stress_i)
= {\bf f}_i + \rho_i {\bf b}_i,
\end{equation}
{\bf partial energy balance}
\begin{equation}
\label{eq:energy-balance}
\pt (\rho_i e_i + \frac{\rho_i}{2} {\bf v}_i^2 )
+ \div ( (\rho_i e_i + \frac{\rho_i}{2} {\bf v}_i^2) {\bf v}_i -  {\bf v}_i \cdot \stress_i + {\bf q}_i)
= h_i + \rho_i {\bf b}_i \cdot {\bf v}_i.
\end{equation}
In \eqref{partial-mass}-\eqref{eq:energy-balance}, the newly introduced quantities are: $r_i$ - mass production due to chemical
reactions, $\stress_i$ - partial stress, ${\bf f}_i$ - momentum
production, ${\bf b}_i$ - body force acting on $A_i$, $e_i$ -
specific internal energy, ${\bf q}_i$ - individual
heat flux, $h_i$ - internal energy production. Energy sources
due to electromagnetic fields (i.e., radiation) are not considered here.

With these notations, the conservation laws for total mass, momentum
and energy read
\begin{equation}
\label{conservation-laws}
\sum_{i=1}^N r_i =0, \qquad
\sum_{i=1}^N {\bf f}_i =0, \qquad
\sum_{i=1}^N h_i =0.
\end{equation}
The mass productions $r_i$ have the structure
\begin{equation}
\ts r_i = \sum\limits_{a=1}^{N_R} M_i \nu_i^a R_a,
\end{equation}
where $R_a=R_a^f-R_a^b$ is the rate of reaction $a$,
given by the rate $R_a^f$ of the forward minus that of the backward path,
and $\nu_i^a := \beta_i^a - \alpha_i^a$.
Because mass is conserved in every single reaction, it holds that
\begin{equation}
\ts \sum\limits_{i=1}^N M_i \nu_i^a =0 \quad \mbox{ for all } a.
\end{equation}
The reaction rates $R_a$ as well as the interaction forces ${\bf f}_i-r_i {\bf v}_i$
are to be modeled. A constitutive model for the $h_i$ is not required, since they do not
appear in the common energy balance.

We define the {\it partial pressures} $P_i$ and the traceless part $\stress_i^\circ$
of the stresses $\stress_i$ by
\begin{equation}
\label{partial-stress}
P_i := - \frac 1 3 \tr (\stress_i ), \qquad
\stress_i = -P_i {\bf I} + \stress_i^\circ.
\end{equation}
We further decompose the partial pressures according to
\begin{equation}
\label{pressure-decomposition}
P_i=p_i + \Pi_i,
\end{equation}
where the $\Pi_i$ vanish in equilibrium. The pressure contribution $p_i$ is often called
{\it hydrostatic pressure} in which case $\Pi_i$ is usually called {\it dynamic pressure}.
For reasons which will become clear below, we call $p_i$ the {\it partial thermodynamic pressure}
and  $\Pi_i$ the {\it irreversible partial pressure contribution}.
To economize notation, we use the abbreviation
\begin{equation}
\label{irr-stress}
\stress_i^{\rm irr} = -\Pi_i {\bf I} + \stress_i^\circ
\end{equation}
to denote the irreversible stress contribution. Consequently, $\stress_i = -p_i {\bf I} + \stress_i^{\rm irr}$.
Finally, the total pressure of the mixture is defined as $P:=\sum_i P_i$. The latter is decomposed as
$P=p+\Pi$ with $p:=\sum_i p_i$ the total thermodynamic pressure.
We also use the short hand notation
$\stress^{\rm irr} = -\Pi \, {\bf I} + \stress^\circ$ for the irreversible part of the mixture stress, where $\stress^\circ$ is the traceless part of $\stress$. Above, the short-hand notation
$\sum_i$ stands for the sum, where $i$ runs from $1$ to $N$. If nothing else is said, the
same applies throughout the entire paper.

In the present work, we only consider {\it non-polar} fluid mixtures,
i.e.\ there are no so-called surface or body couples and the fluid has vanishing spin density,
which implies symmetry of stresses, i.e.\ $\stress_i = \stress_i^{\sf T}$ for all $i$ and
$\stress = \stress^{\sf T}$.\\[-1ex]

{\bf Mixture balances.}
We start by formulating the balances of total mass, momentum and energy, which follow by summation of the corresponding partial balances.
Then appropriate further mixture quantities need to be defined, where the
three meta-rules given by C.~Truesdell (see \cite{CT69}) provide a guidance: {\small
\begin{enumerate}
\item[(I)]
 All properties of the mixture must be mathematical
 properties of the constituents.
\item[(II)]
So as to describe the motion of a constituent, we may in
imagination isolate it from the rest of the mixture, provided
we allow properly for the actions of the other constituents upon it.
\item[(III)]
The motion of the mixture is governed by the same equations as is a single body.
\end{enumerate}}
Following this principle, we introduce the following additional quantities in order to obtain mixture balance equations which resemble the partial balances:
\begin{align*}
& \ts
\rho {\bf b}:= \sum_i \rho_i {\bf b}_i
& \mbox{(total external force),} \\[1ex]
& \ts
{\stress}:= \sum_i \big( \stress_i - \rho_i {\bf u}_i \otimes {\bf u}_i \big)
\hspace{-0.2in}
& \mbox{(mixture stress tensor),}\\[1ex]
& \ts
\rho e_{\rm tot}:= \sum_i  \rho_i (e_i + \frac{{\bf v}_i^2}{2})
=\rho \frac{{\bf v}^2}{2} + \sum_i  \rho_i (e_i + \frac{{\bf u}_i^2}{2})\hspace{-0.4in}
& \mbox{(total energy density),}\\[1ex]
& \ts
{\bf q}_{\rm tot}:= \sum_i  ({\bf q}_i - {\bf u}_i \cdot \stress_i
+ \rho_i ( e_i + \frac{{\bf u}_i^2}{2}) {\bf u}_i )
& \mbox{(total energy flux),}\\[1ex]
& \ts \rho \pi:= \sum_i \rho_i {\bf b}_i \cdot {\bf u}_i
& \mbox{(power of body force due to diffusion)}.
\end{align*}
Summation of the partial balances now yields the balances for total mass, momentum and energy which read as\\[1ex]
{\bf mass balance (continuity equation):}
\begin{equation}
\pt \rho  +  \div ( \rho {\bf v} )=0,
\end{equation}
{\bf momentum balance:}
\begin{equation}
\label{total-momentum-balance}
\pt (\rho {\bf v}) + \div (\rho {\bf v} \otimes {\bf v} - \stress)
= \rho {\bf b},
\end{equation}
{\bf energy balance:}
\begin{equation}
\label{eq:mixture-total-energy-balance}
\ts
\pt (\rho e_{\rm tot})
+ \div \big( \rho e_{\rm tot} {\bf v} -{\bf v} \cdot \stress + {\bf q}_{\rm tot} \big)
= \rho {\bf v} \cdot {\bf b}+\rho \pi.
\end{equation}

For the constitutive modeling, in particular for exploitation of the entropy principle,
we need a balance of internal energy instead of total energy.
Here it is important to observe that different definitions of an internal part of the energy
are possible. If partial energy balances are ignored, like in the case of standard TIP,
the {\it internal energy} is (in the present context of non-polar fluids)
that part of the total energy which cannot be eliminated by a Euclidean transformation,
i.e.\ by a change to a different frame of reference. This leads to the definition of
\begin{equation}
\label{tot-int-energy}
\rho e_{\rm int}:= \rho e_{\rm tot} - \frac 1 2 \rho {\bf v}^2
\qquad \mbox{(mixture internal energy).}
\end{equation}
Subtracting the balance of kinetic energy due to barycentric motion from the balance of total energy,
a straightforward computation yields the\\[1ex]
{\bf mixture internal energy balance:}
\begin{equation}
\label{eq:mixture-internal-energy-balance}
\pt (\rho e_{\rm int})
+ \div ( \rho e_{\rm int} {\bf v} +  {\bf q}_{\rm tot} ) =
\stress  : \nabla {\bf v} +\rho \pi.
\end{equation}

As already noted in Chapter~III, \S4 of \cite{dGM}, the above defined internal energy
contains kinetic energy of diffusion, while ''the internal energy should only contain
contributions from the thermal agitation and the short-range molecular interactions''.
In the present context of class-II modeling, this turns out to be even more relevant and
we base our theory on the use of the {\it thermal part of the internal energy}, defined as
\begin{equation}
\label{def:total-internal-energy}
\ts
\rho e:= \sum_i  \rho_i e_i
\qquad \mbox{(thermal energy)}.
\end{equation}
Evidently, the link between the two quantities is given by
\[
\ts
\rho e_{\rm int}=\rho e + \frac 1 2 \sum_i \rho_i {\bf u}_i^2.
\]

To derive the balance of thermal energy,
we start from the partial internal energy balance, obtained as usual by subtracting the (partial) kinetic energy balance
from \eqref{eq:energy-balance}. This leads to the\\[1ex]
{\bf partial balance of internal energy:}
\begin{equation}
\label{partial-energy-balance}
\pt (\rho_i e_i ) + \div ( \rho_i e_i  {\bf v}_i + {\bf q}_i)
= \stress_i : \nabla {\bf v}_i + h_i - {\bf v}_i \cdot \big( {\bf f}_i - r_i
\frac{{\bf v}_i }{2} \big).
\end{equation}
Summation of \eqref{partial-energy-balance} for all constituents yields
a first version of the {\it thermal energy balance} which reads as
\begin{equation}
\label{eq:total-internal-energy-balance1}
\ts
\pt (\rho e)
+ \div ( \rho e {\bf v} +  \tilde{\bf q} ) =
\sum_i \stress_i : \nabla {\bf v}_i
-  \sum_i {\bf u}_i \cdot \big( {\bf f}_i - r_i  {\bf v}_i +  \frac{r_i}{2} {\bf u}_i \big),
\end{equation}
where
\begin{equation}
\label{def:reduced-heat-flux0}
\ts
\tilde{\bf q}:= \sum_i  ({\bf q}_i + \rho_i  e_i  {\bf u}_i ).
\end{equation}
The form \eqref{eq:total-internal-energy-balance1} of the thermal energy balance will be used
for exploitation of the entropy principle below. It will then turn out that another definition
of the heat flux, containing the equilibrium part of the trace part of the stresses, is
more convenient. Thus, for later use, observe that the identity
\[
\ts \sum_i {\bf S}_i:\nabla {\bf v}_i=
\sum_i {\bf S}^\textrm{irr}_i:\nabla {\bf v}_i - p \, \div {\bf v}
-\div (\sum_i p_i{\bf u}_i) +\sum_i {\bf u}_i \cdot \nabla p_i
\]
generates a further useful alternative to represent the
\\[1ex]
{\bf thermal energy balance:}
\begin{equation}
\label{eq:total-internal-energy-balance} \ts \pt (\rho e) + \div ( \rho e {\bf v} +  {\bf q} )
= \sum_i  \stress_i^{\rm irr} : \nabla {\bf v}_i
-p \, \div {\bf v} -  \sum_i {\bf u}_i \cdot \big( {\bf f}_i - r_i  {\bf v}_i
 + \frac{r_i}{2} {\bf u}_i - \nabla p_i \big),
\end{equation}
where the {\it heat flux} contains non-convective transport of
heat and diffusive transport of the species enthalpy $\rho_i e_i + p_i$ and reads as
\begin{equation}
\label{def:reduced-heat-flux}
\ts
{\bf q}:= \sum_i  \big({\bf q}_i + (\rho_i  e_i + p_i)  {\bf u}_i \big).
\end{equation}
Later, we use the abbreviation $h_i:= \rho_i e_i + p_i$. Since, from here on, we do not consider partial energy balances, this will not lead to confusion with the interspecies energy exchange rate.

As already mentioned, the balance equations given above
are not closed. Together with the entropy principle below, they
form the starting point of our constitutive theory.
\section{Fundamentals of Constitutive Modeling}
\label{sec:fundamentals}
{\bf General Strategy.}
We consider the model classes I and II. Our strategy is to start with the closure procedure
for class-II, followed by a reduction to a class-I model. Since the latter has to be thermodynamically
consistent, we need to know the entropy production for both model classes.
In the next sections, we focus on the model class II, although some considerations could be done
for both classes simultaneously. The reason why we defer considerations on the class-I model to
Section~\ref{sec:model-reduction1} is a subtle difference in the temperature definition. This will be explained in detail at the appropriate place below.

%Recall first that we talk of a {\it class-III model}, if partial
%balances for the three extensive quantities {\it mass, momentum and energy} are used, while in a {\it class-II model} there
%are individual balances for the two quantities {\it mass and momentum} and a single common energy balance, hence
%the same temperature for all components.
%Finally, a {\it class-I model} is made of partial balances for {\it mass},
%but common single momentum and energy balances,
%i.e.\ all constituents have the same local temperature and are advected with the
%same velocity.

Recall the basic variables of the class-II model, which are $\rho_i$, ${\bf v}_i$ and
the common temperature $T$.
The determination of these variables relies on the balance equations \eqref{partial-mass},
 \eqref{partial-mom} and \eqref{eq:total-internal-energy-balance}.
%For convenience, we repeat them here, but with the
%momentum balance in its nonconservative form:
%\begin{equation}\label{CIImass}
%\ts \pt \rho_i + \div (\rho_i {\bf v}_i)= r_i \quad \mbox{ with }
%\quad r_i= \sum\limits_{a=1}^{N_R} M_i \nu_i^a R_a,
%\end{equation}
%\begin{equation}\label{CIImomentum}
%\rho_i \big( \pt  {\bf v}_i +  ({\bf v}_i \cdot \nabla ) {\bf v}_i \big) - \div \stress_i
%= {\bf f}_i - r_i  {\bf v}_i + \rho_i {\bf b}_i,
%\end{equation}
%\begin{equation}\label{CIIenergy}
%\ts \pt (\rho e) + \div ( \rho e {\bf v} +  {\bf q} ) = \sum_i
%\stress_i^{\rm irr} : \nabla {\bf v}_i-  p\, \div {\bf v}
%-  \sum_i {\bf u}_i \cdot \big( {\bf f}_i - r_i  {\bf v}_i - \nabla p_i
%+  \frac{r_i}{2} {\bf u}_i \big).
%\end{equation}
We also need the momentum balance in its nonconservative form, i.e.\
\begin{equation}\label{CIImomentum}
\rho_i \big( \pt  {\bf v}_i +  ({\bf v}_i \cdot \nabla ) {\bf v}_i \big) - \div \stress_i
= {\bf f}_i - r_i  {\bf v}_i + \rho_i {\bf b}_i,
\end{equation}
This system of balance equations becomes a PDE-system for the basic variables,
if these are related to the constitutive quantities
\[
R_a, \quad p_i, \quad \Pi_i, \quad \stress_i^\circ  ,\quad {\bf f}_i- r_i {\bf v}_i,\quad \rho e ,\quad {\bf q}
\]
in a material dependent manner. Here we use the notion ''balance equation''
for the unclosed balance, while talk of ''PDE-system'' refers to the system obtained after inserting the closure relations.\\

\noindent
{\bf Euclidean transformations.} We consider the behavior of the
involved quantities under coordinate transformations. The most
general transformation between two Cartesian coordinate systems,
with coordinates written as $(x_1,x_2,x_3)=(x_i)_{i=1,2,3}$ and
$(x_1^\ast,x_2^\ast,x_3^\ast)=(x_i^\ast)_{i=1,2,3}$, respectively,
is of the form (employing summation convention)
\begin{equation}
t^\ast = t+a, \quad
x_i^\ast = O_{ij}(t) x_j + b_i (t), \quad
\mathbf{O}(t) \mathbf{O}(t)^{\sf T} =\mathbf{I}.
\end{equation}
These are called {\it Euclidean transformations}.
We define the notion of {\it objective scalars, vectors and tensors (of rank two)} if their components transform according to
\begin{align}
s^\ast =s & \quad \mbox{ for scalars},\\
v_i^\ast = O_{ij} v_j & \quad \mbox{ for vectors},\\
T_{ij}^\ast = O_{ik} O_{jl} T_{kl} & \quad \mbox{ for rank two tensors}.
\end{align}
Note that there are two different classes of quantities: those for
which the transformation behavior can be derived are called
{\it kinematic} quantities. Among them are the velocity, the
velocity gradient, the acceleration and the diffusion
velocities. The transformation properties of the remaining,
non-kinematic quantities must be postulated, motivated by
experimental experience. Important examples are the scalars mass
densities and temperature, the vectors interaction forces and heat flux and the %(Cauchy)
stress tensor. Note that a velocity is not an objective vector in the above sense,
but diffusion velocities are, being the difference of two velocities.\\

\noindent
{\bf Parity.} To every physical quantity we introduce an associated {\it parity}, called
{\it positive} (+1) or {\it negative} (-1).
The original concept of parity for a time-dependent
quantity characterizes its behavior under time reversal,
based on the ODE-system
\begin{equation}
\label{Newton-ODE}
m\, \ddot{\bf x} (t) = K ({\bf x} (t))
\end{equation}
of Newtonian mechanics. This system determines the position ${\bf x} (t)$ and the
velocity ${\bf v}(t)=\dot{\bf x} (t)$ of a point mass $m$ under the action
of a prescribed force $K$, given the initial values ${\bf x}(0)$ and ${\bf v} (0)$.
Now, if new position and velocity are defined according to
\begin{equation}
\label{t-reversal}
\tilde{\bf x} (t):={\bf x} (T-t),\qquad
\tilde{\bf v} (t):=- {\bf v} (T-t),
\end{equation}
the equation of motion \eqref{Newton-ODE} is unchanged under this transformation $t \to T-t$.
In the case of Newton's equations this implies a symmetry of the solutions with respect to time reversal.

In the continuum setting, the symmetry of the solutions is lost under time reversal, because the PDEs
describe irreversible processes. But note that the structure of the balance equations, without
considering solutions of the PDE-system, keeps unchanged.
Hence, for studying the structure of the balance equations, the concept of reversibility does not play a role.
This allows to generalize the concept and notion of parity to the balance
equations of continuum mechanics. We first define the parity of the transformations in
\eqref{t-reversal} as {\it positive} for \eqref{t-reversal}$_1$ (with factor +1)
and {\it negative} for \eqref{t-reversal}$_2$ (with factor -1). For the generic balance
\begin{equation}
\label{master-balance}
\partial_t \psi + \div (\psi {\bf v} + {\bf j}) = f,
\end{equation}
the operator $\partial_t$ evidently alters the parity under the transformation $t \to T-t$.
In analogy to \eqref{t-reversal}$_2$, we assume the same change in parity for quantities
in the time reversed balance equation which appear with the factor ${\bf v}$.

It turns out that the parity of a quantity is simply obtained by assigning the factor -1 if the time unit ''second'' appears with an uneven power and +1 if it appears with even power. Here we rely on the SI base units m, kg, s, K, mol, A, cd. Note that the unit \textit{Ampere} (A) for the electric current, which does not occur in the present mixture theory, is also among the SI base units. If the latter unit is involved, we must extend our rule and assign additionally a further factor -1 if the unit Ampere appears with uneven power so that, for example, the combination As leads to even parity. The following examples yield the parity of the densities of mass, momentum and internal energy according to \begin{equation} \label{parity-examples} [\rho]=\frac{\rm kg}{\rm m^3} \to+1,\qquad [\rho {\bf v}]=\frac{\rm kg}{\rm m^2 s}\to -1,\qquad [\rho e]=\frac{\rm kg}{\rm m\, s^2}\to+1.
\end{equation}
Evidently, while time derivatives alter the parity, spatial derivatives keep the parity unchanged.\\[1ex]

\noindent
{\bf Universal constitutive principles.}
The constitutive modeling of {\it fluids} needs to be consistent with, in particular,
the following two universal principles; cf.\ \cite{TN2003}, \cite{M85}.
\begin{itemize}
\item[(i)]
{\it The principle of material frame indifference}
makes a statement on constitutive functions of objective quantities, saying
that constitutive functions remain invariant under Euclidean transformation.
As an example, consider the constitutive relation of an objective tensor ${\bf T}$, i.e.\
$T_{ij}^\ast = O_{ik} O_{jl} T_{kl}$,
where we assume that ${\bf T}$ is a function of $\nabla {\bf v}$. In two different Cartesian frames we have
\begin{equation}
T_{ij}=f_{ij} (\frac{\partial v_k}{\partial x_l}), \quad \mbox{ respectively }
T_{ij}^\ast=f_{ij}^\ast (\frac{\partial v_k^\ast}{\partial x_l^\ast}).
\end{equation}
Then, with $v_i^\ast =O_{ij} (v_j -  \dot{O}_{jl} O_{kl}(x_k^\ast -b_k)-\dot{b}_j)$,
objectivity of ${\bf T}$ requires
\begin{equation}
\label{f-trafo}
f^\ast (\mathbf{O}(t) \nabla \mathbf{v}\, \mathbf{O}(t)^{\sf T} + \dot{\mathbf{O}}(t) \mathbf{O}(t)^{\sf T})
= \mathbf{O}(t) f(\nabla \mathbf{v}) \mathbf{O}(t)^{\sf T}.
\end{equation}
Now the principle of material frame indifference states that $f^\ast=f$, hence -- in this simple case -- it implies that $f$ is an {\it isotropic function}. Moreover, $f$ can only depend on the symmetric part ${\bf D}$ of $\nabla {\bf v}$ which follows
from \eqref{f-trafo} by choosing ${\bf O}=\bf I$ and $\dot{\mathbf{O}}=-{\bf R}$, where ${\bf R}$ is the anti-symmetric part of $\nabla {\bf v}$.
\item[(ii)]
{\it The entropy principle}, which consists of a certain set of axioms given below, further restricts the generality
of the constitutive relations. \vspace{0.1in}
%This principle is also called the 2$^{\rm nd}$ law of thermodynamics.\vspace{0.1in}
\end{itemize}

\noindent
{\bf Formulation of the entropy principle.} Any solution of the above system of partial
differential equations, composed of the balance equations \eqref{partial-mass}, \eqref{CIImomentum},
\eqref{eq:total-internal-energy-balance} for the different model classes, is
called a {\it thermodynamic process}. Here, by a solution we just
mean functions which satisfy the equations in a local sense, i.e.\ they do not form the solution
of a complete initial boundary value problem.
In particular, the value of a quantity and of its spatial derivative can
thus be chosen independently by adjusting the time derivative appropriately.
With this notion, the 2$^{\rm nd}$ law of thermodynamics consists
of the following axioms.\\[-2ex]
\begin{enumerate}
\item[(I)]
There is an entropy/entropy-flux pair $(\rho s, \Phi)$ as a material dependent quantity,
where $\rho s$ is an objective scalar and $\Phi$ is an objective vector.\vspace{0.05in}
\item[(II)]
The pair $(\rho s, \Phi)$ satisfies the balance equation
\begin{equation}
\pt (\rho s) + \div (\rho s {\bf v} + \Phi ) = \zeta.
\end{equation}
\item[(III)]
The entropy has the physical dimension ${\rm J}\, {\rm kg}^{-1}\, {\rm K}^{-1}={\rm m}^2
{\rm s}^{-2} {\rm K}^{-1}$,
hence is of positive parity. The entropy flux and the entropy production thus have negative parity.\vspace{0.05in}
\item[(IV)]
Any admissible entropy flux is such that\\[0.5ex]
%    (i) the entropy production $\zeta$ is non-negative for every thermodynamic process,\\[0.25ex]
    (i) $\zeta$ consists of a sum of binary products according to
    \begin{equation}\label{M15}
    \zeta = \sum_m \mathcal{N}_m \mathcal{P}_m,
    \end{equation}
    where the $\mathcal{N}_m$ denote quantities of negative parity, while $\mathcal{P}_m$
    refers to positive parity.\\[0.25ex]
    (ii) Each binary product describes a dissipative mechanism which must be introduced in advance. Then,
    $\mathcal{N}_m \mathcal{P}_m \geq 0$ for all $m$ and for every thermodynamic process.\vspace{0.05in}
\item[(V)]
     A thermodynamic process for which $\zeta =0$ is said to be in thermodynamic equilibrium. This statement is to be understood in a pointwise
    sense; in particular, this must not hold everywhere, i.e.\ thermodynamic equilibrium can be attained locally.\\
    A thermodynamic process is called reversible if $\zeta =0$ everywhere.\\[1ex]
\hspace{-0.4in} In addition to these universal axioms, we impose two further ones
which refer to the most general constitutive models we are interested in. These are:\vspace{0.1in}
\item[(VI)]
There are the following {\it dissipative mechanisms} for the fluid mixtures under
consideration: {\it mass diffusion}, {\it chemical reaction},
{\it viscous flow} (including bulk and shear viscosity) and {\it heat conduction}.\vspace{0.05in}
\item[(VII)]
For the class of fluid mixtures under consideration, we restrict the dependence of the entropy according to
\begin{equation}
\label{eq:entropy-fct}
\rho s = \rho \tilde s (\rho e,\rho_1, \ldots ,\rho_N),
\end{equation}
where $\rho \tilde s$ is a {\it strictly concave function} which satisfies the principle of material frame indifference.
By means of this function, we define the
{\it (absolute) temperature} $T$ and {\it chemical potentials} $\mu_i$ according to
\begin{equation}\label{defT-mu}
\frac 1 T := \frac{\partial \rho \tilde s}{\partial \rho e}, \qquad
-\frac{\mu_i}{T} := \frac{\partial \rho \tilde s}{\partial \rho_i}.\vspace{0.1in}
\end{equation}
\end{enumerate}
\noindent
\textbf{Remarks on the Entropy Principle.}
1. Axioms (I) and (II) are identical to those versions given by M\"uller in
\cite{M85} and Alt in \cite{Al09}.

2. Our concept of parity employed in axiom (III) and (IV-i) for
generic quantities was stimulated by Meixner's paper
\cite{Meixner73}. In the present work this concept is simplified
to the above rule on how to determine the parity of a given
quantity. This concept of parity is crucial for a proper
introduction of cross-couplings via entropy invariant mixing
between dissipative mechanisms. This will be taken up in
Section~\ref{sec:cross-coupling} below.

3. There is no axiom (IV-i) in the entropy principles of M\"uller in
\cite{M85} or Alt in \cite{Al09}. Instead, the versions of Müller and Alt rely on the
{\it principle of equipresence} whereupon all
constitutive quantities may depend on the same set of variables.
For example, if the velocity gradient appears in the constitutive
law of the stress tensor it should also appear in the constitutive
law of the entropy density. The entropy flux in their theories is completely determined by
the principle of equipresence.
Our version does not use this principle and thus is much simpler to exploit.
In particular for advanced constitutive models the equipresence principle requires an enormous
computational effort, cf.\ \cite{Liu}.
However, without equipresence, more preliminary knowledge is required.

4. Axiom (IV-ii) is a strengthened form of the principal of
detailed balance, which essentially says that in equilibrium all
individual dissipative mechanisms are themselves in equilibrium,
i.e.\ zero entropy production implies that all binary products
vanish. The latter property can be guaranteed if every binary
product has non-negative entropy production for any thermodynamic
process. This is precisely the content of axiom (IV-ii). The valid
use of this strengthened form requires a careful choice of the
decomposition of the entropy production into binary products such
that each product contains a co-factor which can be varied
independently of the others; cf.\ also Remark~6 and observe
that the representation of the entropy production is not uniquely
determined. For example, in the following we will show that
chemical reactions contribute with two binary products to the
entropy production. There is one product due to the generation of
thermal energy and another one, which is new, related to the transfer of momentum.

5. In total, axiom (IV) implies that {\it the entropy inequality} holds,
stating that $\zeta \geq 0$ for all thermodynamic processes.
The entropy inequality is also called {\it the 2$^{\rm nd}$ law of thermodynamics}.

6. Axiom (VI) requires a priori information on the underlying physics, which is to be modeled,
in terms of the dissipative mechanisms. This also includes knowledge of the constitutive quantities (i.e., fluxes or rates) involved in the respective dissipative mechanism. For instance, for the mechanism of multicomponent mass diffusion, the set of $N-1$ diffusive fluxes ${\bf j}_i$ is associated to this mechanism as the constitutive quantity.
This a priori information determines the decomposition of the binary products in \eqref{M15}
into two specific factors.

7. The exploitation of Axiom (V) requires preliminary knowledge on equilibrium thermodynamics,
where those quantities that vanish in thermodynamic equilibrium are to be identified.
For example, we know by experience in advance that in thermodynamic equilibrium we must have
\begin{equation}\label{equi1}
R_a =0 , \quad {\bf u}_i =0 ,\quad \nabla {\bf v}_i + \nabla {\bf v}_i^{\sf T} = 0,
\quad  \nabla T=0.
\end{equation}
Correspondingly, these conditions motivate the constitutive axiom
(VI), saying that the mechanisms that drive a body to equilibrium
are: chemical reactions, mass diffusion, viscous flow and heat
conduction. These mechanisms contribute with $N_R$, $N-1$, $N$,
respectively 1 binary product to the entropy production. Linear
relations between the factors $\mathcal{N}_m$ and $\mathcal{P}_m$
with appropriate coefficients provide the simplest possibility to
guarantee non-negativity of the entropy productions. However, note
that this does not imply linearity with respect to the variables.

8. The assumptions on the constitutive function $\rho
\tilde s$ in Axiom (VII) are crucial since they fix the class of
fluid mixtures under consideration. In particular, the omission of
density gradients in the list of variables for $\rho \tilde s$
excludes nonlocal effects, e.g.\ due to mesoscopic forces, which
are required for so-called phase field models. Here, we consider
single-phase fluid mixtures. The choice of $\rho e$ and the
$\rho_i$ as independent variables in \eqref{eq:entropy-fct} is
generic, since these are the balanced quantities. A further
dependence directly on the ${\bf v}_i$ is not possible due to
material frame indifference. But a dependence on the diffusion
velocities ${\bf u}_i$ is possible. Furthermore, a dependence on
$\div {\bf v}$ and/or $\nabla T$ would also be consistent with the
principle of material frame indifference. However, if we were to
include such dependencies in the entropy function they will be
ruled out by the exploitation of the entropy principle. The proof
of this proposition is left to the reader.
%In any case the restriction of the constitutive class according to
%\eqref{eq:entropy-fct} is called \textit{assumption of local equilibrium}.

9. The concavity postulate (VII) can be motivated as a generalization of a well-known
result of equilibrium thermodynamics for homogeneous bodies. To
this end we consider a fluid mixture in a container with time-dependent volume $V(t)$. The container is closed by a movable
piston subjected to a constant pressure $p_0$, and the outer
surface of the whole device is at constant temperature $T_0$. In
this case the equations of balance for total energy $E$ and
total entropy $S$ read
\begin{equation}\label{conc13}
    \frac{dE}{dt}+p_0\frac{dV}{dt}=-\oint {\bf q}\cdot{\bf
    n}~da\quad\textrm{and}\quad
    \frac{dS}{dt}+\frac{1}{T_0}\oint {\bf q}\cdot{\bf
    n}~da\geq0~.
\end{equation}
Elimination of the heat flux leads to the inequality
\begin{equation}\label{conc14}
    \frac{dA}{dt}\leq 0\quad\textrm{with}\quad
    A=E-T_0 S+p_0V.
\end{equation}
Thus the quantity $A$, which is called available free energy, can
only decrease and assumes its minimum in equilibrium.

Now we study for fixed external values $T_0$ and $p_0$ a homogeneous process inside the
mixture which is characterized by the state
$(T(t),p(t),m_1(t),...,m_N(t))$ and ask for its equilibrium state. Note that $A$ is neither the Gibbs energy $G=E-TS+pV$
nor the Helmholtz free energy $\Psi=E-T S$. Rather, we have $A=G+(T-T_0)S+(p_0-p)V$.
Moreover, in a homogenous system we can write $\Psi = \rho {\psi} V$.

To keep this motivation short we only consider the
non-reactive case, where the masses $m_i(t)$ are independent of
time. Then the relevant variables of $A$ are $T$ and $p$ and the
minima follow from $\partial A/\partial T=0$ and $\partial
A/\partial p=0$. The physically obvious result is that $T=T_0$ and $p=p_0$
give the necessary condition for equilibrium.

Its derivation makes use of three identities which are valid for homogenous systems, viz.\
\begin{equation}\label{conc15}
    \frac{\partial G(T,p,m_1,...,m_N)}{\partial
    T}=-S,\quad \frac{\partial G(T,p,m_1,...,m_N)}{\partial
    p}=V,\quad \frac{\partial G(T,p,m_1,...,m_N)}{\partial
    m_i}=\mu_i.
\end{equation}
The proof of these relations starts from
$G(T,p,m_1,...,m_N)=V(\rho\hat\psi(T,\frac{m_1}{V},...,\frac{m_N}{V})+p)$
with $V=V(T,p,m_1,...,m_N)$. For example, we have
\begin{equation}\label{conc16}
    \frac{\partial G}{\partial m_i}=\frac{\partial V}{\partial
    m_i}(\rho\hat\psi+p)+
    \sum\limits_{i=1}^N
    \frac{\partial\rho\hat\psi}{\partial\rho_j}(\frac{\delta_{ij}}{V}-\frac{m_j}{V^2}
    \frac{\partial V}{\partial
    m_i})=\frac{\partial\rho\hat\psi}{\partial\rho_i}=\mu_i.
\end{equation}
We proceed to characterize the equilibria. The sufficient conditions that $A$
assumes a minimum at $T=T_0$, $p=p_0$ is the positive definiteness of the
matrix of its second derivatives. We obtain, in the simple case
with constant masses, the condition that
\begin{equation}\label{conc17}
    \left(%
\begin{array}{cc}
\ds  \frac{\partial S}{\partial T} & \ds -\frac{\partial V}{\partial T} \\[2ex]
\ds  -\frac{\partial V}{\partial T} & \ds -\frac{\partial V}{\partial p} \\
\end{array}%
\right)\quad\textrm{is positive definite.}
\end{equation}
If chemical reactions and, hence, time-dependent $m_i(t)$ are included above, the
condition that the matrix
\[
\Big( \frac{\partial \mu_i}{\partial \rho_j}\Big)_{i,j\in \{1,\ldots ,N \} }
\quad \mbox{ is positive definite}
\]
is also derived as an additional part of the sufficient conditions for a minimum.
We refrain from giving the details here.

These conditions coincide with corresponding inequalities which will be derived in the following section
 from the concavity postulate.
But they will be more general in two aspects: They do not refer to homogeneous
systems and they also hold in non-equilibrium.

10. The definition of temperature is accompanied by some subtleties. From a theoretical point of view there is a certain freedom how to define temperature. Because a quantity that we might call temperature does not explicitly appear in the equations of balance it cannot be introduced as a primitive quantity in thermodynamics. At first glance our definition \eqref{defT-mu}$_1$ seems natural, because it agrees with the corresponding definition of equilibrium thermodynamics. But the key point of the present approach is the choice of the total internal energy $\rho e$ according to \eqref{def:total-internal-energy} as a variable in \eqref{defT-mu}$_1$. Other choices are possible.
Interestingly, de~Groot and Mazur \cite{dGM} mention our choice \eqref{def:total-internal-energy}
but due to some unidentified reasons they finally prefer to choose $\rho e_{\rm int}$ from
\eqref{tot-int-energy} as a variable in the entropy function. M\"{u}ller proceeds differently in \cite{M85}. He starts with an empirical temperature and after a chain of reasonings M\"{u}ller obtains our definition \eqref{def:total-internal-energy} as a result. Let us also mention here the situation in Boltzmann's kinetic theory. Recall that this theory is restricted to monatomic ideal gases. Here the temperature is defined via $p/\rho$, where pressure and density are directly related to the distribution function. In equilibrium our choice obviously agrees with the kinetic definition, but in nonequilibrium there is a difference, see \cite{IM-Ruggeri} for more details.
However, this fact does not generate any problem. One simply has to deal with two different
temperature measures.

%\section{Entropy flux and entropy production for the viscous Cahn-Hilliard equation}
%\label{sec:VCH}

11. The axioms constituting the above entropy principle do not
lead to a uniquely defined model, even for the same set of independent variables.
The following simple example from \cite{DG13} shows that this non-uniqueness provides useful flexibility
to extend a given constitutive model, in this case leading to the viscous Cahn-Hilliard equation.
To this end we consider a rigid heat conductor
in one space dimension that may develop two coexisting phases. At
first we assume that the phase transition can be described by the
classical Cahn-Hilliard model. The variables are the concentration
$u(t,x)$ of the phases and the energy density $e(t,x)$. The
coupled system of partial differential equations rely on the two
conservation laws
\begin{equation}\label{VCH1}
    \partial_t u+\partial_x J=0,\qquad\partial_t e+\partial_x q=0.
\end{equation}
The diffusion flux $J$ and the heat flux $q$ are the constitutive
quantities which we determine by means of our system of entropy
axioms.

We assume that the free energy function is given by $\psi=F(e,u)+\beta/2 u_x^2$,
$\beta>0$, with corresponding chemical potential $\mu=\partial_u F-\beta u_{xx}$.
According to the procedure from above we calculate the entropy production to obtain
at first
\begin{equation}\label{VCH2}
    \zeta=\partial_x(\varphi-\frac{q-J\mu}{T}-\frac{\beta u_x \partial_t
    u}{T})+q\, \partial_x(\frac{1}{T})-J(\frac{\mu}{T}).
\end{equation}
Next we define an admissible entropy flux $\varphi$ so that the
entropy production $\zeta$ assumes the required binary form:
\begin{equation}\label{VCH3}
\Phi=\frac{q-J\mu}{T}+\frac{\beta u_x \partial_t
    u}{T},\qquad
    \zeta=q\, \partial_x(\frac{1}{T})-J\partial_x(\frac{\mu}{T}).
\end{equation}
Ignoring cross effects the most simple constitutive equations
compatible with a non-negative entropy production read
\begin{equation}\label{VCH4}
    q=a\partial_x(\frac{1}{T}),\qquad J=-M\partial_x(\frac{\mu}{T}),\quad
    a,M>0.
\end{equation}
The constitutive equation \eqref{VCH4}$_1$ is the Fourier law and
the diffusion flux \eqref{VCH4}$_2$ leads in the isothermal case
to the classical Cahn-Hilliard equation. Next we study the
thermodynamic setting of the viscous Cahn-Hilliard equation, which
results from the diffusion law
\begin{equation}\label{VCH5}
    J=-M\partial_x(\frac{\mu+\gamma\partial_t u}{T}),\qquad
    M,\gamma >0.
\end{equation}
At first glance one might think that \eqref{VCH5} is a consequence
of a more general free energy function including time derivatives
of the concentration in the state space of the material. However,
we will show that this is by no means the case. Let us at first
multiply the conservation law \eqref{VCH1}$_2$ by
$\gamma\partial_t u/T$, where $\gamma$ is a positive constant. We
obtain
\begin{equation}\label{VCH6}
    \frac{\gamma}{T}(\partial_t u)^2+\partial_x(\frac{\gamma}{T}J\partial_t
    u)-J\partial_x(\frac{\gamma}{T}\partial_t u)=0.
\end{equation}
Adding this identity to the right-hand side of the entropy
production \eqref{VCH2} yields
\begin{equation}\label{VCH7}
    \zeta=\partial_x(\Phi-\frac{q-J\mu}{T}-\frac{\beta u_x \partial_t u}{T}+\frac{\gamma}{T}J\partial_t u)
    +q\, \partial_x(\frac{1}{T})-J(\frac{\mu+\gamma\partial_t u}{T})+\frac{\gamma}{T}(\partial_t u)^2.
\end{equation}
We now redefine the entropy flux and consequently obtain a new
entropy production:
\begin{equation}\label{VCH8}
    \Phi=\frac{q-J\mu}{T}+\frac{\beta u_x \partial_t
    u}{T}-\frac{\gamma}{T}J\partial_t u~,\qquad
\zeta=q\, \partial_x(\frac{1}{T})-J(\frac{\mu+\gamma\partial_t
u}{T})+\frac{\gamma}{T}(\partial_t u)^2.
    \end{equation}
Obviously the extended diffusion flux \eqref{VCH5} of the
viscous Cahn-Hilliard model leads likewise to a non-negative entropy
production, which is now larger as the one in the classical
Cahn-Hilliard case because the third term in \eqref{VCH8}$_2$ adds a further dissipative
mechanism.
\section{Exploitation of the entropy principle}
\label{sec:exploitation}
{\bf Entropy inequality.}
Starting point for the evaluation of the $2^{\rm nd}$ law is the entropy balance,
which we use in the form
\[
\zeta = \pt (\rho s) + {\bf v}\cdot \nabla (\rho s)+\rho s \, \div
{\bf v} + \div \Phi.
\]
We insert $\rho s = \rho \tilde s (\rho e,\rho_1, \ldots ,\rho_N)$
into the first two terms  and carry out the differentiations. Then
we eliminate the resulting time derivatives by means of the balance equations \eqref{partial-mass} and \eqref{eq:total-internal-energy-balance1}.
We introduce temperature and chemical potentials according to the definitions \eqref{defT-mu}.
Then, after straightforward computations, a first representation of the entropy production is
\begin{align}
\label{entropy-production1}
\zeta =
\div \Big( \Phi - \frac{\tilde{\bf q}}{T} + \sum_i \frac{\rho_i \mu_i {\bf u}_i}{T} \Big)
- \frac 1 T \Big(  \rho e - T \rho s - \sum_i \rho_i \mu_i  \Big)\, \div {\bf v}
- \frac 1 T \sum_i  P_i \, \div {\bf v}_i\\
- \frac 1 T \sum_i {\bf u}_i \cdot \Big( \rho_i T \nabla \frac{\mu_i}{T}
+ {\bf f}_i - r_i {\bf v}_i  + \frac{1}{2} r_i{\bf u}_i \Big)
- \frac 1 T \sum_{a=1}^{N_R} R_a \mathcal{A}_a +
\frac 1 T \sum_i \stress_i^\circ : {\bf D}_i^\circ
+ \tilde{\bf q} \cdot \nabla \frac 1 T. \nonumber
\end{align}
In \eqref{entropy-production1}, we have rewritten the term $\sum_{i=1}^N r_i \mu_i$ as
$\sum_{a=1}^{N_R} R_a \mathcal{A}_a$, introducing the so-called {\it affinities}
%$\mathcal{ A}_a$, viz.\
\begin{equation}
\label{affinities}
\mathcal{ A}_a = \sum_{i=1}^N \nu_i^a M_i \mu_i.
\end{equation}
This reflects the fact that every chemical reaction relates to a dissipative mechanism.

Now, for the derivation of the entropy production in the class-II model one has to account for
the fact that, while for instance $\div {\bf u}_i$ can be chosen independently of ${\bf u}_i$ in exploiting the entropy principle, the defining relations
${\bf u}_i={\bf v}_i-{\bf v}$ lead to the constraints $\div {\bf u}_i=\div {\bf v}_i-\div {\bf v}$.
These constraints are incorporated into the exploitation using Lagrange multipliers $\Lambda_i =\Lambda_i (T, \rho_1, \ldots ,\rho_N)$ and adding the terms
\[
\sum_i \frac{\Lambda_i}{T} (\div {\bf v}_i - \div {\bf v} - \div {\bf u}_i)
\]
to the entropy production; cf.\ Lemma~7.3.1 in \cite{Liu}
concerning a proof that the entropy inequality with this addition and without constraints is equivalent to the original inequality, but evaluated under the constraints.
Finally, we introduce the {\it free energy density} $\rho\psi:=\rho e - T \rho s$.
Then, after rearrangement of terms, the entropy production for a class-II
model reads as
\begin{align}
\zeta =
\div \Big( \Phi - \frac{\tilde{\bf q}}{T} + \sum_i \frac{(\rho_i \mu_i -\Lambda_i) {\bf u}_i}{T} \Big)
- \frac 1 T \Big(  \rho \psi - \sum_i (\rho_i \mu_i -\Lambda_i)  \Big)\, \div {\bf v}
\nonumber
\\
- \frac 1 T \sum_i \big( P_i - \Lambda_i \big)\, \div {\bf v}_i
- \frac 1 T \sum_i {\bf u}_i \cdot \Big( \rho_i T \nabla \frac{\mu_i}{T}
+ {\bf f}_i - r_i {\bf v}_i  + \frac{1}{2} r_i{\bf u}_i  -T \nabla \frac{\Lambda_i}{T} \Big)
 \label{entropy-inequality-II} \\
- \frac 1 T \sum_{a=1}^{N_R} R_a \mathcal{A}_a +
\frac 1 T \sum_i \stress_i^\circ : {\bf D}_i^\circ
+ \tilde{\bf q} \cdot \nabla \frac 1 T. \nonumber
\end{align}
%$\stress_i^{\rm irr}=-\Pi_i {\bf I} + \stress_i^\circ$.
%
%
Now, we proceed as follows.
\begin{enumerate}
\item[(a)]
To satisfy axiom~(IV-i), we choose the entropy flux as
\begin{equation}\label{entropyfluxIIa}
\Phi = \frac{\tilde{\bf q}}{T} - \sum_i \frac{(\rho_i \mu_i -\Lambda_i) {\bf u}_i }{T},
\end{equation}
where the $\Lambda_i$ will be determined below.\vspace{0.05in}
\item[(b)] The entropy production reduces to a sum of six binary products and we first choose
\[
R_a = 0, \quad \nabla T =0,\quad \mathbf{u}_i = 0,\quad \mathbf{D}_i^\circ =0;
\]
recall that ${\rm tr}\, ({\bf D}_i )=\div {\bf v}_i$ and $P_i=p_i + \Pi_i$, where
$\Pi_i =0$ in equilibrium.
In this case the entropy inequality becomes
\[
\ts
\qquad T \zeta =
(\sum_i (\rho_i \mu_i -\Lambda_i) - \rho \psi )\, \div {\bf v}
+  \sum_i ( p_i - \Lambda_i )\, \div {\bf v}_i
- \sum_i \Pi_i \, \div {\bf v}_i \geq 0,
\]
where $\div {\bf v}$ and all $\div {\bf v}_i$ can be varied independently in the exploitation.
For $\div {\bf v}=\div {\bf v}_i =0$ the processes are in equilibrium, where the entropy production assumes a minimum. This implies
\begin{equation}
\label{GDi}
\ts
\rho \psi - \sum_i (\rho_i \mu_i -\Lambda_i) = 0, \qquad  p_i - \Lambda_i  = 0
\quad \mbox{ and } \quad - \sum_i \Pi_i \, \div {\bf v}_i \geq 0.
\end{equation}
Inserting $\Lambda_i=p_i$ from \eqref{GDi}$_2$ into \eqref{GDi}$_1$ yields
\begin{equation}
\label{Gibbs-Duhem}
\ts
\rho \psi + p - \sum_i \rho_i \mu_i = 0.
\end{equation}
Relation \eqref{Gibbs-Duhem} is the {\it Gibbs-Duhem equation} which is well-known and also holds true for class-I models.
Since it  only involves equilibrium quantities, the Gibbs-Duhem equation also holds for general thermodynamic processes. Note that here an important result is the fact that only a {\it single} Gibbs-Duhem equation results.

Inserting $\Lambda_i=p_i$ from \eqref{GDi}$_2$ into \eqref{entropyfluxIIa} yields
\begin{equation}\label{entropyfluxII}
\Phi = \frac{\bf q}{T} - \sum_i \frac{\rho_i \mu_i  {\bf u}_i}{T},
\end{equation}
if we define the {\it heat flux} for the class-II model as we already noted in
\eqref{def:reduced-heat-flux}, namely
\[
\ts
{\bf q}:= \sum_i  \big({\bf q}_i + (\rho_i  e_i + p_i) {\bf u}_i \big).\vspace{0.1in}
\]
\end{enumerate}

\noindent
The remaining part is the sum of four binary products
which cause entropy production. These are the four postulated dissipative
mechanisms, namely (i) mass diffusion, (ii) chemical
reactions, (iii) viscous flow (bulk and shear viscosity) and (iv) heat conduction. The second
law gives restrictions for constitutive equations of these
dissipative mechanisms,
where the axiom~(VI-ii) requires all individual contributions to be non-negative.

With (a) and (b) above, the entropy production of the class-II model assumes its final form:
\begin{align}
 \zeta = &- \sum_i {\bf u}_i \cdot \Big( \rho_i \nabla
 \frac{\mu_i}{T} + \frac 1 T ({\bf f}_i - r_i {\bf v}_i
 + \frac{1}{2} r_i {\bf u}_i -\nabla p_i) - (\rho_i  e_i + p_i) \nabla \frac 1 T \Big)
\nonumber\\
&- \frac 1 T \sum_{a=1}^{N_R} R_a \mathcal{A}_a +
\frac 1 T \sum_i \stress_i^{\rm irr} : {\bf D}_i +
\sum_i {\bf q}_i \cdot \nabla \frac 1 T. \hspace{1in}
\label{entropy-inequalityII}
\end{align}\\

\noindent {\bf Remarks on the derivation of the entropy
production.} 1. The specific form of the intermediate
representation \eqref{entropy-production1} is not unique. Above,
we employed the relation
\[
\div (\rho_i {\bf v}_i) = {\bf v}\cdot \nabla \rho_i
 + {\bf u}_i \cdot \nabla \rho_i + \div (\rho_i {\bf u}_i)
\]
to arrive at \eqref{entropy-production1}.\\[0.5ex]
\indent
2. If the entropy balance is exploited with the thermal
energy balance in the form
\eqref{eq:total-internal-energy-balance} and if, furthermore, the
term $\div {\bf v}$ is eliminated using the relation
\begin{equation}
\label{eq:sep-divv}
\div {\bf v}=\sum_i y_i  \, \div {\bf v}_i +\sum_i{\bf u}_i \cdot \nabla y_i,
\end{equation}
then no Lagrange multipliers are required. This way, the derivation of the Gibbs-Duhem
relation could be somewhat simplified, avoiding the appearance of a zero contribution
in the entropy production formula. Indeed, the entropy production then reads as
\begin{align}
& \qquad \zeta =
 \frac 1 T \sum_i \big( \stress_i^{\rm irr} - y_i (\rho \psi +p - \sum_k \rho_k \mu_k){\bf I} \big) : {\bf D}_i +
\sum_i {\bf q}_i \cdot \nabla \frac 1 T
- \frac 1 T \sum_{a=1}^{N_R} R_a \mathcal{A}_a \nonumber\\
& - \sum_i {\bf u}_i \cdot \Big( \rho_i \nabla
 \frac{\mu_i}{T} + \frac 1 T ({\bf f}_i - r_i {\bf v}_i
 + \frac{r_i}{2}  {\bf u}_i -\nabla p_i + (\rho \psi +p - \sum_k \rho_k \mu_k) \nabla y_i) - (\rho_i  e_i + p_i) \nabla \frac 1 T \Big).
\nonumber
\end{align}
Linear closure for the viscous stresses then yields
\[
 \tr ( \stress_i^{\rm irr})=2 \sum_k \eta_{ik} {\bf D}_k, \qquad
  - \Pi_i =3 y_i (\rho \psi +p - \sum_k \rho_k \mu_k) +2 \sum_k \lambda_{ik} \div {\bf v}_k
\]
with positive (semi-)definite matrices $[\eta_{ik}]$, $[\lambda_{ik}]$.
In equilibrium, we have $\div {\bf v}_i=0$ and $\Pi_i=0$ for all $i$, hence also
$\rho \psi +p - \sum_k \rho_k \mu_k=0$. The latter then necessarily holds independent
of the equilibrium assumption, hence the Gibbs-Duhem equation \eqref{Gibbs-Duhem} is valid.

Despite of the elegance of this variant,
we nevertheless preferred the derivation given above, using in particular $\tilde{\bf q}$ instead of ${\bf q}$
because the form of the heat flux according
to \eqref{def:reduced-heat-flux} is not a priori clear. In fact, this definition for ${\bf q}$
is only motivated by the final form which it implies for the entropy flux, i.e.\
\eqref{entropyfluxII}, which resembles the classical one from TIP for mixtures.

Apparently, relation \eqref{eq:sep-divv} is the
only form in which the interdependencies between the diverse
velocity divergences are not interfering with the exploitation.
Note that the Lagrange multipliers could still be used in which
case $\Lambda_i =0$ for all $i$.\\[0.5ex]
\indent
3. It is interesting to observe that the entropy flux
can be written as
\[
\Phi = \sum_i \Big( \frac{{\bf q}_i}{T} + \frac{\rho_i  e_i + p_i - \rho_i \mu_i}{T}  {\bf u}_i \Big).
\]
Let us note that in the case of so-called simple mixtures,
introduced in detail in section~\ref{sec:free-energies} below, the
quantity $(\rho_i  e_i + p_i - \rho_i \mu_i)/T$ is nothing but the
partial entropy $\rho_i s_i$. This nicely shows the two different
contributions to the entropy flux in a class-II mixture.  Note
also that the total flux of entropy then attains the form
\[
\rho s {\bf v} + \Phi = \sum_i \big( \rho_i  s_i  {\bf v}_i +  \frac{{\bf q}_i}{T} \big),
\]
thus being just the sum of the individual total entropy fluxes of
the constituents,
 each having $\Phi_i = {\bf q}_i / T$ as the non-convective flux of partial entropy.\\[0.5ex]
\indent
4. Note that in \eqref{entropy-inequalityII} the total
heat flux according to \eqref{def:reduced-heat-flux} contributes
with two terms: The first term involves heat transport due to
conduction, while the second term describes diffusive heat
transport and is proportional to the diffusion velocities.
Therefore, in order to collect all co-factors of the ${\bf u}_i$
which, hence, belong to the mechanism of diffusion, we shuffled
those terms to the first
binary product.\\[1ex]
\noindent
Based on the entropy production \eqref{entropy-inequalityII} we are going to derive mixture
models for fluid mixtures of increasing physical complexity.\\[1ex]
\noindent
{\bf Helmholtz free energy and its derivatives.} Recall that absolute
temperature and chemical potentials are introduced according to
the definitions \eqref{defT-mu}. The specific (Helmholtz) free energy
$\psi=e -T s$ is useful if we change the variables from
$\{\rho e,\rho_1,...,\rho_N\}$ to $\{T,\rho_1,...,\rho_N\}$ with
$\rho e=\rho \hat e (T,\rho_1,\ldots ,\rho_N)$ by solving
\eqref{defT-mu}$_1$ for $\rho e$. Then, for
$\psi=\hat\psi(T,\rho_1,...,\rho_N)$ we obtain from
\eqref{defT-mu} the relations
\begin{equation}\label{rho-psi-partial-derivatives}
 \rho s=-\frac{\partial \rho
\hat\psi}{\partial T},\qquad \mu_i=\frac{\partial \rho
\hat\psi}{\partial \rho_i}\quad\textrm{implying}\quad
e=-T^2\frac{\partial}{\partial T}(\frac{\hat\psi}{T})~.
\end{equation}
The Gibbs-Duhem equation \eqref{Gibbs-Duhem} serves to determine
the pressure $p$ from the free energy density and for this reason we
write
\begin{equation}\label{HFE1}
    p=-\rho\psi+\sum\limits_{i=1}^N \rho_i \mu_i.
\end{equation}
Two further sets of variables are in common use. Using
$\rho_i=\rho y_i$ we may change the variables from
$\{T,\rho_1,...,\rho_N\}$ to $\{T,\rho,y_1,...,y_{N-1}\}$ and
indicate this by writing, say,
\begin{equation}\label{HFE2}
    \rho\psi=\rho\bar\psi(T,\rho,y_1,...,y_{N-1}).
\end{equation}
The generic function $F(T,\rho,y_1,...,y_{N-1})=\hat F(T,\rho_1,...,\rho_N)$ satisfies
the simple rule $F_\rho :=\frac{\partial F}{\partial \rho}
=\sum_{j=1}^N  y_j \hat F_{\rho_j}$ which is used to obtain from \eqref{HFE1}
\begin{equation}\label{HFE3}
    p=\bar p(T,\rho,y_1,...,y_{N-1})=\rho^2\frac{\partial\bar\psi}{\partial\rho}.
\end{equation}
Next we solve \eqref{HFE3} for $\rho=\check{\rho}(T,p,y_1,...,y_{N-1})$
and calculate from $e=\bar e(T,\rho,y_1,...,y_{N-1})$, obtained
from \eqref{rho-psi-partial-derivatives}$_3$, the function
$e=\check e(T,p,y_1,...,y_{N-1})$. Finally we introduce the specific heat
at constant densities, $c_v$, and the specific heat at constant
pressure, $c_p$, by
\begin{equation}\label{HFE4}
    c_v=\frac{\partial \hat e}{\partial T}\quad\textrm{and}\quad
    c_p=\frac{\partial \check e}{\partial T}+p\frac{\partial}{\partial
    T}(\frac{1}{\check \rho}).
\end{equation}
The specific heat $c_p$ is related to $c_v$ by
\begin{equation}\label{HFE5}
    c_p=c_v+T
   \frac{\partial (1/\check \rho)}{\partial T} \frac{\partial \bar p}{\partial T}=
   c_v+\frac{T}{\rho^2}(\frac{\partial \check \rho}{\partial T})^2 \frac{\partial \bar p}{\partial \rho},
\end{equation}
The proof of \eqref{HFE5}$_1$ starts from the two definitions of
$c_p$ and $c_v$ and proceeds via $e=\psi+Ts$ and the relation
$\partial_T \check s=1/T \partial_T \check e+ \partial_T (1/\check \rho) \partial_T \bar p$. Equation \eqref{HFE5}$_2$ holds due to
the identity $\partial_T \bar p=- \partial_T \check \rho / \partial_p \check \rho$.\\[1ex]
\noindent
{\bf Concavity of the entropy function.}
This paragraph addresses the exploitation of the concavity postulate.
Here derivatives are indicated by indices, except in
the main results. Moreover, note that two different
representations of the entropy function will be used, viz.
\begin{equation}\label{con1}
    \rho s=\rho \tilde s(\rho e, \rho_1,\rho_2,...,\rho_{N})=
    \rho\hat s(T, \rho_1,\rho_2,...,\rho_N).
\end{equation}

The entropy postulate (VII) states that the entropy function
\eqref{con1}$_1$ is strictly concave with respect to all variables. Thus
the $(N+1)\times(N+1)$ matrix $-\bf M$ of its second derivatives
has the property that
\begin{equation}\label{conc2}
    \bf M=-\left(%
\begin{array}{cccccc}
  (\rho \tilde s)_{\rho e\rho e} & (\rho \tilde s)_{\rho e\rho_1} & \ldots & (\rho \tilde s)_{\rho e\rho_N} \\[0.5ex]
  (\rho \tilde s)_{\rho_1\rho e} & (\rho \tilde s)_{\rho_1 \rho_1} & \ldots & (\rho \tilde s)_{\rho_1\rho_N}  \\
  . & . & \ldots & . \\
  . & . & \ldots & . \\
  . & . & \ldots & . \\
  (\rho \tilde s)_{\rho_N\rho e} & (\rho \tilde s)_{\rho_N \rho_1}
  & \ldots & (\rho \tilde s)_{\rho_N\rho_N}\\
\end{array}%
\right)\qquad \textrm{is positive definite},
\end{equation}
implying important inequalities. The positive definiteness of $\bf
M$ is exploited by Sylvester's criterion: All principal
determinants must be positive, i.e.
\begin{equation}\label{conc3}
    -(\rho \tilde s)_{\rho e\rho e}>0,\;\;
    -((\rho \tilde s)_{\rho e\rho e}(\rho \tilde
    s)_{\rho_1\rho_1}-((\rho \tilde s)_{\rho e\rho_1})^2)>0,\;\;
    -\left|\begin{array}{cccccc}
  (\rho \tilde s)_{\rho e\rho e} & (\rho \tilde s)_{\rho e\rho_1} &
  (\rho \tilde s)_{\rho e\rho_2} \\[0.5ex]
  (\rho \tilde s)_{\rho_1\rho e} & (\rho \tilde s)_{\rho_1 \rho_1} &
   (\rho \tilde s)_{\rho_1\rho_2} \\[0.5ex]
    (\rho \tilde s)_{\rho_2\rho e} & (\rho \tilde s)_{\rho_2 \rho_1}
  & (\rho \tilde s)_{\rho_2\rho_2}
\end{array}\right|>0,
    \end{equation}
and so on.
These conditions are necessary and sufficient for the strict concavity of
the entropy function. By means of \eqref{defT-mu} we can also
write
\begin{equation}\label{conc4}
    -(\frac{1}{\tilde T})_{\rho e}>0,\quad
    (\frac{1}{\tilde T})_{\rho e}(\frac{\tilde\mu_1}{\tilde T})_{\rho_1}
    -((\frac{1}{\tilde T})_{\rho_1})^2>0,\quad
    \left|\begin{array}{cccccc}
  (-\frac{1}{\tilde T})_{\rho e} & (-\frac{1}{\tilde T})_{\rho_1} &
  (-\frac{1}{\tilde T})_{\rho_2} \\[0.5ex]
  (-\frac{1}{\tilde T})_{\rho_1} & (\frac{\tilde\mu_1}{\tilde T})_{\rho_1} &
   (\frac{\tilde\mu_1}{\tilde T})_{\rho_2} \\[0.5ex]
    (-\frac{1}{\tilde T})_{\rho_2} & (\frac{\tilde\mu_1}{\tilde T})_{\rho_2}
  & (\frac{\tilde\mu_2}{\tilde T})_{\rho_2}
\end{array}\right|>0,~...
    \end{equation}
Next, we rewrite these inequalities in the variables $T,
\rho_1,\rho_2,...,\rho_N$. To this end we consider a generic
relation $\tilde F(\rho e, \rho_1,\rho_2,...,\rho_N)=\hat F(T,
\rho_1,\rho_2,...,\rho_N)$ with $\rho e=\rho \hat e(T,
\rho_1,\rho_2,...,\rho_N)$. The corresponding derivatives are
related to each other by
\begin{equation}\label{con5}
    \hat F_{T}=\tilde F_{\rho e}(\rho\hat e)_{T},\qquad
    \hat F_{\rho_i}=\tilde F_{\rho e}(\rho\hat e)_{\rho_i}+\tilde
    F_{\rho_i}.
\end{equation}
We insert $(\rho\hat e)_{T}=\rho\hat e_{T}=\rho c_v$ from \eqref{HFE4}$_1$.
Furthermore, using the identity $(\rho\hat e)_{\rho_i}=\hat\mu_i-T\hat\mu_{i T}$ which
follows by differentiation of \eqref{rho-psi-partial-derivatives}, we obtain
\begin{equation}\label{con6}
    \tilde F_{\rho e}=\frac{1}{\rho c_v}\hat F_{T},\qquad
    \tilde F_{\rho_i}=\hat F_{\rho_i}-\frac{1}{\rho c_v}(\hat\mu_i-T\hat\mu_{i T})\hat F_{T}.
\end{equation}
Choosing $\tilde F=-1/\tilde T$ and $\tilde F=\tilde \mu_i/\tilde
T$, respectively, we derive
\begin{equation}\label{con7}
    (-\frac{1}{\tilde T})_{\rho e}=\frac{1}{\rho T^2 c_v},\;\;
    (-\frac{1}{\tilde T})_{\rho_i}=-\frac{\hat\mu_i-T\hat\mu_{i T}}{\rho T^2
    c_v},\;\;
    (\frac{\tilde\mu_i}{\tilde T})_{\rho_j}=\frac{\hat \mu_{i \rho_j}}{T}
    +\frac{(\hat\mu_i-T\hat\mu_{i T})(\hat\mu_j-T\hat\mu_{j T})}{\rho T^2 c_v}.
\end{equation}
Due to these representations we finally may rewrite the
inequalities \eqref{conc4} as the simple statement
\begin{equation}\label{conc8}
    c_v>0\qquad\textrm{and}\qquad \left(\frac{\partial
    \hat\mu_i}{\partial\rho_j}\right)_{i,j\in\{1,...,N\}}
    \;\;\textrm{ is positive definite}.
\end{equation}
From \eqref{conc8} we may derive further important inequalities
concerning the constitutive functions for the specific thermal
energy $e$ and the pressure $p$. We again use $\rho_i=\rho y_i$ to
change the variables from $\{T,\rho_1,...,\rho_N\}$ to
$\{T,\rho,y_1,...,y_{N-1}\}$. Recall that the generic function $\bar
F(T,\rho,y_1,...,y_{N-1})=\hat F(T,\rho_1,...,\rho_N)$ satisfies
the simple rule $\bar F_\rho=\sum_{j=1}^N y_j \hat F_{\rho_j}$ which
is used to obtain from the Gibbs-Duhem equation \eqref{HFE1} the relations
\begin{equation}\label{conc11}
     \bar p_\rho=-\bar\psi-\rho\bar\psi_\rho+
    \sum\limits_{i=1}^N y_i \mu_i+\rho\sum\limits_{i=1}^N y_i \mu_{i\rho}
    =\rho\sum\limits_{i,j=1}^N y_i y_j \hat\mu_{i\rho_j}.
\end{equation}
Employing the relations \eqref{HFE5}$_2$ between the specific heats and \eqref{conc11}$_2$,
we obtain from \eqref{conc8} the additional inequalities
\begin{equation}\label{conc8a}
    \frac{\partial \bar p}{\partial \rho}>0,\qquad
    c_p>c_v\quad\textrm{and}\quad
    (\frac{\partial \check \rho}{\partial T})^2<\frac{c_p\rho^2}{T}\frac{\partial \check \rho}{\partial p}.
\end{equation}
In particular the inequality \eqref{conc8a}$_3$ plays an
important role in the limiting case of incompressibility, which
will be the subject of Section~\ref{sec:incomp}.
%The resulting conditions coincide with the inequalities that
%follow from the concavity postulate, except of course the
%statement on the density derivatives of the chemical potentials.
%These can also be derived in the homogeneous setting, but we refrain from doing this here.
%Note, finally, that the conditions guaranteeing the concavity postulate are
%more general in two aspects: They do not refer to homogeneous
%systems and they also hold in non-equilibrium.
%
%
%
%
%
%
%
%
\section{Entropy invariant cross-couplings and the Onsager relations.}
\label{sec:cross-coupling}
By the above entropy principle, the entropy production has the form \eqref{M15}, i.e.\
\[
\zeta = \sum_m \mathcal{N}_m \mathcal{P}_m,
\]
where the sum runs through all dissipative mechanisms and
 the factors $\mathcal{N}_m$, $\mathcal{P}_m$ have negative, respectively positive parity.
  But the precise form of this decomposition of the entropy production into binary products
   is by no means unique! Indeed, cross-effects between the
dissipative mechanisms can be introduced
by mixing inside the parity-groups. Such cross-effects leave both the parity structure
 \eqref{M15} and the
entropy production invariant.
To be more specific, let us write the entropy production as a scalar product in an appropriate
 finite dimensional space, i.e.\
\begin{equation}
\label{entropy-form}
\zeta = \langle {\bf N} , {\bf P} \rangle,
\end{equation}
where all fluxes, respectively forces are placed inside the single vectors ${\bf N}$, respectively
${\bf P}$. Considering this representation, axiom~(IV-ii) enforces a diagonal closure,
corresponding to let
\begin{equation}
\label{diagonal-closure}
{\bf N} = \Lambda \, {\bf P} \quad \mbox{ with } \quad
\Lambda = {\rm diag} (\lambda_k), \, \lambda_k \geq 0.
\end{equation}
Let us note in passing that we may want to impose several restrictions on the coefficients $\lambda_k$ in \eqref{diagonal-closure} due to the assumption of isotropy; e.g., we assume the
same coefficient for closing different Cartesian components of the same vector. However,
for the subsequent explanation this does not play a role.

Now consider, instead of \eqref{entropy-form}, the form
\begin{equation}
\label{eq:mixed-entropy}
\zeta = \langle \tilde{\bf N} , \tilde{\bf P} \rangle
\quad \mbox{ with } \tilde{\bf N}= A \, {\bf N} ,\;  \tilde{\bf P}=B \, {\bf P},
\end{equation}
where $A$ and $B$ are regular matrices.
In order that, for arbitrary thermodynamic processes, the form $\zeta$ does not change
we must have
\begin{equation}
\langle {\bf N} , {\bf P} \rangle =\langle \tilde{\bf N} , \tilde{\bf P} \rangle =
 \langle A \, {\bf N} , B \, {\bf P} \rangle
\quad \mbox{ for all } {\bf N}, {\bf P},
\end{equation}
implying $A^{\sf T} \, B = {\bf I}$.
Axiom~(IV-ii) also applies to this representation of $\zeta$, i.e.\
$\tilde{\mathcal{N}}_m \tilde{\mathcal{P}}_m \geq 0$ for all $m$ which again enforces a diagonal closure.
This leads to the constitutive relations
\begin{equation}
\tilde{\bf N} = \tilde{\Lambda}\, \tilde{\bf P}
\quad \mbox{ hence } \quad
{\bf N} = A^{-1}\, \tilde{\Lambda} \, B \, {\bf P} = B^{\sf T} \, \tilde{\Lambda} \, B \,{\bf P}
\end{equation}
with a diagonal matrix $\tilde{\Lambda}$ with non-negative entries.
Hence
\begin{equation}
\label{Onsager-closure}
{\bf N} = L\, {\bf P} \quad \mbox{ with $L$ being symmetric and positive
semi-definite.}
\end{equation}
Consequently, the entropy principle above, together with a linear (in non-equilibrium quantities) closure after entropy invariant mixing exactly
leads to a full cross-coupling with a positive semi-definite and {\it symmetric} matrix of phenomenological coefficients. Hence the Onsager reciprocity relations are satisfied.

Let us note in passing that the classical origin of Onsager symmetry for ODE systems relies on the notion of microscopic reversibility. A distinct presentation of the subject is found in Mauri's textbook \cite{Mauri}. However, the ODE treatment cannot be transferred to PDE systems.

The above observation leading to \eqref{Onsager-closure}
is related to known considerations of the invariance of Onsager's relations under transformations of the system of fluxes and driving forces; cf.\ \cite{dGM}.
Note also that the entropy production invariance is to be understood for the form of the unclosed entropy production, not for the production rate in the final PDE-system.
The considered cross-coupling via entropy neutral mixing between mechanisms is given above
for fixed number of dissipative mechanisms. It is also possible to apply such a mixing
in a more general way, also adding new dissipative mechanisms. This way, also the entropy
flux will be altered. An example is provided below in Section~\ref{sec:model-reduction1}, where we reduce our class-II model to obtain improved class-I closure relations; cf.\ also Remark~10 in Section~\ref{sec:fundamentals}.

In the classical literature on irreversible processes, the factors building the binary products in
the entropy production are called {\it fluxes} and {\it driving forces}. This notion can be rather
misleading in the context of cross-effects, since the assignment of the label ''flux'' to a quantity is
usually guided by its appearance under a divergence, independently whether it has positive or negative parity. In fact this can lead to an apparent anti-symmetric coupling. To understand this, assume a
second classification of the set of all factors appearing as components in ${\bf N}$ and ${\bf P}$ into fluxes and driving forces such that
\[
{\bf N} = \left[ \begin{array}{c} {\bf F}_1\\ {\bf D}_2 \end{array} \right], \quad
{\bf P} = \left[ \begin{array}{c} {\bf D}_1\\ {\bf F}_2 \end{array} \right],
\]
where the ${\bf F}_i$ and ${\bf D}_i$ are blocks entirely composed of fluxes or driving forces.
Applying entropy invariant mixing and linear closure as above, we arrive at the constitutive relation
\[
\left[ \begin{array}{c} {\bf F}_1\\ {\bf D}_2 \end{array} \right]
 = L\, \left[ \begin{array}{c} {\bf D}_1\\ {\bf F}_2 \end{array} \right]
 \quad \mbox{ with } \;
 L=\left[ \begin{array}{cc} L_{11} & L_{12}\\ L_{21} & L_{22}\end{array} \right]\geq 0,\;
 L_{ii}^{\sf T}=L_{ii},\; L_{12}^{\sf T}=L_{21}.
\]
Rewritten as a relation between fluxes and driving forces, this becomes
\[
\left[ \begin{array}{c} {\bf F}_1\\ {\bf F}_2 \end{array} \right]
 = \mathcal{L}\, \left[ \begin{array}{c} {\bf D}_1\\ {\bf D}_2 \end{array} \right]
 \quad \mbox{ with } \;
 \mathcal{L}=\left[ \begin{array}{cc} \mathcal{L}_{11} & \mathcal{L}_{12}\\
 \mathcal{L}_{21} & \mathcal{L}_{22}\end{array} \right],
\]
where
\[
 \mathcal{L}_{11}=L_{11}-L_{12} L_{22}^{-1} L_{21},\;\;
 \mathcal{L}_{12}=L_{12} L_{22}^{-1}=-\mathcal{L}_{21}^{\sf T},\;\;
 \mathcal{L}_{22}=L_{22}.
\]
Evidently, the matrix $\mathcal{L}$ is not symmetric, but consists of symmetric blocks on the block-diagonal and skew-symmetric off-diagonal blocks.
Note also that $\mathcal{L}$ inherits the positive semi-definiteness of $L$; in particular,
the block $L_{11}-L_{12} L_{22}^{-1} L_{21}$ itself is positive semi-definite
which follows from
\[
\langle L_{11} \, x , x \rangle +
\langle L_{12} \, y , x \rangle +
\langle L_{21} \, x , y \rangle +
\langle L_{22} \, y , y \rangle \geq 0
\quad \mbox{ for all } x,\, y
\]
applied with $y=- L_{22}^{-1} L_{21}x$.

The above considerations on the one hand show how to introduce cross-effects on the entropy production level. On the other hand, the replacement of the notions ''flux / driving force'' by ''co-factors of
positive and negative parity'' resolves the apparent anti-symmetry of certain cross-coefficients.
This will be taken up again in Section~\ref{sec:cross-effects}.
\section{Inviscid, non-reactive fluid mixtures}
\label{section4}
We are going to employ the entropy inequality for obtaining constitutive laws for the
thermo-mechanical interactions, i.e.\ for the structure of the
${\bf f}_i$. Specialized to fluid mixtures without chemical
reactions ($R_a=0$) and with vanishing viscosity ($\stress_i^\circ
=0$, $\Pi_i =0$), the entropy production is reduced to the two dissipative mechanisms of
diffusion and heat conduction and becomes
\begin{equation}
\label{zeta-nc-nv}
\ts \zeta = - \sum_i {\bf u}_i
\cdot \Big( \rho_i \nabla \frac{\mu_i}{T} + \frac 1 T ({\bf f}_i -\nabla p_i
-h_i T \nabla \frac 1 T) \Big)
 + \sum_i {\bf q}_i \cdot \nabla \frac 1 T
\end{equation}
with the partial enthalpies $h_i= \rho_i e_i + p_i $.
For an efficient notation,we proceed to write \eqref{zeta-nc-nv} as
\begin{equation}
\label{entropy-prod1}
\ts
\zeta =
\zeta_{\rm diff} +  \zeta_{\rm heat}
\quad \mbox{ with } \quad
\zeta_{\rm heat} = \sum_i {\bf q}_i \cdot \nabla \frac 1 T.
\end{equation}
We impose linear relations between the conjugate factors in the entropy production; but note that the resulting constitutive equations are of course nonlinear relations between the variables.
For simplicity, we start with the case in which the two principal mechanisms are treated independently, without explicitly introducing an additinal thermo-diffusion coupling.
Full thermo-diffusion will be considered in Section~\ref{thermo-diffusion} below.

We now exploit axiom~(IV-ii). The term $\zeta_{\rm heat}$ is non-negative if we assume
{\it Fourier's law} for the thermal part $\sum_i {\bf q}_i$ of the heat flux ${\bf q}$ from \eqref{def:reduced-heat-flux}, i.e.\
\[
\ts
\sum_i {\bf q}_i = \alpha \nabla \frac 1 T \quad \mbox{ with } \; \alpha=\alpha (T,\rho_1,\ldots ,\rho_N) \geq 0.
\]
In this case the heat flux itself is given by
\begin{equation}
\label{Fouriers-law}
\ts
{\bf q}=\sum_i {\bf q}_i + \sum_i h_i {\bf u}_i= \alpha \nabla \frac 1 T + \sum_i h_i {\bf u}_i.
\end{equation}
Hence, despite the simple Fourier law, some cross-effects are already included in ${\bf q}$.

The remaining part $\zeta_{\rm diff}$ is of the form
\begin{equation}
%\label{entropy-prod1}
\label{def:Bi}
\ts
\zeta_{\rm diff} =
- \sum_i {\bf u}_i \cdot \Big( {\bf B}_i  + \frac 1 T {\bf f}_i \Big)
\quad \mbox{ with } \quad {\bf B}_i:= \rho_i \nabla \frac{\mu_i}{T}
- \frac 1 T  \nabla p_i - h_i \nabla \frac 1 T.
\end{equation}
This shows a further advantage of the separation into the principal mechanisms,
since now the ${\bf B}_i$ turn out to sum up to zero which is basic for the exploitation of axiom~(IV-ii) below. We indeed have
\begin{equation}
\ts
\sum_i {\bf B}_i  =
\frac 1 T \Big( \sum_i \rho_i \nabla \mu_i - \nabla p \Big)
+ \Big( \sum_i \rho_i \mu_i - (\rho e + p) \Big) \nabla \frac 1 T
\end{equation}
and the Gibbs-Duhem equation \eqref{Gibbs-Duhem} implies
\[
\ts
\nabla p = \sum_{i=1}^N \rho_i  \nabla \mu_i + \rho s \nabla T,
\]
hence $\sum_i {\bf B}_i = 0.$
Axiom~(IV-ii) of the entropy principle now requires
\begin{equation}
\label{inequ-BF}
\ts
- \sum_{i=1}^N {\bf u}_i \cdot \Big( {\bf B}_i + \frac 1 T {\bf f}_i \Big) \geq 0.
\end{equation}
To incorporate the constraints
\begin{equation}
\ts
\sum_{i=1}^N {\bf B}_i  =0, \quad \sum_{i=1}^N {\bf f}_i  =0,
\end{equation}
these are built into \eqref{inequ-BF} by eliminating ${\bf B}_N$ and ${\bf f}_N$. Thus, the necessary condition becomes
\begin{equation}
\label{B+F}
\ts
- \sum_{i=1}^{N-1} ({\bf u}_i - {\bf u}_N )\cdot ( {\bf B}_i + \frac 1 T {\bf f}_i ) \geq 0.
\end{equation}

The corresponding linear Ansatz for ${\bf B}_i + \frac 1 T {\bf f}_i$, in order to guarantee \eqref{B+F}, is
\begin{equation}
\label{closure-fi}
\ts
{\bf B}_i + \frac 1 T {\bf f}_i = - \sum_{j=1}^{N-1} \tau_{ij}\, ( {\bf u}_j  - {\bf u}_N )
\quad \quad (\mbox{for } i=1,\ldots ,N-1)
\end{equation}
with a positive definite $(N-1)\times (N-1)$-matrix $[\tau_{ij} ]$. This yields the
constitutive laws for the momentum productions ${\bf f}_i$.

The representation \eqref{closure-fi} can be made symmetric with respect to the constituents
by extending the {\it Maxwell-Stefan matrix} $[\tau_{ij}]$ to $N\times N$-format according to
\begin{equation}
\label{tau-NN}
\ts
\tau_{Nj} = - \sum_{i=1}^{N-1} \tau_{ij} \;\; (j=1,\ldots ,N-1),\quad
\tau_{iN} = - \sum_{j=1}^{N-1} \tau_{ij} \;\; (i=1,\ldots ,N).
\end{equation}
Note that the definition of the $\tau_{Nj}$ is required to obtain

\[
\ts
{\bf B}_i + \frac 1 T {\bf f}_i = - \sum_{j=1}^{N} \tau_{ij}\, ( {\bf u}_j  - {\bf u}_N )
\quad \mbox{ for  } i=1,\ldots ,N,
\]
including $i=N$, while the definition of the $\tau_{iN}$ yields
$\sum_{j=1}^{N} \tau_{ij} =0$ for all $i$, hence ${\bf u}_N$ above can be replaced by ${\bf u}_i$
to attain the symmetric form
\begin{equation}
\label{Bi+fi-sym}
\ts
{\bf B}_i + \frac 1 T {\bf f}_i = \sum_{j=1}^{N} \tau_{ij}\, ( {\bf u}_i  - {\bf u}_j )
\quad \mbox{ for  } i=1,\ldots ,N.
%\quad  = \sum_{j=1}^{N} \tau_{ij}\, ( {\bf v}_i  - {\bf v}_j )
\end{equation}

The extended matrix will be positive semi-definite. We record this observation as\\[1ex]
{\bf Proposition.} The extended $N\times N$-matrix $[\tau_{ij} ]_1^N$ is positive semi-definite
if the $(N-1)\times (N-1)$-matrix $[\tau_{ij} ]_1^{N-1}$ is positive (semi-)definite.\\[1ex]
{\it Proof.} By the criterion of Sylvester, a quadratic matrix $M$ is positive semi-definite iff
the determinant of every upper left sub-matrix is non-negative. Here, we only need to check
this for the determinant of the full matrix.
The latter vanishes since, by definition, the $N$-th column
is a linear combination of the first $N-1$ columns.\hfill $\Box$\\[2ex]
\noindent
In the present case of {\it non-reactive} multicomponent fluids, we add the following
assumption of {\bf binary-type mechanical interactions}:
\begin{equation}
\label{binary-interaction}
\ts
\tau_{ij}=\tau_{ij}(T, \rho_i, \rho_j) \to 0 \quad \mbox{ if } \; \rho_i \rho_j \to 0+ .
\end{equation}
Let us emphasize that this assumption is not true for chemically reacting mixtures -- this will
become clear in section~\ref{sec:reacting-class-II} below.
Following Truesdell \cite{CT62}, binary-type interactions imply the symmetry of $[\tau_{ij}]$ as follows: evaluate
$\sum_{i,j=1}^N \tau_{ij}\, ( {\bf u}_i  - {\bf u}_j ) =0$. In case $N=2$, one obtains
$(\tau_{12} - \tau_{21})\, ( {\bf u}_1  - {\bf u}_2 )=0$ for any thermodynamical process, hence
$\tau_{12} =\tau_{21}$. Note that for $N=2$ the system can only have binary interactions.
In case $N=3$, the summation gives
\[
(\tau_{12} - \tau_{21})\, ( {\bf u}_1  - {\bf u}_2 )
+ \;(\tau_{13} - \tau_{31})\, ( {\bf u}_1  - {\bf u}_3 )
+ \;(\tau_{23} - \tau_{32})\, ( {\bf u}_2  - {\bf u}_3 )=0.
\]
Exploitation of \eqref{binary-interaction} yields:
\[
\rho_3 \to 0+ \; \Rightarrow \; \tau_{12} = \tau_{21};\quad
\rho_2 \to 0+ \; \Rightarrow \; \tau_{13} = \tau_{31};\quad
\rho_1 \to 0+ \; \Rightarrow \; \tau_{23} = \tau_{32}.
\]
The general case $N>3$ follows in an analogous manner. Consequently, under the assumption \eqref{binary-interaction}, the full matrix $[\tau_{ij}]$ is symmetric and positive semi-definite.
Let us note in passing, that this derivation of the symmetry of the mechanical interactions
works exclusively in the considered case of binary (non-reactive) interactions and it exploits
conservation of total momentum.
%Newton's third law, viz.\ ''actio = reactio''.

Exploiting the symmetry of  $[\tau_{ij}]$  and using \eqref{Bi+fi-sym},
the diffusive entropy production is
\begin{equation}
\ts
\zeta_\textrm{diff}=
- \sum_{i=1}^N {\bf u}_i \cdot \big( {\bf B}_i + \frac 1 T {\bf f}_i \big)
 = - \frac 1 2  \sum_{i,j=1}^{N} \tau_{ij} ( {\bf u}_i  - {\bf u}_j )^2,
\end{equation}
showing that, necessarily,
\[
\tau_{ij} \leq 0 \quad \mbox{ for all } i\neq j.
\]
Hence the natural Ansatz to incorporate both symmetry and binary-type interactions is
\begin{equation}
\label{fij-def}
\tau_{ij}=-\, f_{ij} \rho_i \rho_j
\quad \mbox{ with } f_{ij}=f_{ji},\; f_{ij}=f_{ij} (T, \rho_i ,\rho_j)\geq 0
\quad (i\neq j).
\end{equation}
From a molecular dynamics viewpoint, the $f_{ij}$ can be interpreted as "friction factors".

Thus, for class-II mixtures, we obtain
\begin{equation}
\label{fi-law}
\ts
{\bf f}_i = - \rho_i T \nabla \frac{\mu_i}{T} + \nabla p_i +
T h_i \nabla \frac 1 T - T \sum_j f_{ij}  \rho_i \rho_j ( {\bf v}_i  - {\bf v}_j )
\end{equation}
as the constitutive law for the thermo-mechanical interactions.
Inserting this into the partial momentum balances, we obtain their final form in the case of inviscid, non-reactive fluid mixtures with binary-type interactions as
\begin{equation}
\label{class-II-momentum}
\ts
\rho_i \big( \pt  {\bf v}_i + {\bf v}_i \cdot \nabla {\bf v}_i \big)
= - \rho_i T \nabla \frac{\mu_i}{T} + T h_i \nabla \frac 1 T
- T \sum_j f_{ij}  \rho_i \rho_j ( {\bf v}_i  - {\bf v}_j )
+ \rho_i {\bf b}_i.
\end{equation}
In the {\it isothermal case}, i.e.\ in the limit when the approach to thermal equilibrium is infinitely fast so that no temperature gradients persist,
this simplifies according to
\begin{equation}
\label{class-II-momentum-isotherm}
\ts
\rho_i \big( \pt  {\bf v}_i + {\bf v}_i \cdot \nabla {\bf v}_i \big)
= - \rho_i \nabla \mu_i  - T \sum_j f_{ij}  \rho_i \rho_j ( {\bf v}_i  - {\bf v}_j )
+ \rho_i {\bf b}_i.
\end{equation}
In the non-isothermal case, the resulting class-II model for inviscid,
non-reactive fluids with binary-type thermo-mechanical interactions in conservative form reads as
\begin{align*}
{\bf mass:} & \; \pt \rho_i + \div (\rho_i {\bf v}_i)= 0 \\[0.5ex]
{\bf mom.:} &
\ts \; \pt (\rho_i {\bf v}_i) + \div (\rho_i {\bf v}_i \otimes {\bf v}_i )
= - \rho_i \nabla \mu_i + T (h_i - \rho_i \mu_i) \nabla \frac 1 T
 - T \sum_j f_{ij}  \rho_i \rho_j ( {\bf v}_i  - {\bf v}_j )
+ \rho_i {\bf b}_i \\[1ex]
{\bf energy:} &
\; \pt (\rho e ) + \div ( \rho e {\bf v} + {\bf q})
= \ts
-p \, \div {\bf v}  -  \sum_i {\bf u}_i \cdot ({\bf f}_i - \nabla p_i )
\end{align*}
with $\bf q$ and ${\bf f}_i$ from \eqref{Fouriers-law} and \eqref{fi-law}, respectively.

To derive a mixture heat equation as the temperature form of the energy balance, we choose the independent variables as $(T,\rho,y_1, \ldots ,y_{N-1})$.
Relying on \eqref{rho-psi-partial-derivatives} and the Gibbs-Duhem equation \eqref{Gibbs-Duhem}, we obtain
\begin{equation}
\rho c_v \dot{T} + \div {\bf q} =
-  \sum_{i=1}^N {\bf u}_i \cdot ({\bf f}_i - \nabla p_i ) - \sum_{i=1}^N \big( \mu_i - \mu_N - T \frac{\partial (\mu_i - \mu_N)}{\partial T} \big) \rho \dot{y}_i
-\frac T \rho \frac{\partial p}{\partial T} \dot{\rho}
\end{equation}
with $c_v$ the specific heat capacity at constant density from \eqref{HFE4}. Furthermore,
the dot denotes the Lagrangean derivative w.r.\ to the barycentric velocity,
e.g.\ $\dot T = \partial_t T + {\bf v}\cdot \nabla T$.

Let us briefly mention the special case of {\it simple mixtures}
(cf.\ section~\ref{sec:free-energies} below for more details) for which
\begin{equation}
\label{simple-mixture}
\rho \psi = \sum_{i=1}^N \rho_i \psi_i (T,\rho_i)
\quad \mbox{ and } \quad
\rho_i \psi_i +p_i = \rho_i \mu_i.
\end{equation}
In this case
the constitutive law for the thermo-mechanical interactions \eqref{fi-law} reduces to
\begin{equation}
\label{fi-law-sm}
\ts
{\bf f}_i =  - T \sum_j f_{ij}  \rho_i \rho_j ( {\bf v}_i  - {\bf v}_j ),
\end{equation}
hence the partial momentum balance simplifies to
\begin{equation}
\ts \; \pt (\rho_i {\bf v}_i) + \div (\rho_i {\bf v}_i \otimes {\bf v}_i )
= - \nabla p_i  - T \sum_j f_{ij}  \rho_i \rho_j ( {\bf v}_i  - {\bf v}_j ) + \rho_i {\bf b}_i.
\end{equation}
Note that the term $-\sum_i {\bf u}_i \cdot {\bf f}_i$ in the heat equation becomes $\frac T 2 \sum_j f_{ij}  \rho_i \rho_j ( {\bf v}_i  - {\bf v}_j )^2$ in case
of a simple mixture.
\section{Thermo-Diffusion Revisited}
\label{thermo-diffusion}
It is well known that temperature gradients can generate interaction forces and
diffusion fluxes contribute to the heat flux. In class-I models,
these phenomena are called Soret and Dufour effect, respectively.
In fact there is a contribution to the heat flux in \eqref{Fouriers-law} that is proportional to the diffusion velocity and in \eqref{fi-law} we observe the corresponding contribution $T h_i
\nabla \frac 1 T$ to the interaction force. However, for simple
mixtures the interaction forces becomes independent of the temperature gradient.

In order to prevent the disappearance of such phenomena, in
particular for simple mixtures, we now apply the entropy invariant mixing. Usually,
cross-effects are introduced into the thermo-fluid mechanical
modeling via full coupling in the constitutive relations. Given
$M$ binary products in the entropy production, this leads to $M^2$
phenomenological coefficients. If linear constitutive relations
are employed, symmetry between the phenomenological coefficients
are assumed by postulating the Onsager reciprocity relations.
We proceed in a different manner, employing the entropy invariant mixing
 from Section~\ref{sec:cross-coupling}.
The key point is the equivalent reformulation of the entropy production rate,
which here can be simply based on a zero addition, introducing a mixing between
the mechanisms ''diffusion'' and ''heat conduction''.
For this purpose we rewrite $\zeta$ from \eqref{zeta-nc-nv} as
\begin{equation}
\label{entropy-prod2}
\ts
\zeta =
- \sum_i {\bf u}_i \cdot \Big( {\bf B}_i  + \frac 1 T {\bf f}_i - d_i^T \nabla \frac 1 T \Big)
 + \sum_i ({\bf q}_i - d_i^T {\bf u}_i ) \cdot \nabla \frac 1 T.
\end{equation}
Note that $\nabla \frac 1 T$ has been introduced as an additional
driving mechanism for diffusion, while the new term is compensated
by an appropriate change in the binary product involving the heat
flux. The newly introduced phenomenological {\it thermal diffusivities} $d_i^T$
are functions of $(T,\rho_1, \ldots ,\rho_N)$, they can attain positive or negative values.
This fact is well-known in kinetic theories, cf.\ \cite{FK}.

We now intend to repeat the arguments of Section~\ref{section4}, but
$\sum_i  \big( {\bf B}_i  + \frac 1 T {\bf f}_i - d_i^T \nabla \frac 1 T \big)$
need not be zero which was crucial in the derivation of the class-II model.
This defect can be removed by introduction of a Lagrange multiplier. Indeed, given any $\lambda \in \R$, the entropy production
equals
\begin{equation}
\label{entropy-prod3}
\ts
\zeta =
- \sum_i {\bf u}_i \cdot \Big( {\bf B}_i  + \frac 1 T {\bf f}_i - (d_i^T - y_i \lambda) \nabla \frac 1 T \Big)
 + \sum_i ({\bf q}_i - d_i^T {\bf u}_i ) \cdot \nabla \frac 1 T,
\end{equation}
since $\sum_i y_i {\bf u}_i =0$. With the choice of $\lambda = \sum_j d_j^T$, the co-factors of ${\bf u}_i$ sum up to zero as before.
Let $\delta_i^T = d_i^T - y_i \sum_j d_j^T$ and observe that the second binary product in \eqref{entropy-prod3}
remains unchanged if $d_i^T$ is replaced by $\delta_i^T$. It is therefore no restriction to consider the entropy production
given in \eqref{entropy-prod2} with the additional constraint
\begin{equation}
\label{diT-sum}
\sum_{i=1}^N d_i^T = 0.
\end{equation}
Now, the arguments from Section~\ref{section4} apply without changes and yield the closure relations
\begin{equation}
\label{fi-law-Soret} \ts {\bf f}_i = - \rho_i T \nabla
\frac{\mu_i}{T} + \nabla p_i + (h_i+ d_i^T) T \nabla \frac 1 T - T
\sum_j f_{ij}  \rho_i \rho_j ( {\bf v}_i  - {\bf v}_j ) =:{\bf
f}_i^{\rm M} ,
\end{equation}
\begin{equation}
\label{Fouriers-law-Dufour} \ts {\bf q}=\sum_i{\bf q}_i+\sum_i h_i
{\bf u}_i= \alpha \nabla \frac 1 T + \sum_i (h_i + d_i^T) {\bf
u}_i.
\end{equation}
Insertion of \eqref{fi-law-Soret} into the momentum balance yields
\begin{equation}
\label{eq:momentum-classII-Soret}
\ts
\rho_i \big( \pt  {\bf v}_i + {\bf v}_i \cdot \nabla {\bf v}_i \big)
= - \rho_i \nabla \mu_i + T (h_i + d_i^T - \rho_i \mu_i ) \nabla \frac 1 T
 - T \sum_j f_{ij}  \rho_i \rho_j ( {\bf v}_i  - {\bf v}_j )
+ \rho_i {\bf b}_i,
\end{equation}
which is complemented by the partial mass balances and the mixture internal energy balance to obtain the class-II model for an inviscid, non-reactive mixture
with thermo-diffusion and diffusion-thermo effects.
In case of simple mixtures, the partial momentum balances become
\begin{equation}
\label{eq:momentum-classII-Soret-simple}
\ts
\rho_i \big( \pt  {\bf v}_i + {\bf v}_i \cdot \nabla {\bf v}_i \big)
= - \nabla p_i - d_i^T \nabla \ln T
 - T \sum_j f_{ij}  \rho_i \rho_j ( {\bf v}_i  - {\bf v}_j )
+ \rho_i {\bf b}_i.
\end{equation}
Note that now $\nabla T$ remains in the momentum balances even
for simple mixtures.

Later on we will generate class-I models by a reduction procedure applied to class-II models. Then we will see that all cross-effects which are introduced by entropy invariant mixing in class-II automatically come with the usually postulated Onsager reciprocity relations
into the class-I model. This is the true rationale behind the Onsager symmetry:
We have Onsager reciprocity relations if the corresponding cross-effect does not generate further entropy production. Further examples of this statement are found in Section
\ref{sec:cross-effects}, where we take up the subject of entropy invariant mixing once more.

For subsequent sections, in which chemical reactions and viscous
effects are to be included, it is helpful to observe the overall
effect of thermo-diffusion: instead of the partial enthalpies
$h_i$, one simply has to write $\mathcal{D}_i^T\equiv h_i + d_i^T$, where
one may impose \eqref{diT-sum}.
\section{inviscid, chemically reactive fluid mixtures}
\label{sec:reacting-class-II} Chemical reaction appear in the
model with three contribution: (i) mass production in the partial
mass balances, (ii) momentum production in the partial momentum
balances and (iii) heat production in the energy balance.
With chemical reactions, the entropy production reads as\\[-1ex]
\begin{equation}
\label{entropy-prod-chem1} \ts \zeta = \sum_i ({\bf q}_i-d_i^T{\bf
u}_i )\cdot \nabla \frac 1 T - \sum_i {\bf u}_i \cdot \Big( {\bf
B}_i -d_i^T\nabla \frac 1 T + \frac 1 T ({\bf f}_i - r_i  {\bf
v}_i +\frac 1 2 r_i {\bf u}_i) \Big) - \frac 1 T \sum_{a=1}^{N_R}
R_a \mathcal{A}_a
\end{equation}
with the affinities $\mathcal{A}_a$ from \eqref{affinities} and
the ${\bf B}_i$ from \eqref{def:Bi}. Recall that $\sum_i {\bf
B}_i=0$ and $\sum_i d_i^T=0$.

Here we note another ambiguity in the decomposition of $\zeta$
into a sum of binary products. The term $\frac 1 2 r_i {\bf u}_i$
appeared in the energy production because of chemical reactions
via the mass balances. Similar to diffusive contributions to the
heat flux in Section~\ref{section4}, we prefer to shuffle the term
$\frac 1 2 r_i {\bf u}_i$ to become a co-factor of ${\bf u}_i$ in
order to collect all terms related to the mechanism of diffusion.
If we were to leave this term attached to the chemical reactions'
entropy production, we then would have
\begin{equation}
\ts
-\frac 1 T \sum_{a=1}^{N_R} R_a \sum_{n=1}^N M_i \nu_i^a (\mu_i + \frac 1 2 {\bf u}_i^2 ).
\end{equation}
While this change would not modify the value of $\zeta$, it has tremendous implications
after employing (linear) constitutive laws. In fact, the latter choice would contradict
axiom~(IV-ii) and a linear diagonal closure would result in a non-wellposed PDE-system.
This subtle point will be taken up in more detail below.

We decompose the interaction force ${\bf f}_i - r_i {\bf v}_i$ according to
\begin{equation}
\label{force-decomposition}
{\bf f}_i - r_i {\bf v}_i= {\bf f}_i^{\rm M} + {\bf f}_i^{\rm C}- r_i {\bf v}_i,
\end{equation}
where we assume that the mechanical part ${\bf f}_i^{\rm M}$ has the same structure as for a non-reactive mixture; i.e.,
we model the ${\bf f}_i^{\rm M}$ according to \eqref{fi-law-Soret}.
The chemical interaction part ${\bf f}_i^{\rm C}- r_i {\bf v}_i$ corresponds to exchange of momentum because in a chemical reaction
the reacting constituents inherit their momentum to the products.
We now compute ${\bf f}_i^{\rm C}- r_i {\bf v}_i$ via balancing of partial momenta.
At first, for simplicity, we only consider the forward path, say, of the $a^{\rm th}$ chemical reaction, viz.\
\[
\alpha_1^a \, A_1 + \ldots + \alpha_N^a \, A_N
\to
\beta_1^a \, A_1 + \ldots + \beta_N^a \, A_N
\]
with rate $R_a^f$. The rate of change of momentum of constituent $A_i$ is
\[
-R_a^f \alpha_i^a M_i {\bf v}_i + R_a^f \beta_i^a M_i {\bf v}_i^{a,f}.
\]
Here ${\bf v}_i^{a,f}$ denotes the average velocity of molecules of type $A_i$ {\it after}
they are formed in a reactive collision of the involved reactants,
i.e.\ those for which $\beta_i^a >0$.
The only obstacle is that ${\bf v}_i^{a,f}$ is not know, but needs to be modeled based on a micro-theory.
At this point it is helpful to notice that mass and momentum are conserved during chemical reactions, i.e.\
\begin{equation}
\label{reactive-mom-cons}
\ts
\sum_{i=1}^N \alpha_i^a M_i = \sum_{i=1}^N \beta_i^a M_i
\; \mbox{ and } \;
\sum_{i=1}^N \alpha_i^a M_i {\bf v}_i = \sum_{i=1}^N \beta_i^a M_i {\bf v}_i^{a,f}.
\end{equation}
Now, if only one product is present in the $a^{\rm th}$ reaction,
say $A_k$, this alone is sufficient to determine the velocity
${\bf v}_k^{a,f}$ from \eqref{reactive-mom-cons}. In the more
complicated general case it is sufficient to think of a reaction
with only one reactant, i.e.\ the decomposition of $A_k$, say,
into several parts. This is no restriction of generality, because
any chemical reaction proceeds via a so-called transition complex
where the interacting reactants form a (very short lived)
intermediate species. An observer which follows this transition
complex sees a stagnant particle which decomposes into several
parts which then move away from the steady barycenter. In any new
instance of this reaction, the products will move in different
directions. But since we assume isotropic media in our theory, no
direction is preferred. Consequently, the (ensemble) averaged
velocity is zero. In other words, all products from one reaction path have {\it the same
velocity} which coincides with the one of the transition complex,
which in turn is nothing but the velocity of the center of mass of
the reacting constituents.

Moreover, for the backward reaction the momentum conservation reads,
by a similar reasoning, $ \sum_{i=1}^N \alpha_i^a M_i
{\bf v}_i^{b,f} = \sum_{i=1}^N \beta_i^a M_i {\bf v}_i$. Including
also the analogous condition on ${\bf v}_i^{b,f}$
for the backward reactions, this leads to the following\\[0.5ex]
\noindent
{\bf fundamental assumption on reactive collisions}:
\begin{equation}
\label{fundamental-ass} {\bf v}_i^{a,f} = {\bf v}^f_a
~\;\mbox{and} \quad {\bf v}_i^{a,b} = {\bf v}^b_a \;\mbox{ for
forward and backward reactions, respectively.}
\end{equation}
From the conservation of momentum in chemical reactions, we obtain
\begin{equation}
\ts
{\bf v}^f_a =\sum_i \alpha_i^a M_i {\bf v}_i / \sum_i \beta_i^a M_i
\;\mbox{ as well as } \;
{\bf v}^b_a =\sum_i \beta_i^a M_i {\bf v}_i / \sum_i \alpha_i^a M_i.
\end{equation}
In imposing this assumption, we follow Burgers \cite{Burgers} who used the same proposition for a study of the
reactive Boltzmann equations.

Using the above assumptions and considerations, balancing of the chemical exchange of partial momentum leads to the representation
\begin{equation}
\label{fiC} \ts {\bf f}_i^{\rm C} - r_i {\bf v}_i = -\sum_{j=1}^N
C_{ij} ({\bf v}_i - {\bf v}_j),
\end{equation}
where the {\it chemical interaction matrix} has the entries
\begin{equation}
\label{def:Cij}
C_{ij} = \sum_{a=1}^{N_R} \frac{M_i M_j}{\ts \sum_k \alpha_k^a M_k} \big( R_a^f \beta_i^a \alpha_j^a + R_a^b \alpha_i^a \beta_j^a \big).
\end{equation}
Note that the matrix $[C_{ij}]$ is, in general, {\it not} symmetric.
But in the so-called detailed chemical equilibrium, i.e.\ in case $R_a^f=R_a^b$
for {\it all} reactions $a=1,\ldots , N_R$, it attains a symmetric form.
Indeed, the anti-symmetric part of the chemical interaction matrix is
\begin{equation}
\label{Cij-Cji}
C_{ij}-C_{ji} = \sum_{a=1}^{N_R} R_a \frac{M_i M_j}{\ts \sum_k \alpha_k^a M_k}
\big( \beta_i^a \alpha_j^a - \alpha_i^a \beta_j^a \big).
\end{equation}
Summation over $i$ yields
\begin{equation}
\label{Cij-Cji-sum}
\sum_{i=1}^N \big( C_{ij}-C_{ji} \big)=
\sum_{a=1}^{N_R} R_a M_j \big( \alpha_j^a - \beta_j^a \big)=-r_j,
\end{equation}
showing that $r_j=0$ for all $j=1,\ldots ,N$ is required for the symmetry of $[C_{ij}]$.

The decomposition into a chemical part and the mechanical
interaction, the latter without changes compared to the
non-reactive case, has to be justified by checking the
thermodynamic consistency. At this point, insertion of ${\bf
f}_i^{\rm M} + {\bf f}_i^{\rm C} -r_i {\bf v}_i$ into
\eqref{entropy-prod-chem1} leads to an interesting observation:
if the term $\frac 1 2 r_i {\bf u}_i$ were to remain in the binary
product which corresponds to entropy production via chemical
reactions in \eqref{entropy-prod-chem1}, the entropy production
due to diffusion can, in general, become negative! As an example,
in the simplest case $2 A_1 \to A_2$, the contribution of ${\bf
f}_i^{\rm C} -r_i {\bf v}_i$ is $2 R^f M_1 {\bf u}_2 ({\bf u}_2 -
{\bf u}_1)$ which evidently can attain negative values; note that
forward and backward reactions separately have to generate the
right sign in the entropy production since no chemical equilibrium
is assumed. For this reason we have shuffled the term $\frac 1 2
r_i {\bf u}_i$ to the contribution corresponding to diffusive mass
transport.

We proceed to check the fulfilment of the 2$^{\rm nd}$ law, where we already know that the entropy production due to the mechanical part is non-negative.
For the remainder, the contribution of the forward path, say, of the $a^{\rm th}$ reaction is
\begin{equation}
\label{entropy-prod-chem3}
\zeta_{\rm diff}^{a,f}=
R^f_a \big( \sum_i M_i \frac{\alpha_i^a + \beta_i^a}{2} {\bf u}_i^2 - \sum_{i,j} \frac{M_i M_j \beta_i^a \alpha_j^a}{\sum_k \alpha_k^a M_k} {\bf u}_i \cdot {\bf u}_j \big).
\end{equation}
Note that $\sum_k \alpha_k^a M_k = (\sum_k \alpha_k^a M_k)^{1/2} (\sum_k \beta_k^a M_k)^{1/2}$,
hence the contribution
$\zeta_{\rm diff}^{a,f}$ to the entropy production is symmetric in $\alpha_i^a$, $\beta_i^a$.
This only holds if the term $\frac 1 2 r_i {\bf u}_i^2$ is included in the diffusive contribution.
Now, to see that the entropy inequality holds, we fix the reaction index $a$,
abbreviate $M_i \alpha_i^a, M_i \, \beta_i^a$ as $a_i,\, b_i$ and
make use of the Cauchy-Schwarz inequality in the form
\[
\ts
\sum_i a_i z_i = \sum_i a_i^\frac{1}{2} \cdot a_i^\frac{1}{2} z_i \leq \big( \sum_i a_i \big)^\frac{1}{2}
\big( \sum_i a_i z_i^2 \big)^\frac{1}{2}.
\]
Applying this to $\sum_{i,j} b_i a_j {\bf u}_{i,k} \, {\bf u}_{j,k}$, and summing over the velocity components $k=1,2,3$,
we obtain
\[
\ts
\sum_{i,j} b_i a_j {\bf u}_i \cdot {\bf u}_j
= \sum_i b_i {\bf u}_i  \cdot \sum_i a_i {\bf u}_i
\leq \big( \sum_i a_i \big)^\frac{1}{2} \big( \sum_i b_i \big)^\frac{1}{2}
 \big( \sum_i a_i {\bf u}_i^2 \big)^\frac{1}{2} \big( \sum_i b_i {\bf u}_i^2 \big)^\frac{1}{2}.
\]
Employing mass conservation, this yields
\[
\sum_{i,j} \frac{b_i a_j}{\sum_k \! a_k} {\bf u}_i \cdot {\bf u}_j
%= \sum_i \frac{b_i {\bf u}_i}{(\sum_k \! b_k)^{1/2}}  \cdot \sum_i \frac{a_i {\bf u}_i}{(\sum_k \! a_k)^{1/2}}
\leq \big( \sum_i b_i {\bf u}_i^2 \big)^\frac{1}{2} \big( \sum_i a_i {\bf u}_i^2 \big)^\frac{1}{2}
\leq \frac 1 2 \sum_i (a_i + b_i) {\bf u}_i^2.
\]
Consequently, $\zeta_{\rm diff}^{a,f} \geq 0$ and, hence, the Ansatz
\eqref{force-decomposition} with ${\bf f}_i^{\rm M}$ according to \eqref{fi-law-Soret}
and ${\bf f}_i^{\rm C}- r_i {\bf v}_i$ from \eqref{fiC} is consistent with the second law.
Note that the above estimation only works due to the symmetric form of the sums in \eqref{entropy-prod-chem3} and
the latter only results if {\it precisely} the term $\frac 1 2 r_i {\bf u}_i^2$ is shuffled within the entropy production.

To sum up, we obtained the closure law for the interaction force to read as
\begin{equation}
\label{interaction-force-reactive}
{\bf f}_i - r_i {\bf v}_i =
- \rho_i T \nabla \frac{\mu_i}{T} + \nabla p_i + (h_i+ d_i^T) T
\nabla \frac 1 T - \sum_{j=1}^N (T f_{ij} \rho_i \rho_j+C_{ij})
({\bf v}_i - {\bf v}_j)
\end{equation}
with symmetric $f_{ij}(T,\rho_1,\ldots ,\rho_N)>0$, and $C_{ij}$
from \eqref{def:Cij}.

Let us note in passing that the co-factors of the ${\bf u}_i$ in \eqref{entropy-prod-chem1} do not sum up to zero. As in Section~\ref{section4}, the latter can easily be achieved by addition of a Lagrange multiplier. But it is not needed for checking the entropy inequality, and this is all we had to do,
since the closure relation for the mechanical part has already been obtained before (where it
was important to employ a formulation in which the co-factors of ${\bf u}_i$ sum up to zero)
and the chemical part follows
from the balance of momentum transfer in chemical reactions. But for the reduction to a class-I
model below, the addition of a Lagrange multiplier will again be required.\\[1ex]
{\bf Chemical Reaction Kinetics}.
It remains to close the mass production rates $R_a=R_a^f -R_a^b$ which have to respect the entropy inequality as well.
The relevant reactive dissipation term is
\[
\zeta_{\rm chem} = - \frac 1 T \sum_{a=1}^{N_R} (R_a^f -R_a^b) \, \mathcal{ A}_a
\quad \mbox{ with the affinities } \mathcal{ A}_a = \sum_{i=1}^N \nu_i^a M_i \mu_i.
\]
Since chemical reactions are often far from equilibrium, a linear (in the driving forces) closure for $R_a$ is not appropriate.
Guided by the proof of Boltzmann's H-theorem as well as by experimental knowledge on reaction kinetics,
we use the nonlinear closure
\begin{equation}
\label{closure-chem-rates}
\ln \frac{R_a^f}{R_a^b}= - \gamma_a \frac{\mathcal{ A}_a}{RT}
\quad \mbox{ with } \gamma_a >0
\end{equation}
which implies
\[
\zeta_{\rm chem} = R \sum_{a=1}^{N_R} \frac{1}{\gamma_a}
(R_a^f - R_a^b) (\ln R_a^f - \ln {R_a^b}) \geq 0,
\]
since the logarithm is monotone increasing. Because of the strict
monotonicity of $\ln x$, the reactive contribution to the entropy
production only vanishes if {\it all reaction are separately in
equilibrium}, i.e.\ all forward and corresponding backward rates
coincide. This is an instance of the {\it principle of detailed
balance}, called Wegscheider's condition in the context of
chemical reaction kinetics. Notice that still one of the rates --
either for the forward or the backward path -- needs to be
modeled, while the other one then follows from
\eqref{closure-chem-rates}. Here, the approach via rigorous
continuum thermodynamics fixes the set of variables on which the
rates can depend. Note also that the factors $\gamma_a$ in
\eqref{closure-chem-rates} are often set to 1, but a general
$\gamma_a >0$ is required in the Butler-Volmer equation in
electrochemistry \cite{DGM13}. Note that the $\gamma_a$ can also
depend on the system's variables.

The standard closure for chemical reaction rates, used in so-called {\it elementary reactions} which do not involve intermediate steps,
is referred to as {\it mass action kinetics}, where the rate for the forward reaction, say, is modeled as
$R_a^f = k_a^f  \prod_i x_i^{\alpha_i}$ with ''rate constants'' $k_a ^f >0$
which depend on temperature $T$ and pressure $p$.
To obtain the backward rate via \eqref{closure-chem-rates}, the chemical potentials need to be modeled by material functions
since they appear in the affinities.
For the important class of so-called {\it ideal mixtures}, the chemical potentials obey the relations
\begin{equation}
\label{ideal-mixture}
\mu_i (T,p,x_1, \ldots , x_{N-1}) = g_i (T,p) + \frac{RT}{M_i} \log x_i \quad \mbox{ for } i=1,\ldots , N \;\mbox{ (with $\sum_i x_i =1$)},
\end{equation}
where $g_i (T,p)$ is the Gibbs energy of the pure component $A_i$
under the temperature and pressure of the mixture. In particular,
\eqref{ideal-mixture} holds for a mixture of ideal gases, where
$g_i$ is explicitly given as a function of $T,p$. Define $X_i$ for
$i=1,\ldots ,N$ such that $0< X_i < 1$,
\[
\prod_i X_i^{\nu_i^a} = \exp \big( -\frac{1}{RT} \sum_k M_k g_k (T,p) \nu_k^a \big)
\quad \mbox{ and } \quad \sum_i X_i=1;
\]
for details on the (nonempty) solution sets for this system see, e.g., section~2 in \cite{Bo-JDE}.
Then
\[
R_a = R_a^f -R_a^b = R_a^f \big( 1 - \prod_i \Big( \frac{x_i}{X_i}\Big)^{\nu_i^a} \big),
\]
hence if the forward rate is modeled as $R_a^f = k_a^f  \prod_i x_i^{\alpha_i}$, this yields
\begin{equation}
\label{full-rate}
R_a = k_a^f  \prod_i x_i^{\alpha_i} - k_a^b \prod_i x_i^{\beta_i}
\quad \mbox{ with } \; k_a^b=k_a^f / \prod_i {X_i}^{\nu_i^a}.
\end{equation}
Note that the ''rate constants'' $k_a^f, k_a^b >0$ are of course not constant but depend on the primitive variables. In practice, they especially depend strongly on the temperature.
%Let us note in passing that the latter dependence is often modeled by the {\it Arrhenius law} ''$k(T)=A \exp (-\frac{E_a}{RT})$''
%with $R$ the universal gas constant.
It is important to observe that only one of $k_a^f$ or $k_a^b$ can be modeled in this manner,
while the other one is then determined by \eqref{full-rate}.
However, due to \eqref{ideal-mixture} the result \eqref{full-rate} is restricted to the class of ideal
mixtures.

In the case of non-ideal mixtures, the literature usually starts also from \eqref{full-rate}, but with activities instead of molar
fractions; for a rather exhaustive review see \cite{Pekar}.
In the context of our representation of the chemical reaction rates it is obvious that the
corresponding chemical potentials are of the form
\[
\mu_i (T,p,x_1, \ldots , x_{N}) = g_i (T,p) + \frac{RT}{M_i} \log a_i \quad \mbox{ for } i=1,\ldots , N,
\]
where the so-called activities $a_i$ of course depend on all primitive variables.
Note that this just means replacing one unknown function by
another one with the same set of variables; cf.\ also \cite{M+M}. There is no advantage in this procedure, hence we prefer to employ \eqref{closure-chem-rates}, keeping the chemical potentials.
%
%
%
%
%
%
%%%%%%%%%%%%%%%%%%%%%%%%%%%%%%%%
\section{Viscous, chemically reactive fluid mixtures}
\label{sec:viscous-mixtures}
%{\bf Class-II model with partial viscosities.}
The next task is the closure of the viscous stresses ${\bf S}_i^{\rm irr}=-\Pi_i {\bf I} + {\bf S}_i^\circ$.
The viscous part of the entropy production in the class-II model is
\begin{equation}
\ts
\zeta_{\rm visc}= \frac 1 T \sum_i (\stress_i^\circ -\Pi_i \, {\bf I}): ({\bf D}_i^\circ +\frac 1 3 \, (\div {\bf v}_i) \, {\bf I})
= \frac 1 T \sum_i  \stress_i^\circ : {\bf D}_i^\circ -  \frac 1 T \sum_i \Pi_i \, \div {\bf v}_i.
\end{equation}
Employing linear constitutive theory, we obtain the constitutive relations
\begin{equation}
\label{stressesII}
\stress_i^\circ = \sum_j 2 \eta_{ij} {\bf D}_j^\circ
\quad \mbox{ and } \quad
\Pi_i = - \sum_j \lambda_{ij} \div {\bf v}_j
\end{equation}
with positive (semi-)definite matrices $[\eta_{ij}]$ and $[\lambda_{ij}]$,
where the shear and bulk viscosities $\eta_{ij}$ and  $\lambda_{ij}$, respectively, are functions of the primitive thermodynamic variables.
%The $\eta_{ii}\geq 0$ are the {\it dynamic partial viscosities}, the $\eta_{ij}$ for $i\neq j$ are the {\it dynamic cross-viscosities} and the $\lambda_{ij}$ are the {\it partial bulk viscosities} (also called partial volume or second viscosities), again with diagonal {\it partial bulk viscosities} $\lambda_{ii}\geq 0$ and {\it cross bulk viscosities} $\lambda_{ij}$ for $i\neq j$.

Let us note in passing that in our theory, $\stress_i^\circ$ is exclusively related to viscous
mechanisms; this is in contrast to {\it Extended Thermodynamics}
(cf.\ \cite{IM-Ruggeri}), where $\stress_i^\circ$ is itself a primitive variable
for which a balance equation holds.

Inserting the constitutive relations derived in Sections~\ref{section4}, \ref{sec:reacting-class-II}
and \ref{sec:viscous-mixtures}, we arrive at the following
class-II model for a reactive, viscous mixture with binary type mechanical interactions.\\[1ex]
{\bf Class-II model of a reactive, viscous fluid mixture:}
\begin{align}
& \pt \rho_i + \div (\rho_i {\bf v}_i)=\ts \sum_a M_i \nu_i^a (R_a^f -R_a^b)
\label{CIIa}
\\[1ex]
& \rho_i (\pt {\bf v}_i + {\bf v}_i \! \cdot \! \nabla {\bf v}_i)
= \div {\bf S}_i^{\rm irr} \! - \! \rho_i T \nabla \frac{\mu_i}{T}
+ \mathcal{D}_i^T T \nabla \frac 1 T \ts \! - \! \sum_j \! \big(T f_{ij} \rho_i
\rho_j + C_{ij}\big) ( {\bf v}_i \! - \! {\bf v}_j )
+ \rho_i {\bf b}_i \!\!
\label{CIIb}
 \\[1.5ex]
& \ts \pt (\rho e ) + \div ( \rho e {\bf v} + \alpha \nabla \frac
1 T + \sum_i \mathcal{D}_i^T ({\bf v}_i-{\bf v})) =
 -p \, \div {\bf v}+ \sum_i {\bf S}_i^{\rm irr} : \nabla {\bf v}_i \nonumber \\[1ex] \ts
& \qquad \ts -  \sum_i ({\bf v}_i-{\bf v}) \cdot \Big(\nabla p_i -
\rho_i T \nabla \frac{\mu_i}{T} + \mathcal{D}_i^T T \nabla \frac 1 T \ts  -
\sum_j \big( Tf_{ij}  \rho_i \rho_j + C_{ij}\big) ( {\bf v}_i  -
{\bf v}_j ) \Big), \label{CIIc}
\end{align}
where the ratio of reaction rates and the chemical matrix are given by
\begin{align*}
 R_a^f/R_a^b = \exp \big(- \frac 1 {RT} \sum_{k=1}^N \mu_k M_k\nu_k^a \big), & \qquad
C_{ij} = \sum_{a=1}^{N_a} \frac{M_i M_j}{\sum_k \alpha_k^a M_k} \big( R_a^f \beta_i^a \alpha_j^a + R_a^b \alpha_i^a \beta_j^a \big),
\end{align*}
and the irreversible stress parts $\stress_i^{\rm irr}=-\Pi_i
\,{\bf I} + \stress_i^\circ$ are determined by \eqref{stressesII}.
Moreover, $\mathcal{D}_i^T=h_i+d_i^T$ with the thermo-diffusivities $d_i^T$ from above.
%If full thermo-diffusion is to be considered, the partial
%enthalpies $h_i$ in \eqref{CIIb} and \eqref{CIIc} have to be
%replaced by $h_i+d_i^T$ with the thermo-diffusivities $d_i^T$ from
%above.

This system becomes a closed PDE-system for the primitive variables, if the following constitutive functions are given:
(i) the free energy function $\rho \psi (T,\rho_1,\ldots, \rho_N)$ to calculate $\rho e$ and $\mu_i$ according to
\eqref{rho-psi-partial-derivatives}$_{2,3}$; (ii) the partial energy $e_i(T,\rho_1,\ldots, \rho_N)$ and partial pressure
$p_i(T,\rho_1,\ldots, \rho_N)$ such that the constraints $\rho e=\sum_i \rho_i e_i$ and $p=\sum_i p_i$ are satisfied,
where $p$ follows from \eqref{HFE1}; (iii) constitutive models for, say, the forward rates $R_a^f$.

The class-II model from above is new.
In particular, the appearance of the chemical matrix for the reactive contribution to the interaction forces,
as well as the nonlinear and thermodynamically consistent closure of the chemical reaction rates are, to the best of
our knowledge, not present in the existing literature. Furthermore, our exploitation of the entropy principle
based on $\div {\bf v}_i$ instead of $\div {\bf v}$ leads to the simple derivation of partial and cross viscosities in case of viscous
fluid mixtures. Class-II models with partial and cross viscosities, but based on kinetic theory and without chemistry,
can be found in \cite{M68} and in \cite{Kerk}, \cite{Kerk-AicheJ}.
More details concerning related work will be given in Section~\ref{sec:literature-cII}
for models including thermo-diffusion.
\section{Cross-Effects Revisited}
\label{sec:cross-effects}
We already considered several cross-effects, namely cross-diffusion and thermo-diffusion.
In the present section we introduce additional couplings in which chemical reactions are involved. Before doing so, we first show how the cross-diffusion can also be derived as an entropy invariant mixing between the individual diffusion mechanisms.\\[0.5ex]
\noindent {\bf Cross-Diffusion}. In Section \ref{section4}, the
symmetry of the phenomenological matrix $[\tau_{ij}]$ relies on
the conservation law of barycentric momentum and on an assumption
of pairwise interactions. Now we give a further proof that relies
on the method of entropy invariant mixing. To this end we
reconsider the entropy production \eqref{entropy-prod1} due to
pure diffusion. To simplify the notation we abbreviate ${\bf U}_i={\bf u}_i-{\bf u}_N$
and ${\bf F}_i={\bf B}_i + \frac 1 T {\bf f}_i$
and may write
\begin{equation}\label{CER1}
    \zeta_{\rm diff} = - \sum_{i=1}^{N-1} {\bf U}_i \cdot {\bf
    F}_i~.
\end{equation}
For illustration of the method we do not use the general abstract setting
of Section \ref{sec:cross-coupling}, but introduce a reduced mixing between the individual
diffusive contributions ''by hand''.
For this purpose, we consider four
constituents only and have $\zeta_{\rm diff} = - ({\bf U}_1 \cdot
{\bf F}_1+{\bf U}_2 \cdot {\bf F}_2+{\bf U}_3 \cdot {\bf F}_3)$.
Here we introduce new terms without any effect on the entropy
production, viz.
\begin{align}
\zeta_{\rm diff} = -\Big(
({\bf U}_1+\lambda_{12}{\bf U}_2+\lambda_{13}{\bf U}_3 )\cdot {\bf F}_1
\hspace{1in}\nonumber \\
+({\bf U}_2 +\lambda_{23}{\bf U}_3)\cdot ({\bf F}_2-\lambda_{12}{\bf F}_1)
+{\bf U}_3 \cdot( {\bf F}_3 -\lambda_{13}{\bf F}_1-\lambda_{23}{\bf F}_2+\lambda_{23}\lambda_{12}{\bf F}_1)\Big).
\label{CER2}
\end{align}
According to Axiom~(IV-ii), a diagonal closure is required. Thus with
$\lambda_{11}\geq 0$, $\lambda_{22}\geq 0$ and $\lambda_{33}\geq
0$ we may let
\begin{align}
{\bf F}_1=-\lambda_{11}({\bf U}_1+\lambda_{12}{\bf
U}_2+\lambda_{13}{\bf U}_3 ),\quad {\bf F}_2-\lambda_{12}{\bf
F}_1=-\lambda_{22}({\bf U}_2 +\lambda_{23}{\bf
U}_3),\nonumber \\
%\label{CER3}
\label{CER4}
{\bf F}_3-\lambda_{13}{\bf F}_1-\lambda_{23}{\bf F}_2+\lambda_{12}\lambda_{23}{\bf F}_1=-\lambda_{33}{\bf U}_3.
\end{align}
This can be rewritten as ${\bf F}_i=-\sum_j \tau_{ij}{\bf U}_j$,
where the matrix of coefficients is given by
\begin{equation}\label{CER5}
    [\tau_{ij}]=\left(\begin{array}{cccccc}
   \lambda_{11} & \lambda_{11} \lambda_{12} &  \lambda_{11} \lambda_{13}\\[0.5ex]
\lambda_{11} \lambda_{12} & \lambda_{22} + \lambda_{11} \lambda_{12}^2 &
\lambda_{11} \lambda_{12} \lambda_{13} + \lambda_{22} \lambda_{23}\\[0.5ex]
\lambda_{11} \lambda_{13} & \lambda_{11} \lambda_{12} \lambda_{13} + \lambda_{22} \lambda_{23} &\lambda_{11} \lambda_{13}^2 + \lambda_{22} \lambda_{23}^2 + \lambda_{33}
\end{array}\right)
\end{equation}
and is symmetric as well as positive definite.
The above introduction of reduced cross-diffusion via \eqref{CER2}
corresponds to choosing an upper triangular matrix $A$ in \eqref{eq:mixed-entropy},
such that also $A_{ii}=1$ for all $i$. This does not cover all possibilities, which
are included if a general symmetric and positive definite matrix $[\tau_{ij}]$ is allowed
as in section~\ref{section4}.

The fact that the cross-diffusion in case of binary interactions was shown to be governed by
a symmetric matrix of interaction coefficients can now be interpreted as follows:
for binary interactions between species,
the barycentric momentum balance enforces entropy-neutrality of the cross-diffusion coupling.\\[1ex]
\noindent
{\bf Volume variation/chemical reaction cross-effect.}
The volume-variation/chemical reaction cross-effect refers to a possible coupling between
the following binary products in \eqref{entropy-inequalityII}:
\begin{equation}
\zeta_{\rm vol} = - \frac 1 T \sum_{i=1}^N \Pi_i \, \div {\bf v}_i
\quad \mbox{ and } \quad
\zeta_{\rm chem} = - \frac 1 T \sum_{a=1}^{N_R} R_a \, \mathcal{A}_a.
\end{equation}
The parity of the involved quantities are:
\begin{equation}
\label{paritiesII}
T\to+1,\qquad \Pi_i \to 1,\qquad \div {\bf v}_i \to -1,\qquad R_a \to -1,\qquad \mathcal{A}_a \to +1.
\end{equation}
The mixing inside the parity-classes can most easily be introduced via a zero addition according to
\begin{equation}
\zeta_{\rm chem} + \zeta_{\rm vol} = - \frac 1 T
\Big( \sum_{a=1}^{N_R} \big( R_a + \sum_{i=1}^N l_{ai}  \div {\bf v}_i \big) \, \mathcal{A}_a
\; + \;
\sum_{i=1}^N \div {\bf v}_i \, \big( \Pi_i - \sum_{a=1}^{N_R} l_{ai}  \mathcal{A}_a \big),
\Big).
\end{equation}
where the cross-coefficients $l_{ai}$ have arbitrary signs.
For simplicity we here rest content with linear constitutive laws.
Axiom~(IV-ii) and the procedure leading to \eqref{Onsager-closure} imply a
closure with symmetry between the coefficients, resulting in
\begin{equation}
R_a + \sum_{i=1}^N l_{ai}  \div {\bf v}_i = - \sum_{b=1}^{N_R} l_{ab} \mathcal{A}_b,
\qquad
\div {\bf v}_i = - \sum_{j=1}^N \lambda_{ij}^{-1}
\big( \Pi_j - \sum_{a=1}^{N_R} l_{aj}  \mathcal{A}_a \big)
\end{equation}
with positive (semi-)definite and symmetric matrices $[l_{ab}]$ and
$[\lambda_{ij}^{-1}]$, where we assume the latter to be invertible
with $[\lambda_{ij}^{-1}]=[\lambda_{ij}]^{-1}$.
This finally leads to the closure
\begin{equation}
\label{anti-symm-closure}
R_a  = - \sum_{b=1}^{N_R} l_{ab} \mathcal{A}_b -  \sum_{i=1}^N l_{ai}  \div {\bf v}_i,
\qquad
\Pi_i= - \sum_{j=1}^N \lambda_{ij} \div {\bf v}_j   + \sum_{a=1}^{N_R} l_{ai}  \mathcal{A}_a;
\end{equation}
cf.\ also \eqref{example-closure3} and \eqref{example-closure4} below.
Evidently, the mismatch between the standard ''flux/driving force'' notion and the parity-classes
explains the apparent anti-symmetry of the volume variation/chemical reaction cross-effect in \eqref{anti-symm-closure}.\\[1ex]
\noindent
{\bf Cross-effects between chemical reaction with nonlinear closure.}
Let us finally show how to introduce cross-effects also between different chemical reactions with the much more appropriate nonlinear closure for the rate functions from \eqref{closure-chem-rates}. The original contribution in the entropy production corresponding to the chemical reaction rates is $\zeta_{\rm R} = - \frac 1 T \sum_a R_a \mathcal{A}_a$.
After entropy neutral mixing between the different reactions, this becomes
\begin{equation}
\zeta_{\rm R} = - \frac 1 T \langle B {\bf R}, B^{-\sf T} \, {\bf A} \rangle
\quad \mbox{ with } \; {\bf R}=(R_1, \ldots ,R_{N_R})^{\sf T}
\mbox{ and } {\bf A}=(\mathcal{A}_1, \ldots ,\mathcal{A}_{N_R})^{\sf T}.
\end{equation}
Splitting ${\bf R}$ according to
\begin{equation}
{\bf R}={\bf R}^f-{\bf R}^b=(R_1^f-R_1^b, \ldots ,R_{N_R}^f -R_{N_R}^b)^{\sf T},
\end{equation}
we perform the diagonal but nonlinear closure
\begin{equation}
\ln \frac{(B {\bf R}^f )_a}{(B {\bf R}^b )_a}
= - \frac{\beta_a}{RT} (B^{- \sf T} {\bf A})_a
\quad \mbox{ with } \; \beta_a >0,
\end{equation}
where $B$ denotes a regular $N_R \times N_R$-matrix which is also positive, i.e.\ it has non-negative
entries. Note that the latter is required for exploiting the monotonicity of the logarithm to fulfil
the entropy inequality; it also means that $B$ leaves the positive cone invariant and
the mixing hence does not exchange the role of forward and backward reactions.

This leads to the closure relation
\begin{equation}
{\bf R}^b = B^{- 1} \Lambda B {\bf R}^f \quad \mbox{ with } \;
\Lambda = \mbox{diag}\, \big( \exp \big( \frac{\beta_a}{RT} (B^{-\sf T} {\bf A} )_a \big) \big)_{a=1,\ldots ,N_R},
\end{equation}
hence
\begin{equation}
\label{nonlinear-chemistry-coupling}
{\bf R}={\bf R}^f-{\bf R}^b  = (I - B^{- 1} \Lambda B ) {\bf R}^f,
\end{equation}
where the vector of forward rates ${\bf R}^f$ still is to be modeled based on a micro-theory or
on experimental experience. It actually suffices to have $\beta_a =1$ for all $a$, i.e.\ to
let
\begin{equation}
\Lambda = \mbox{diag}\, \big( e^{(B^{-\sf T} {\bf A} )_1/RT} , \ldots ,
e^{(B^{-\sf T} {\bf A} )_{N_R} /RT} \big)
\end{equation}
in \eqref{nonlinear-chemistry-coupling}, since {\it any} regular and positive
$N_R \times N_R$-matrix is admissible for $B$.
Therefore, letting $M=B^{-1}$ and $\alpha_b = (B {\bf R}^f )_b$,
this closure finally leads to
\begin{equation}
\label{cross-chem-CG}
R_a = \sum_{b=1}^{N_R} \alpha_b M_{ab} \Big( 1 - \exp( \sum_c M_{cb} \frac{\mathcal{A}_c}{RT}) \Big)
\end{equation}
with non-negative coefficients $\alpha_b$ and $M_{ab}$, being functions of $(T,\rho_1, \ldots ,\rho_N)$.
Note that the nonlinear closure \eqref{cross-chem-CG} does not exhibit the symmetry of the phenomenological matrix. However, in the limiting case $\mathcal{A}_c /RT \ll 1$ the matrix becomes symmetric, because we then have
\begin{equation}
R_a=-\sum_{b,c=1}^{N_R} \alpha_b M_{ab} M_{cb}\frac{\mathcal{A}_c}{RT}.
\end{equation}
The coupling of the chemical reaction rates according to \eqref{cross-chem-CG}
has first been proposed in \cite{Diss-Guhlke}. The above derivation shows how it fits
into our general scheme to introduce cross-effects via entropy invariant mixing.
\section{Comparison with Class-II models from the literature}
\label{sec:literature-cII}
As early as in 1868, J.\ C.\ Maxwell in his paper \cite{Max} ''On the dynamical theory
of gases'', building on and improving his own work, and three years later
J.\ Stefan in the paper \cite{Stef} ''\"{U}ber das Gleichgewicht und die Bewegung,
insbesondere die Diffusion von Gasgemengen'' were the first to formulate continuum mechanical
partial momentum balances, i.e.\ they derived the first known class-II models.
Maxwell considered binary gas mixtures and used kinetic theory to obtain the friction
force between two different moving gas species. Since, as Stefan wrote\footnote{\;''Das Studium der Maxwell'schen Abhandlung ist nicht leicht.'' From \cite{Stef}, p.\ 72.},
the paper by Maxwell is not easy to digest, he himself
gave a phenomenological derivation of essentially
the same equations, but generalized to the case of $N$ constituents.
Written in our notation, his final partial momentum balances read as
\begin{equation}
\label{Stefan}
\rho_i (\pt {\bf v}_i + ({\bf v}_i \cdot \nabla) {\bf v}_i) =
-\nabla p_i +\rho_i {\bf b}_i -\sum_{j\neq i} f_{ij} ({\bf v}_i - {\bf v}_j).
\end{equation}
He also derived a formula for the $f_{ij}$ in the binary case and noted
that these coefficients are symmetric,
depend weakly on the temperature and are otherwise constant for given types of gases.
Without rationale, he also used the symmetry for more than two constituents, thus
implicitly assuming binary interactions.
Stefan did not comment on equations of state for the partial pressures, but he successfully applied his model, in the stationary case and of course including partial mass balances, to describe experimental data for binary and also ternary systems which, as Stefan mentions
in \cite{Stef}, had been obtained by Loschmidt and by Wretschko.
So, at least for an isothermal mixture of ideal gases, one can say that
J.\ Stefan in \cite{Max} obtained the first closed class-II model for a fluid mixture
composed of $N$ constitutents.

Given these significant early contributions, it is somewhat
surprising that further developments on class-II models for fluid
mixtures are rarely found in the literature. Apparently, the next
notable and in a sense very modern contribution was given by C.\
Truesdell in 1957 in \cite{CT57}. In this and his later work
\cite{CT62} and \cite{CT69}, Truesdell made very important
progress in uniting the continuum mechanical approach started by
Stefan with thermodynamics, thus enabling a consistent derivation
within a single theory. Moreover, Truesdell showed that the
symmetry of $f_{ij}$ is a consequence of the conservation of total
momentum combined with binary interactions.
This was developed
considerably further by I.\ M\"{u}ller in particular in
\cite{M68}; see also \cite{M75}. Except for the momentum production due to chemical reactions, i.e.\ the appearance of the $C_{ij}$ which is a central part of our model,
M\"{u}ller already obtained the right-hand side of \eqref{CIIb},
however, with a different arrangement of terms.
At the same time, the phenomenological coefficients in this work are not related
to those that are available from the experimental side.
This might be the reason why this most important line of work, has been largely
ignored especially in the engineering community - unfortunately, even until today.
Let us note in passing that in this period quite some papers on rational thermodynamics
of mixtures appeared, but some of those contained significant misconceptions; cf.\ the
literature survey given in \cite{Bowen}.

In the engineering science, more practical mixture models where developed, sometimes
including partial momentum balances. An early example is \cite{Snell}, showing also
the main problem in this line of research: Partial momentum balances are set up in a
naive way, by adopting formulas from different theories to describe various force
contributions and to insert them without a rationale behind. In particular, the Gibbs-Duhem
relations are used in a non rigorous way. There are several more such contributions, but instead
of quoting them individually we only cite the recent paper \cite{Datta} which contains
a nice overview of the literature on partial momentum balances. Unfortunately, this paper
also employs the Gibbs-Duhem relation for isothermal systems in order to describe the
force due to chemical potential gradients which is then assumed to hold in all situations.

Still another line of research uses approximate solutions of the Boltzmann equations to
obtain formulas describing the interspecies forces which are then represented in primitive
variables of a continuum mechanical theory in which they are inserted.
This approach is especially used for the derivation of multicomponent diffusion fluxes
within a class-I framework and will thus be discussed further in Section~\ref{Sec:Maxwell-Stefan} on the Maxwell-Stefan equations.
A closed class-II model is derived in this way by Kerkhof in \cite{Kerk} and \cite{Kerk-AicheJ}, where he introduces partial momentum balances of the form
\begin{align*}
\rho_i \big( \pt  {\bf v}_i + {\bf v}_i \cdot \nabla {\bf v}_i \big) =
& -c_i \nabla_T \mu_i^m  - c R\hspace{0.25pt} T \sum_{j\neq i} \frac{ x_i \, x_j}{\D_{ij}} ({\bf v}_i -{\bf v}_j)
 + \rho_i {\bf b}_i\\
& - c R\hspace{0.25pt} T \sum_{j\neq i} \frac{ x_i \, x_j}{\D_{ij}}
\Big( \frac{D_i^T}{\rho_i} - \frac{D_j^T}{\rho_j} \Big) \nabla \ln T.
\end{align*}
Here $\mu_i^m$ denotes the molar-based chemical potentials which satisfy $\mu_i^m = M_i \mu_i$, and $\nabla_T$ denotes
the ''gradient taken at constant $T$''. For a sound comparison, note first that the expression by Kerkhof is obtained
from approximate solutions to the Boltzmann equations.
It is well-known that the multi-species Boltzmann equations are a model for mixture of ideal gases,
which represents a special simple mixture.
In particular, if $(T,p_i)$ are chosen
as independent variables, then $c_i \nabla_T \mu_i^m = \rho_i \nabla_T \mu_i = \rho_i \frac{\partial \mu_i}{\partial p_i} \nabla p_i
=\nabla p_i$. If we further let $\D_{ij} = R / (c M_i M_j f_{ij})$, then the first three terms on the
right-hand side coincide with those in \eqref{eq:momentum-classII-Soret-simple}.
To compare the thermo-diffusion term, recall that any function $d_i^T \hspace{-1pt} (T,\rho_1, \ldots ,\rho_N)$ with $\sum_i d_i^T =0$ is
admissible in \eqref{eq:momentum-classII-Soret} or \eqref{eq:momentum-classII-Soret-simple}. In particular,
\[
d_i^T = - c R\hspace{0.25pt} T \sum_{j\neq i} \frac{ x_i \, x_j}{\D_{ij}}
\Big( \frac{D_i^T}{\rho_i} - \frac{D_j^T}{\rho_j} \Big)
\]
is a valid choice, hence the class-II momentum balances of Kerkhof above are consistent to our result, but
specialized to the case of simple mixtures.

If the above form for the $d_i^T$ is used, the two sums in the partial momentum balance can be combined and become
\[
 - c R\hspace{0.25pt} T \sum_{j\neq i} \frac{ x_i \, x_j}{\D_{ij}}
\Big({\bf v}_i + \frac{D_i^T}{\rho_i T} \nabla T - ({\bf v}_j + \frac{D_j^T}{\rho_j T} \nabla T) \Big).
\]
This form might call for a suggestive interpretation in the framework of a kinetic gas theory: two colliding molecules had their last collision at slightly different positions, hence at different temperatures. Indeed, the term
${\bf u}_i + \frac{D_i^T}{\rho_i} \nabla \log T$ can be viewed as a corrected diffusion velocity
 in case of an inhomogeneous temperature field. But an explanation of
thermal diffusion along such a  line of arguments is only possible if the temperature
dependence of the collision frequencies is taken into account, i.e.\ it is a second-order
transport effect.
Instead of going into further details, we recommend Chapter~6 in \cite{FK}
and the original literature mentioned there.
A more thorough derivation via the Boltzmann equation gives the correct form and also leads to the above structure of the $d_i^T$.
Note that, at this point and for the case of a simple mixture of ideal gases,
the kinetic theory provides additional information not available in the pure phenomenological
continuum thermodynamics.
But also note that the Boltzmann equation only describes thermal diffusion if sufficiently realistic collision models are employed. The simple model of so-called {\it Maxwell molecules}, a favored choice since it allows for closed form computations of the collision integrals, does not include these effects,
since then $d_i^T=0$.
For more details on the Boltzmann equation we again refer to \cite{FK} and, in the case of
mixtures, however for Maxwell molecules, see \cite{HecklM}.

One main achievement of the present work is the consistent and coherent derivation of
closed class-II models within a continuum thermodynamical theory, including the
proper extension to chemically reactive fluid mixtures.
To the best of our knowledge, up to now no contribution treats the case of chemically
reacting fluid mixtures in such a way that the momentum change due to chemistry is
appropriately accounted for. One of the very few papers which explicitly include
a chemical contribution is \cite{Whitaker}, where the partial momentum balance
reads as
\begin{equation}
\label{partial-mom-Whit} \rho_i ( \pt  {\bf v}_i + {\bf v}_i \cdot
\nabla {\bf v}_i ) = \rho_i {\bf b}_i + \stress_i + \sum_{i=1}^N
{\bf P}_{ij} + r_i (\overline{\bf v}_i - {\bf v}_i).
\end{equation}
Here ${\bf P}_{ij}$ are the interaction forces between species
$A_i$ and $A_j$, assuming ${\bf P}_{ij}=-{\bf P}_{ji}$ without
rationale, and $\overline{\bf v}_i$ is introduced to be the
velocity of species $A_i$ produced by chemical reaction. Later on,
this velocity is not considered anymore. A similar case is
\cite{Datta}, where $\overline{\bf v}_i$ is actually assumed to
equal the partial bulk velocity ${\bf v}_i$ such that the effect
of chemical reactions is in fact eliminated. Note the difference
to our modeling in which every chemical reaction is considered
separately, since the (continuum averaged) velocity of $A_i$ produced by reaction $a$
depends on $a$, while for any particular reaction it is
independent of the species!
\section{Class-II $\rightarrow$ Class-I model reduction}
\label{sec:model-reduction1}
We introduce a new method for the derivation of the class-I
model via reduction of the corresponding class-II versions, avoiding scale-separation
arguments which are usually employed; cf.\ section~\ref{Sec:Maxwell-Stefan}
concerning the latter.\\[0.5ex]
\noindent
{\bf Reduction principle and its exploitation}. Recall first that
the basic variables of the class-I model are $\rho_i$, ${\bf v}$
and $T$. The relevant balance equations of the class-I model rely
on \eqref{partial-mass}, \eqref{total-momentum-balance} and
\eqref{tot-int-energy}. Here the inner structure of stress, energy
density and heat flux are ignored, and they are considered as
primitive quantities in the class-I setting.
%In particular, the characterizations as $e_{\rm int}$ and ${\bf q}_{\rm tot}$
%have no meaning and thus are suppressed.
Exclusively the inner structure of the external body force $\rho
{\bf b}$  and its internal power are taken into account. Thus the
relevant balance equations read
\begin{equation}\label{CImassA}
\ts \pt \rho_i + \div (\rho_i {\bf v} + {\bf j}^\textrm{I}_i)=
 \sum\limits_{a=1}^{N_R} M_i \nu_i^a R^\textrm{I}_a,\vspace{-0.1in}
\end{equation}
\begin{equation}\label{CImomentumA}
\rho(\pt {\bf v} + {\bf v}\cdot\nabla {\bf v} )-
\div \stress^\textrm{I} = \sum_{i=1}^N \rho_i {\bf b}_i,\vspace{-0.1in}
\end{equation}
\begin{equation}
\label{CIenergyA} \ts \pt (\rho e^\textrm{I}) + \div ( \rho
e^\textrm{I} {\bf v} +  {\bf q}^\textrm{I} ) =\stress^\textrm{I}:
\nabla {\bf v} + \sum\limits_{i=1}^N {\bf j}^\textrm{I}_i \cdot
{\bf b}_i.\vspace{-0.05in}
\end{equation}
Note the difference between \eqref{eq:total-internal-energy-balance} and
\eqref{CIenergyA}. In particular, the external body
forces do not explicitly occur in \eqref{eq:total-internal-energy-balance}.

The constitutive quantities in the class-I model are
\[
R^\textrm{I}_a, \quad p^\textrm{I}, \quad \stress^{\rm irr, I}=-\Pi^{\rm I}\, {\bf I} +
\stress^{\circ, {\rm I}},
\quad {\bf j}^\textrm{I}_i, \quad \rho e^\textrm{I} , \quad {\bf
q}^\textrm{I},
\]
and they must be related in a material dependent manner to the
variables.

We now compare the equations \eqref{CImassA}--\eqref{CIenergyA}
with the relevant equations of balance of the class-II model, viz.\
\begin{equation}\label{CIImassA}
\ts \pt \rho_i + \div (\rho_i {\bf v}+\rho_i {\bf u}_i)=
\sum\limits_{a=1}^{N_R} M_i \nu_i^a R_a,
\end{equation}
\begin{equation}\label{CIImomentumA}
\rho_i \big( \pt  {\bf v}_i +  {\bf v}_i \cdot \nabla {\bf v}_i
\big) - \div \stress_i = {\bf f}_i - r_i  {\bf v}_i + \rho_i {\bf
b}_i,
\end{equation}
\begin{equation}\label{CIIenergyA}
\ts \pt (\rho e) + \div ( \rho e {\bf v} +  {\bf q} ) = \sum_i
\stress_i^{\rm irr} : \nabla {\bf v}_i-  p \, \div {\bf v}
-  \sum_i {\bf u}_i \cdot \big( {\bf f}_i - r_i  {\bf v}_i -
\nabla p_i +  \frac{r_i}{2} {\bf u}_i \big).
\end{equation}
The mass balances of the species and the energy balance of the
mixture should be same in both models. Exclusively the partial
momentum balances are ignored in class-I, where they are replaced
by the momentum balance of the mixture.

Comparing the mass balances \eqref{CImassA} and \eqref{CIImassA},
we identify
\begin{equation}\label{Comp1}
    {\bf j}^\textrm{I}_i=\rho_i{\bf u}_i\quad\textrm{and}\quad
    R^\textrm{I}_a=R_a.
\end{equation}
To compare the energy balances \eqref{CIenergyA} and
\eqref{CIIenergyA}, we substitute the first term on the right-hand side
of \eqref{CIIenergyA} by the identity
\begin{equation}
    \sum_i
\stress_i^{\rm irr} : \nabla {\bf v}_i= \sum_i \stress_i^{\rm
irr}: {\bf D} +\div (\sum_i{\bf u}_i\cdot\stress_i^{\rm irr})-
\sum_i{\bf u}_i\cdot\div(\stress_i^{\rm irr}).
\end{equation}
Then we use \eqref{def:total-internal-energy} and
\eqref{def:reduced-heat-flux} and identify energies, heat fluxes
and stresses of both models:
\begin{equation}\label{Comp2}
    \rho e^\textrm{I}=\sum_i\rho_i e_i,\quad
{\bf q}^\textrm{I}=
\sum_i \big({\bf q}_i+(\rho_i e_i+p_i){\bf u}_i - {\bf u}_i\cdot\stress_i^{\rm irr} \big),
\quad\stress^{\rm I}=\sum_i\stress_i.
\end{equation}
Observe that $\stress^{\rm I}$ does not equal the mixture stress $\stress$ obtained
within class-II via Truesdell's meta-principle.
Next, the equivalence of the energy balances \eqref{CIenergyA}
and \eqref{CIIenergyA} requires the relation
\begin{equation}\label{Comp3}
    \rho_i{\bf b}_i=-{\bf f}_i+r_i{\bf v}_i
    -\frac{1}{2}r_i{\bf u}_i-\div \stress_i+\rho_i\lambda,
\end{equation}
where the Lagrange parameter $\lambda$ is introduced here to consider the constraints
$\sum_i{\bf f}_i=0$ and $\sum_i r_i=0$ and is thus given by
\begin{equation}\label{Comp4}
\rho\lambda=\rho{\bf b}-\frac{1}{2}\sum_i r_i{\bf u}_i+\div \sum_i \stress_i.
\end{equation}
We obtain
\begin{equation}\label{Comp5}
    {\bf f}_i-r_i{\bf v}_i = - \frac{1}{2}(r_i{\bf u}_i+y_i\sum_k r_k{\bf u}_k)-
    \div {\bf S}_i + y_i \, \div (\sum_k {\bf S}_k)
    -\rho_i({\bf b}_i-{\bf b}).
\end{equation}
Below, this relation will serve to eliminate the momentum
production ${\bf f}_i$ in the class-II consti\-tu\-tive law
\eqref{interaction-force-reactive} which then becomes a
constitutive equation for the class-I diffusion flux ${\bf j}^\textrm{I}_i$.

Now we discuss the status of the class-I momentum balance
\eqref{CImomentumA}. Recall at first that the mass balances of the
species and the energy balance of the mixture occur in class-II
and class-I as well. The essential difference of the two models
concerns the momentum balances. While we consider the $N$ partial
momentum balances of the species in class-II, we only have a
single momentum balance in class-I serving to determine the barycentric
velocity of the mixture. It is thus natural to attain equivalent
versions of the partial mass balances and the energy balance in
the two models. However, then the equivalence of the barycentric
momentum balance of the class-II model with the class-I momentum
balance can only be reached in an approximative sense.
If we were to have the relation ${\bf S}^\textrm{I}=\sum_i({\bf
S}_i-\rho_i{\bf u}_i\otimes{\bf u}_i)$, then \eqref{CImomentumA}
is a consequence of the class-II partial momentum balances
\eqref{CIImomentumA}. However, we have attained ${\bf
S}^\textrm{I}=\sum_i{\bf S}_i$ from the equivalence of the energy
balances. Thus our reduction class-II $\rightarrow$ class-I leads
to a class-I momentum balance that is an approximation because
terms of order $|{\bf u}|^2$ are ignored from the class-II point
of view.

Finally, we compare the entropy balances of both model classes.
The class-I entropy principle can be exploited in an analogous manner
to the above treatment in the model class-II. In particular, the class-I entropy balance
\begin{equation}\label{EIQ-I}
    \ts \pt (\rho s^\textrm{I}) + \div ( \rho s^\textrm{I} {\bf v} +  {\bf \Phi^\textrm{I}} )
    =\zeta^\textrm{I}
\end{equation}
implies
\begin{align}
\zeta^\textrm{I} \, =\,
\div (\Phi^{\rm I} - \frac{{\bf q}^{\rm I}}{T} +\sum_i \frac{\mu_i {\bf j}_i^{\rm I}}{T} )
- \frac 1 T \Big(  \rho e - T \rho s^\textrm{I} + p - \sum_i \rho_i \mu_i  \Big)\, \div {\bf v}
\nonumber
\\
\label{entropy-representation-I}
- \sum_i {\bf j}^\textrm{I}_i \cdot \Big( \nabla \frac{\mu_i}{T}
-\frac{{\bf b}_i}{T} \Big) - \frac 1 T \sum_{a=1}^{N_R} R_a^{\rm
I} \mathcal{A}_a + \frac 1 T  \stress^{\rm {I,irr}} : {\bf D} +
{\bf q}^{\rm I} \cdot \nabla \frac 1 T,
\end{align}
corresponding to the first representation of the class-II entropy production in
\eqref{entropy-production1}. As in the class-II case, the second term on the right-hand side
has to vanish, which again gives rise to the Gibbs-Duhem relation \eqref{Gibbs-Duhem},
now in the class-I context.
Next, the above identifications according to \eqref{Comp2} and \eqref{Comp5} are used
to compare with \eqref{entropy-production1}. This shows that the identity
\begin{equation}
\label{Phi+zeta}
\zeta^{\rm I}-\div \Phi^{\rm I} =\zeta^{\rm II}-\div \Phi^{\rm II}
\end{equation}
is valid which implies that the desired final identification of the specific entropies,
$s^{\rm I}=s^{\rm II}$, is consistent with those from above - a very satisfactory result,
since it also yields $T^{\rm I}=T^{\rm II}$ as well as $\mu_i^{\rm I}=\mu_i^{\rm II}$!
Note that we did not use different notations for the latter quantities in the different model
classes, although their identification was not a priori possible.

Now we can start with the model reduction, based on all identifications and on relation \eqref{Phi+zeta}. Before we continue, it is important to recall from the remarks on the
entropy principle that the representation of the entropy production as a sum of binary
products in not unique. Correspondingly, the choice of the entropy flux is not unique
and can be modified due to a zero addition to $\zeta$, adding also new dissipative mechanisms
and changing co-factors in binary products which are already present.
Actually, it is precisely this non-uniqueness which allows for improved class-I closure
relations building on the additional knowledge from class-II.
Moreover, since the system of axioms from our entropy principle applies also to the model class-I, the representation
$\zeta^{\rm I} =\zeta^{\rm II}+ \div \big( \Phi^{\rm I} - \Phi^{\rm II}\big)$,
as a consequence of \eqref{Phi+zeta}, and the fact that the entropy flux $\Phi^{\rm I}$
has to be chosen in such a way that $\zeta^{\rm I}$ is a sum of binary products, yields the
fundamental requirement
\begin{equation}\label{Comp6}
\zeta^{\rm I}\equiv \zeta^{\rm II}
\quad \mbox{ for any thermodynamic process}.
\end{equation}
Hence the only consistent way for a class-II $\to $ class-I model reduction is by means of
an {\it entropy invariant model reduction}.
Note also that the equality of the class-I and class-II entropy productions
in particular guarantees that the reduced class-I model is thermodynamically consistent,
i.e.\ the class-I closure relation obtained from the class-II constitutive laws always
respect the class-I entropy inequality - this is by no means a trivial property.\\[1ex]
%where the irreversible part $\stress^{\rm {irr,I}}=-\Pi \, {\bf I}
%+ \stress^\circ$  of the stress, and the affinities $\mathcal{
%A}_a$ appear. Note that ${\bf j}_i=\rho_i {\bf u}_i$ in the
%class-I model and we use both ways of writing this term.
%
%At a first glance it seems to be possible to employ the definition of the mixture quantities from
%Section~\ref{sec:balances}, such as ${\bf j}_i = \rho_i {\bf u}_i$, to compute the fluxes in the
%class-I model from the class-II quantities. But this does not lead to a closed class-I model,
%hence certain approximations - usually based on scale arguments - are then required to obtain
%a closed PDE-system. Then, it is not obvious at all whether the resulting fluxes satisfy the
%entropy inequality on the class-I level. In fact, while this works in the non-reactive inviscid case,
%it is in general wrong if chemical reactions are present.
\noindent
{\bf Class-I constitutive model}. For later comparison, let us start with a brief recap
of the standard closure relations from classical TIP.
Exploitation of \eqref{entropy-representation-I}, applying the axioms from the entropy principle in the class-I context and assuming as dissipative mechanisms the same as for the class-II model, except for using only a single viscous mechanism,
one obtains the TIP-version of the entropy flux as
\begin{equation}\label{EF-I}
    {\bf \Phi^\textrm{TIP}}=\frac{{\bf q^\textrm{I}}}{T}-\frac{\sum_i\mu_i{\bf
    j}_i^\textrm{I}}{T},
\end{equation}
and the corresponding entropy production as
\begin{equation}\label{entropy-inequality-I}
    \zeta^\textrm{TIP} \, =\,- \sum_i {\bf j}^\textrm{I}_i \cdot \Big( \nabla \frac{\mu_i}{T}
-\frac{1}{T}{\bf b}_i \Big) - \frac 1 T \sum_{a=1}^{N_R} R_a^{\rm
I} \mathcal{A}_a + \frac 1 T  \stress^{\rm {I,irr}} : {\bf D} +
{\bf q}^{\rm I} \cdot \nabla \frac 1 T.
\end{equation}
Based on
\eqref{entropy-inequality-I}, the simplest closure relations for
the constitutive quantities that vanish in equilibrium are
obtained by assuming {\it linear relations} between so called
fluxes and driving forces in the binary products of the entropy
production. This yields Fourier's law for the heat flux, Fick's
law for mass diffusion and Newton's law for the (deviatoric)
stress in viscous fluids. It is known from experimental
observations that cross-effects between the different dissipative
mechanisms like the Soret effect (thermal diffusion) and the
Dufour (diffusion thermo) effect occur, such that--in
principle--all possible couplings should be accounted for. This
introduces a large number of so-called phenomenological
coefficients which are often reduced by postulating the famous
Onsager reciprocity (i.e.\ symmetry) relations (cf.\ \cite{dGM}). For example, for
isotropic fluid mixtures one has the well-known relations
\begin{align}
\label{example-closure2} {\bf j}^\textrm{TIP}_i \; = \; & L_{i0}
\nabla \frac 1 T \; - \;
\sum_{j=1}^{N-1} L_{ij}\big( \nabla (\frac{\mu_j - \mu_N}{T})
-\frac{{\bf b}_j-{\bf b}_N}{T} \big),\\[0.5ex]
\label{example-closure1} {\bf q}^\textrm{TIP} \; = \; & L_{00}
\nabla \frac 1 T \; - \; \sum_{i=1}^{N-1} L_{0i}
\big( \nabla (\frac{\mu_i - \mu_N}{T})
 -\frac{{\bf b}_i-{\bf b}_N}{T} \big),\qquad\\[0.5ex]
 \label{example-closure5} \stress^{\textrm{TIP},\circ} \; = \; & \eta
\, {\bf D}^\circ.\\[0.5ex]
 \label{example-closure3} \Pi^\textrm{TIP} \; =
\; & - l_{00} \, \div {\bf v} \; - \;
 \sum_{a=1}^{N_R} l_{0a}\mathcal{A}_a,\\[0.5ex]
\label{example-closure4}
R^\textrm{TIP}_a \; = \; & - l_{a0} \, \div {\bf v} \; - \;
 \sum_{b=1}^{N_R} l_{ab}\mathcal{A}_b.
\end{align}
$\mbox{ }$\\[-1ex]
\noindent Further couplings are possible like via a term $\nabla
\frac 1 T \cdot {\bf D}$ with ${\bf D}$ the symmetric part of the
velocity gradient. However, those terms are excluded by assuming
linearity between fluxes and driving forces. In the class-I
setting within TIP this restriction is usually referred to as Curie's
principle. The Onsager-Casimir reciprocity relations for
\eqref{example-closure1}--\eqref{example-closure5} impose the
following symmetry properties:
\begin{itemize}
\item symmetry of the mobilities $[L_{ij}]$, including
$L_{0i}=L_{i0}$\vspace{0.05in} \item symmetry of the reaction
cross-couplings $[l_{ab}]$ for $a, b\geq 1$\vspace{0.05in} \item
anti-symmetry of the volume variation/chemical reaction coupling, i.e.\
$l_{0a}=\, - \, l_{a0}$\vspace{0.05in}
\end{itemize}
The last coupling with its apparently anti-symmetric character has
caused quite some confusion, while this issue is fully clarified in our present theory.
Note also that, according to \cite{IM-history}, a convincing proof of the Onsager
symmetry relations applied to transport phenomena is missing. Finally, the linear closure in
\eqref{example-closure4} might be inappropriate for
chemically reacting flows due to significant deviations from chemical equilibrium.\\[1ex]
\noindent
{\bf Class-I constitutive model via class-II $\rightarrow$ class-I reduction}.
In the next sections we derive the class-I constitutive equations for the diffusion fluxes,
heat fluxes, stresses and reaction rates that follow from the corresponding class-II equations.

We start with the two representations of the interaction forces
${\bf f}_i -r_i{\bf v}_i$. There is the class-II constitutive
equation \eqref{interaction-force-reactive} and we have
\eqref{Comp5} which is among the conditions guaranteeing the
equivalence of class-I and class-II energy balances. Eliminating
${\bf f}_i -r_i{\bf v}_i$ yields the constitutive equations for the
diffusion fluxes of class-I. At first we obtain
\begin{equation}
\label{jB}
    \sum_{j=1}^N (T \tau_{ij} - C_{ij} )({\bf u}_i-{\bf u}_j)
    +\frac{1}{2}\big( r_i{\bf u}_i + y_i \sum_{j=1}^N r_j{\bf u}_j \big) =
    {\bf d}_i
\end{equation}
with the abbreviation
\begin{equation}
\label{di}
    {\bf d}_i:=\rho_i T \nabla\frac{\mu_i}{T}-\rho_i ({\bf b}_i - {\bf b})
    -T \mathcal{D}_i^T\nabla\frac{1}{T}-\div {\bf S}^\textrm{irr}_i
    +y_i \,\div (\sum_{j=1}^N {\bf S}^\textrm{irr}_j ),
\end{equation}
where $\mathcal{D}_i^T:=h_i+d_i^T$; recall also that $y_i = \rho_i / \rho$ are the mass fractions.
Employing $\sum_{j=1}^N \tau_{ij}=0$ from \eqref{tau-NN} and setting
\begin{equation}
\label{def:gammaij}
\gamma_{ij} = -C_{ij} + \sum_{k=1}^N C_{ik} \delta_{ij} -\frac 1 2 r_i \delta_{ij}
-\frac 1 2 y_i r_j, \qquad
F_{ij} = T \tau_{ij} + \gamma_{ij},
\end{equation}
the system \eqref{jB} simplifies to read
\begin{equation}
\label{jB1}
   - \sum_{j=1}^N F_{ij} {\bf u}_j =  {\bf d}_i.
\end{equation}
Now note that $\sum_{i=1}^N {\bf d}_i=T \sum_{i=1}^N {\bf B}_i=0$ and also $\sum_{i=1}^N F_{ij}=0$ for all $j=1,\ldots ,N$. In order to obtain the class-I diffusion fluxes
${\bf j}^\textrm{I}_i=\rho_i{\bf u}_i$,
the system \eqref{jB1}, together with the constraint $\sum_{i=1}^N y_i {\bf u}_i=0$,
has to be solved for ${\bf u}_i$, given ${\bf d}_1,\ldots ,{\bf d}_N$ with $\sum_{i=1}^N {\bf d}_i=0$. There are essentially two ways to proceed: in the first variant, one keeps
\eqref{jB1} and studies the linear map ${\bf F}:\{ {\bf y} \}^\perp \to \{ {\bf 1} \}^\perp$
induced by the matrix $[F_{ij}]$, where ${\bf y}=(y_1, \ldots ,y_N)$ and
${\bf 1}=(1,\ldots ,1)\in \R^N$. The second variant, which will be employed here and
leads to the Fickean form of the diffusion fluxes, eliminates
${\bf u}_N$ in order to incorporate the constraint.
In both approaches it is crucial to know that the new chemical interaction matrix
$[\gamma_{ij}]$ is positive semi-definite on $\{ {\bf y} \}^\perp$.
The latter holds due to the estimates following equation \eqref{entropy-prod-chem3};
note that the positive semi-definiteness is only required on $\{ {\bf y} \}^\perp$ and, on this set, the $\gamma_{ij}$ can be replaced by
\begin{equation}
c_{ij} = -C_{ij} + \sum_{k=1}^N C_{ik} \delta_{ij} -\frac 1 2 r_i \delta_{ij}.
\end{equation}
Inserting the $C_{ij}$, it follows that
\begin{equation}\label{kleinC}
c_{ij}=\sum_{a=1}^{N_\textrm{R}}\big(
R^f_a(\delta_{ij}\frac{M_j\beta_j+M_j\alpha_j}{2}-\frac{M_i\beta_iM_j\alpha_j}{\sum_k
M_k\alpha_k})+
R^b_a(\delta_{ij}\frac{M_j\beta_j+M_j\alpha_j}{2}-\frac{M_j\beta_jM_i\alpha_i}{\sum_k
M_k\alpha_k})\big),
\end{equation}
hence $[c_{ij}]$ is positive semi-definite. Consequently, the matrix $[F_{ij}]$
is positive definite on $\{ {\bf y} \}^\perp$,
given that the $f_{ij}$ from \eqref{interaction-force-reactive}
are strictly positive for all $i\neq j$.
The matrix  $[F_{ij}]$ plays a central role in the Maxwell-Stefan form of the multicomponent
diffusion modeling; further information will be provided in Section~\ref{Sec:Maxwell-Stefan} below.\\[1ex]
\noindent
{\bf Diffusion fluxes.} To calculate the diffusion fluxes ${\bf
j}^\textrm{I}_i$ from \eqref{jB}, four steps are taken: (i) we
introduce in \eqref{jB} ${\bf j}^\textrm{I}_i=\rho_i{\bf u}_i$ and
consider the constraint $\sum_i{\bf j}^\textrm{I}_i=0$; (ii) we divide \eqref{jB} by $\rho_i T$;
(iii) we subtract the $N^\textrm{th}$ equation; (iv) we define the inverse mobility
matrix as
\begin{equation}\label{mobility}
    m^{-1}_{ij}=\tilde{F}_{ij}
    -\tilde{F}_{Nj}
    -\tilde{F}_{iN}
    +\tilde{F}_{NN}
\quad \mbox{ with } \;  \tilde{F}_{ij}= \frac{F_{ij}}{T\rho_i \rho_j}.
\end{equation}
Evidently, the matrix $[\tilde{F}_{ij}]$ is positive definite on $\{ {\bf y} \}^\perp$.
Given ${\bf w}\in \R^{N-1}$ with ${\bf w}\neq 0$, a direct computation shows that
\begin{equation}
\label{mobility-pd}
\sum_{i,j=1}^{N-1} m^{-1}_{ij} w_i w_j = \sum_{i,j=1}^N \tilde{F}_{ij} \tilde{w}_i \tilde{w}_j,
\end{equation}
if we let $\tilde{w}_i=w_i$ for $i<N$ and $\tilde{w}_N=-\sum_{i<N} w_i$.
Since $\tilde{\bf w} \perp {\bf y}$, the right-hand side in \eqref{mobility-pd} is
non-negative, hence $[m^{-1}_{ij}]$ is positive definite on $\R^{N-1}$.
Thus we have for $i=1,...,N-1$:
\begin{equation}\label{jB2}
    {\bf j}^\textrm{I}_i=-\sum_{j=1}^{N-1}m_{ij}
    \Big(\nabla \frac{\mu_j-\mu_N}{T} -\frac{{\bf b}_j-{\bf b}_N}{T}
    -(\frac{\mathcal{D}_j^T}{\rho_j}-\frac{\mathcal{D}_N^T}{\rho_N})\nabla\frac{1}{T}
    -\frac{\div({\bf S}^\textrm{irr}_j)}{T\rho_j}
    +\frac{\div({\bf S}^\textrm{irr}_N)}{T\rho_N}\Big).
\end{equation}
Comparison with the classical result
\eqref{example-closure2} from TIP shows that the class-II $\rightarrow$
class-I model reduction leads to a third driving force, viz.\
$\frac{1}{\rho_j}\div({\bf S}^\textrm{irr}_j)- \frac{1}{\rho_N}\div({\bf S}^\textrm{irr}_N)$
due to the viscous stresses; cf.\ the next section for a comparison to the literature
in the context of the Maxwell-Stefan equations.
Furthermore, we may identify
\begin{equation}\label{Onsager1}
    L_{ij}=m_{ij}\quad\textrm{and}\quad L_{i0}=
    \sum_{j=1}^{N-1} m_{ij}\big(\frac{\mathcal{D}_j^T}{\rho_j}-\frac{\mathcal{D}_N^T}{\rho_N} \big).
\end{equation}
Now observe that the $m_{ij}$ are symmetric if and only if the
$c_{ij}$ are, i.e.\ precisely if all chemical reactions are in
equilibrium ($R_a^f = R_a ^b$ for all $a=1,\ldots ,N_R$). In this
situation, the Onsager symmetry of $L_{ij}$, which is introduced
in the class-I model as a postulate, is implied by the class-II
$\rightarrow$ class-I reduction. We will take up this point again
below when we discuss thermo-diffusion.

%However, recall that the symmetry of $m_{ij}$ is based on a very
%simple assumption concerning binary interactions and on the
%conservation of the barycentric momentum. \BR ODER ALS
%ENTROPY-NEUTRALE MISCHUNG ! \ER \sout{This phenomenon was first
%observed by Truesdell, \cite{CT69}.} \BR HABEN WIR DAS NICHT VORNE
%SCHON? PASST M.E. HIER NICHT GUT HIN \ER

Note that the new driving force leads to a new and intricate PDE
structure. Due to the class-II constitutive equations
\eqref{stressesII}$_{1,2}$ for the viscous stresses, $\div{\bf j}_i^{\rm I}$ will appear on the right-hand side of \eqref{jB2} and thus there is no algebraic representation
of the diffusion fluxes.
However, this phenomenon can be avoided by a specific assumption on the partial viscosities.
A further, more detailed discussion will be carried out below.\\[1ex]
\noindent
{\bf Heat flux.} The determination of the class-I heat flux ${\bf q}^{\rm I}$
starts from \eqref{Comp2}$_2$, where we insert the class-II constitutive
equations \eqref{Fouriers-law-Dufour} for $\sum_i {\bf q}_i$ to arrive at
\begin{equation}
{\bf q}^\textrm{I} = \alpha \nabla \frac 1 T
+\sum_{i=1}^{N-1} \Big( \frac{\mathcal{D}_i^T}{\rho_i} - \frac{\mathcal{D}_N^T}{\rho_N}
+ \frac{\stress_i^{\rm irr}}{\rho_i} - \frac{\stress_N^{\rm irr}}{\rho_N}\Big)
{\bf j}_i^{\rm I}.
\end{equation}
Inserting the diffusion fluxes \eqref{jB2} and rearranging terms, we obtain
\begin{eqnarray}\label{qB1}
{\bf q}^\textrm{I} &=&
\bigg( \alpha + \sum_{i,j=1}^{N-1} m_{ij}
   \Big( \frac{\mathcal{D}_i^T}{\rho_i} - \frac{\mathcal{D}_N^T}{\rho_N}
+ \frac{\stress_i^{\rm irr}}{\rho_i} - \frac{\stress_N^{\rm irr}}{\rho_N}\Big)
\Big( \frac{\mathcal{D}_j^T}{\rho_j}-\frac{\mathcal{D}_N^T}{\rho_N} \Big) \bigg)
    \nabla\frac{1}{T}\nonumber\\
    &-&\sum_{i,j=1}^{N-1} m_{ij}
    \Big( \frac{\mathcal{D}_i^T}{\rho_i} - \frac{\mathcal{D}_N^T}{\rho_N}
+ \frac{\stress_i^{\rm irr}}{\rho_i} - \frac{\stress_N^{\rm irr}}{\rho_N}\Big)
        \Big(\nabla(\frac{\mu_j-\mu_N}{T})+\frac{{\bf b}_j-{\bf
    b}_N}{T}\Big)\\
    &-&\sum_{i,j=1}^{N-1} m_{ij}\Big( \frac{\mathcal{D}_i^T}{\rho_i} - \frac{\mathcal{D}_N^T}{\rho_N}
+ \frac{\stress_i^{\rm irr}}{\rho_i} - \frac{\stress_N^{\rm irr}}{\rho_N}\Big)
\Big(\frac{\div \stress_j^{\rm irr}}{T \rho_j}-
\frac{\div \stress_N^{\rm irr}}{T \rho_N}\Big).\nonumber
\end{eqnarray}
Recall that both $\alpha$ and the $m_{ij}$ are functions of $T$ and all $\rho_i$.
If we keep only linear terms of those quantities that vanish in equilibrium, we finally get
 \begin{eqnarray}\label{qB2}
{\bf q}^\textrm{I} &=&
\bigg( \alpha + \sum_{i,j=1}^{N-1} m_{ij}
   \Big( \frac{\mathcal{D}_i^T}{\rho_i} - \frac{\mathcal{D}_N^T}{\rho_N} \Big)
\Big( \frac{\mathcal{D}_j^T}{\rho_j}-\frac{\mathcal{D}_N^T}{\rho_N} \Big) \bigg)
    \nabla\frac{1}{T}\nonumber\\
    &-&\sum_{i,j=1}^{N-1} m_{ij}
    \Big( \frac{\mathcal{D}_i^T}{\rho_i} - \frac{\mathcal{D}_N^T}{\rho_N} \Big)
        \Big(\nabla(\frac{\mu_j-\mu_N}{T})+\frac{{\bf b}_j-{\bf
    b}_N}{T}\Big)\\
    &-&\sum_{i,j=1}^{N-1} m_{ij}\Big( \frac{\mathcal{D}_i^T}{\rho_i} - \frac{\mathcal{D}_N^T}{\rho_N} \Big)
\Big(\frac{\div \stress_j^{\rm irr}}{T \rho_j}-
\frac{\div \stress_N^{\rm irr}}{T \rho_N}\Big).\nonumber
\end{eqnarray}
This is the constitutive equation for the class-I heat flux
derived by a class-II $\rightarrow$ class-I reduction.
By comparing \eqref{qB2} with \eqref{example-closure1} we
again observe the further driving force $\frac{1}{\rho_j}\div({\bf S}^\textrm{irr}_j)-
\frac{1}{\rho_N}\div({\bf S}^\textrm{irr}_N)$ due to the viscous stresses.

Furthermore, we identify
\begin{equation}\label{Onsager2}
    L_{00}=\alpha+\sum_{i,j}m_{ij}
    (\frac{\mathcal{D}_i^T}{\rho_i}-\frac{\mathcal{D}_N^T}{\rho_N})
    (\frac{\mathcal{D}_j^T}{\rho_j}-\frac{\mathcal{D}_N^T}{\rho_N})
    \quad\textrm{and}\quad L_{0i}=
    \sum_{j} m_{ji}(\frac{\mathcal{D}_j^T}{\rho_j}-\frac{\mathcal{D}_N^T}{\rho_N}).
\end{equation}
A comparison with \eqref{Onsager1} immediately implies that
$L_{0i}=L_{i0}$ only if there is the symmetry of the mobility
matrix $[m_{ij}]$. Recall its definition by \eqref{mobility} and
\eqref{def:gammaij} and observe that $m_{ij}=m_{ji}$ only holds if
the chemical contribution to \eqref{def:gammaij} is symmetric,
i.e. $C_{ij}=C_{ji}$. However, according to \eqref{def:Cij} the
latter symmetry only holds in the case of chemical equilibrium,
i.e. $R^f_a=R^b_a$  for all $a=1,\ldots ,N_R$.  Thus if there are
chemical reactions outside of equilibrium the Onsager reciprocity
relations $L_{0i}=L_{i0}$ do not hold. Moreover, in this case the
cross-effect of \textit{thermo-diffusion} cannot be introduced by
entropy neutral mixing.\\[1ex]
%
%the chemical contribution to if the Recall that this symmetry must
%be introduced in the class-I model as a postulate, while via the
%class-II $\rightarrow$ class-I model reduction, the Onsager
%reciprocity relations turn out as a consequence.
%\BR COMPARISON MIT TIP HINZUFUEGEN? \ER
%
\noindent
{\bf Viscous stress and viscous pressure.} We start from the
identification of the class-I stress \eqref{Comp2}$_3$. There we
insert the class-II results \eqref{stressesII} and use
${\bf v}_i={\bf v}+{\bf j}_i^{\rm I}/\rho_i$ to introduce the diffusion fluxes.
Then we obtain
\begin{equation}
\label{stressCII} \stress^{\textrm{I},\circ} = \sum_{i,j=1}^N 2\,
\eta_{ij} \Big( {\bf D}^\circ + \frac 1 2 \big( \nabla \frac{{\bf
j}_j^{\rm I}}{\rho_j} + (\nabla \frac{{\bf j}_j^{\rm
I}}{\rho_j})^{\sf T} \big) \Big), \qquad \Pi^\textrm{I} =
-\sum_{i,j=1}^N \lambda_{ij} \big( \div {\bf v} + \div \frac{{\bf
j}_j^{\rm I}}{\rho_j} \big)
\end{equation}
with the material parameters $\eta_{ij}$ and  $\lambda_{ij}$ from
the closure of the class-II model.
This result implies a new and very intricate PDE structure. Due to
\eqref{jB2} and \eqref{stressesII}, at least third-order
derivatives of the velocities appear.

A first simplification of the model is achieved if we consider the special case in which
the class-II viscosities in \eqref{stressesII} and, hence, in \eqref{stressCII}
satisfy the relation
\begin{equation}\label{SC1}
\eta_{ij}= y_i \, \eta_{j} \quad \mbox{ and } \quad \lambda_{ij} =
y_i \, \lambda_{j}.
\end{equation}
The point is that this implies
\begin{equation}
\stress_i^\circ=y_i\, \stress^{\textrm{I},\circ}
\quad \textrm{and}\quad
\Pi_i=y_i\, \Pi^\textrm{I}.
\end{equation}
Consequently, the new driving force--which originates the intricate
structure--simplifies in this case according to
\begin{equation}
\frac{1}{\rho_j} \div \stress^\textrm{irr}_j-
\frac{1}{\rho_N} \div \stress^\textrm{irr}_N=
(\frac{1}{\rho_j}\nabla y_j -\frac{1}{\rho_N}\nabla y_N)
{\bf S}^\textrm{irr}.
\end{equation}
Then the diffusion fluxes and the heat flux, \eqref{jB2} and
\eqref{qB2}, algebraically depend on $\stress^{\textrm{I},\circ}$
and the viscous pressure $\Pi^{\textrm{I}}$.
But note that the stresses still depend on $\div({\bf S}^\textrm{irr}_j)$.

If the material parameters are further restricted to satisfy
\begin{equation}
\eta_{ij}= y_i \, \eta / N \quad \mbox{ and } \quad
\lambda_{ij} = y_i \, \lambda / N
\end{equation}
with common functions $\eta \geq 0$ and $\lambda \geq 0$, we
obtain
\begin{equation}
\stress^\circ
= 2 \eta {\bf D}^\circ -
 \eta /N \sum_{j=1}^N \big( \nabla y_j \otimes \frac{{\bf j}^\textrm{I}_j}{\rho_j}
  + \frac{{\bf j}^\textrm{I}_j}{\rho_j} \otimes \nabla y_j \big),
 \quad
\Pi = -\lambda \, \div {\bf v} + \frac \lambda N \, \div \big( \sum_{j=1}^N
\frac{{\bf j}^\textrm{I}_j}{\rho_j} \big).
\end{equation}
In this case the viscous stresses algebraically depend on the diffusion fluxes.
But observe that insertion of these stresses into the class-I momentum balance still
leads to a PDE of third order. Here it is important to note that the PDE-system has a triangular
structure, where the mass densities, up to perturbations of lower
order, can be solved prior to insertion into the constitutive
stress relation. This allows to solve the system in a regularity
setting, where the mass fluxes in the stress do not give rise to
the leading order terms concerning the existence analysis.

Finally note that all new terms are formed by binary products of
$\nabla y_i$, ${\bf S}^\textrm{irr}_i$ and ${\bf j}^\textrm{I}_i$, respectively.
Thus when $\nabla y_i=0$ in equilibrium, which typically is the case in the setting of this
model, we have nonlinear terms in driving forces and fluxes.
However, those products are usually ignored in class-I models.
In this case we have agrement with TIP and may identify
$l_{00}=\lambda$ and $l_{0a}=0$ according to \eqref{example-closure3}.
There results no cross-coupling with the chemical production rates in the
class-II $\rightarrow$ class-I reduction because we have not considered those couplings
in the class-II setting.\\[1ex]
%\BR SOLLEN WIR NOCH ETWAS ZUM VOLL GEKOPPELTEN FALL SCHREIBEN? NEIN. WD\ER
%
%
%\noindent
%{\bf Reaction rates.} According to \eqref{Comp1}$_2$, the ratio of
%the reaction rates is the same in both models. We have
% \begin{equation}
%R_a^{\rm I}= {R_a^f} \Big( 1 - \exp \big(\gamma_a \frac{\mathcal{
%A}_a}{RT}\big) \Big) \quad \mbox{ with } \gamma_a >0,
%\end{equation}
%where the forward rate ${R_a^f}$ has to be modeled as shown in
%Section~\ref{sec:reacting-class-II}.
%
%If $\frac{\mathcal{ A}_a}{RT}\ll 1$, i.e. near to chemical
%equilibrium, we may expand and obtain
%\begin{equation}
%R_a^{\rm I}= -\frac{\gamma_a{R_a^f}}{RT}\mathcal{ A}_a.
%\end{equation}\BA
%By comparison with the class-I results \eqref{example-closure4} we
%identify $l_{ab}=\gamma_a{R_a^f}/RT\delta_{ab}$ and $l_{a0}=0$.\\[1ex]\EA
%There are again no cross-effects here because they are not introduced in the class-II model.
%
\noindent
{\bf Entropy flux.} Let us finally compare the entropy fluxes. Inspection of
\eqref{entropy-production1}, \eqref{entropy-representation-I} and \eqref{Comp6}
shows that
\begin{equation}
\Phi^{\rm I} = \frac{{\bf q^\textrm{I}}}{T}-\frac{\sum_i\mu_i{\bf j}_i^\textrm{I}}{T}
+ \frac{\sum_i {\bf u}_i \cdot \stress_i^\textrm{I}}{T},
\end{equation}
hence $\Phi^{\rm I}$ does not coincide with the entropy flux from TIP, but
\begin{equation}
\Phi^{\rm I} = \Phi^{\rm TIP}
+ \frac{\sum_i {\bf u}_i \cdot \stress_i^\textrm{I}}{T}.
\end{equation}
Note, however, that $\Phi^{\rm I}=\Phi^{\rm TIP}$ does hold if the partial viscosities are restricted to satisfy \eqref{SC1}.
In total and compared to classical TIP,
the modifications of the entropy flux and the constitutive relations obtained
from the class-II $\to$ class-I model reduction correspond to the addition of
\begin{equation}
\label{zero-addition}
0=\frac 1 T \sum_i \nabla {\bf u}_i : \stress_i^\textrm{irr}
+ \frac 1 T \sum_i {\bf u}_i \cdot \div \stress_i^\textrm{irr}
+ \sum_i ({\bf u}_i \cdot \stress_i^\textrm{irr}) \cdot \nabla \frac 1 T
- \div (\frac{\sum_i {\bf u}_i \cdot \stress_i^\textrm{I}}{T}).
\end{equation}
to $\zeta^{\rm I}$ from \eqref{entropy-representation-I}. This corresponds to the addition
of the new dissipative mechanism related to diffusion against partial stresses and this
additional entropy producing term modifies
the diffusion fluxes, the heat flux and the entropy flux.
\section{The Maxwell-Stefan equations for reactive fluid mixtures}
\label{Sec:Maxwell-Stefan} In the previous section, inversion of
the system \eqref{jB1} has led to the Fickean form of the
diffusivities according to \eqref{jB2}, sometimes also referred to
as the generalized Fick-Onsager form. Let us note that after
insertion of particular forms of the chemical potentials,
respectively the choice of a specific free energy function, the
representation from \eqref{jB2} is usually written in terms of the
partial mass densities or, equivalently, the molar concentrations.
In simplest cases, this leads to relations of the form
\begin{equation}
\label{mcFick}
{\bf j}_i = - \sum_j D_{ij} \nabla \rho_j
\end{equation}
with so-called Fickean diffusivities $D_{ij}$. There are two
disadvantages of the representation \eqref{jB2}, respectively \eqref{mcFick}:
(i) the symmetry between the different species
$A_i$ has been broken by eliminating one particular flux, namely
${\bf j}_N$; (ii) the mobilities $m_{ij}$ show complicated
dependencies on, especially, the composition of the mixture. The
first point is relevant for a mathematical analysis of the final
PDE-system, where the symmetry of the system \eqref{jB1} in the
${\bf u}_j$ is helpful. Besides binary mixtures, the Fickean form
is appropriate in case the mixture is a solution with one major
component (the solvent $A_N$, say) and several minor constituents
(the solutes). Then, if the solution is dilute, i.e.\ if $y_i \ll
1$ for all $i\neq N$, the fluxes even reduce to the classical form
of ${\bf j}_i = - D_{i} \nabla \rho_i$ as discovered by Adolf
Fick; cf.\ \cite{Fick}. The second point is much more severe, due
to the following mathematical fact. A coupled system of diffusion
equations of the form which results if \eqref{mcFick} is inserted
into the partial mass balances, say with ${\bf v}=0$ for
simplicity (which is irrelevant for the local-in-time existence
properties), the solutions for positive initial values will in
general \underline{not} stay positive in case of constant
diffusivities $D_{ij}$, even if the diffusion operator is
elliptic. This underlines the fact that the diffusivities must
depend on the composition and the form of this dependency is
decisive concerning the well-posedness as well as qualitative
properties of the system. Therefore, an alternative description
which includes structural information on the diffusivities such
that the final PDE-system becomes solvable is strongly demanded.
Of course, if such a description is given, the associated Fickean
fluxes can, in principle, be calculated, since the different
representations are equivalent. But let us also note in
passing, that if the final form according to \eqref{mcFick} is
used, the $D_{ij}$ are in general not even symmetric.

Motivated by the above facts, we provide the Maxwell-Stefan form of the multicomponent
diffusion description which, essentially, corresponds to \eqref{jB}. This representation is,
in the absence of chemical reaction, fully symmetric in the constituents and contains additional structural information which solves the problem above as will be explained below.
Supported by statistical mechanics, it even turns out that the ''friction coefficients''
$f_{ij}$ from \eqref{fij-def} are only weakly dependent on the mixture composition.
Furthermore, experiments have shown that this
weak dependence can often be described via affine functions; cf. \cite{TK-book} where also further references are given.

Right from the beginning, there where two different derivations of what is now called the Maxwell-Stefan form. Maxwell in his classical paper \cite{Max} used kinetic theory of gases
to derive a relation for the diffusion velocity of a binary mixture of simple gases.
In order to be able to provide final results for the diffusion coefficient by analytical formulas, he assumed the dependence as $r^{-5}$ of the intermolecular force fields, thus
introducing the nowadays so-called Maxwellian molecules; recall that there is no thermo-diffusion in this case. Shortly after this, Stefan essentially gave a continuum mechanical
derivation in \cite{Stef}, valid for a system of $N$ constituents. He employed the assumption
that\footnote{''In einem Gemenge erf\"{a}hrt jedes einzelne Theilchen eines Gases, wenn es sich bewegt, von jedem anderen Gase einen Widerstand proportional der Dichte dieses Gases und der
relativen Geschwindigkeit beider.'' From \cite{Stef}, p.\ 64.}
every particle of a gas, if it is moving, encounters a
resistive force by every other gas, being proportional to the density of that gas and to the
relative velocity between the two. As already noted above, he used this to formulate the partial momentum balance
\[
\rho_i (\pt {\bf v}_i + ({\bf v}_i \cdot \nabla) {\bf v}_i) =
-\nabla p_i +\rho_i {\bf b}_i -\sum_{j\neq i} f_{ij} ({\bf v}_i - {\bf v}_j).
\]
He actually was also aware that, contrary to Maxwell's derivation, his force balance
can be applied to liquid mixtures as well.

Later on, building on the work by Boltzmann and his famous
equation, use of statistical mechanics gave rise to several types
of approximate solutions to the multi-species Boltzmann equations.
Hirschfelder, Curtiss, and Bird in particular obtained the
so-called generalized driving forces ${\bf d}_i$ which were to
replace the partial pressure $p_i$ in \eqref{Stefan}; see
\cite{Hirsch} and cf.\ also \cite{Bird}. These have then been used
to formulate the reduced force balances
\begin{equation}
\label{MS-naiv}
-\sum_{j\neq i} f_{ij} \rho_i \rho_j ({\bf u}_i -{\bf u}_j) = {\bf d}_i,
\end{equation}
or other equivalent versions thereof like
\begin{equation}
\label{MS-naiv2}
-\sum_{j\neq i} \frac{x_j {\bf J}_i - x_i {\bf J}_j}{c\, \D_{ij}} = {\bf d}_i.
\end{equation}
The latter form is found in most textbooks on multicomponent diffusion, where
$c=\sum_i c_i$ with the molar concentrations $c_i=\rho_i / M_i$, $x_i=c_i/c$ are
the molar fractions, ${\bf J}_i=c_i {\bf u}_i$ the molar mass diffusion fluxes and
$\D_{ij}$ denote the so-called Maxwell-Stefan diffusivities.
But neither a rigorous derivation of \eqref{MS-naiv}
by a time-scale separation argument--except for \cite{Whitaker}--is given,
nor a consistent derivation of the driving forces; the latter have been transferred to the
continuum mechanical force balance from approximate solutions to the Boltzmann equations. Fortunately, it turned out that the resulting diffusive fluxes obey the second law of thermodynamics. While in the engineering literature this is often attributed to \cite{Standart}, it has already been shown much earlier by Truesdell in \cite{CT62}. In the latter paper,
it has also been shown that the Onsager relations of irreversible thermodynamics are in case
of multicomponent diffusion processes nothing but the symmetry of the $f_{ij}$ from above
and that the latter rigorously follows in case of binary interactions from the
total momentum balance.
The system of type \eqref{MS-naiv} or \eqref{MS-naiv2} is nowadays referred to
as the Maxwell-Stefan equations.

Recall that the main advantage of the
Maxwell-Stefan form \eqref{MS-naiv2} is its additional structural information
and the fact that the $\D_{ij}$ turn out to depend only
weakly on the composition. For binary systems, the MS-diffusivity $\D_{12}$ is actually
independent of the composition $(x_1,x_2)$ of the mixture; cf.\ \cite{FK}.
Hence the same is true for the  MS-diffusivities in a dilute mixture, since only binary
interactions between each solute and the solvent occur.

We are now in position to briefly give three different derivations of the Maxwell-Stefan
equations. This is helpful in order to place our new extension of the Maxwell-Stefan
equations to the case of chemically reacting mixtures into the proper context.\\[1ex]
\noindent {\bf Derivation of the Maxwell-Stefan equations within
TIP.} Once the two above-mentioned alternatives are present, it is observed
that--at least in the isothermal case--one can indeed obtain the Maxwell-Stefan equations together
with a set of thermodynamically consistent driving forces by means
of the ''resistance form'' of the closure within TIP, as opposed
to the standard ''conductivity form''; cf.\ \cite{KB-book}. To
briefly illustrate this, we rewrite the TIP entropy production
from \eqref{entropy-inequality-I} for isothermal processes according to
\begin{equation}\label{entropy-inequality-Ib}
    \zeta^\textrm{TIP}_\textrm{isotherm} \, =\,- \sum_i {\bf u}_i \cdot \Big( \rho_i \nabla \frac{\mu_i}{T}
-\frac{\rho_i {\bf b}_i}{T}+y_i \lambda \Big) - \frac 1 T \sum_{a=1}^{N_R} R_a \mathcal{A}_a
+ \frac 1 T  \stress^{\rm {irr}} : {\bf D}
%+ {\bf q} \cdot \nabla \frac 1 T
,
\end{equation}
where we dropped the superscript indicating model class-I and introduced a Lagrange parameter
$\lambda$. This parameter is now used to obtain
\begin{equation}\label{entropy-inequality-Ic}
    \zeta^\textrm{TIP}_\textrm{isotherm} \, =\,-\frac 1 T \sum_i {\bf u}_i \cdot {\bf d}_i
    - \frac 1 T \sum_{a=1}^{N_R} R_a \mathcal{A}_a
+ \frac 1 T  \stress^{\rm {irr}} : {\bf D}
%+ {\bf q} \cdot \nabla \frac 1 T
,
\end{equation}
such that the driving forces ${\bf d}_i$ satisfy $\sum_i {\bf d}_i =0$. Employing the Gibbs-Duhem relation, the latter implies
\begin{equation}
\label{di-naiv}
{\bf d}_i =\rho_i T \nabla \frac{\mu_i}{T} -y_i \nabla p -\rho_i ( {\bf b}_i-{\bf b}).
\end{equation}
Incorporating the constraint on the ${\bf d}_i$, the entropy production becomes
\begin{equation}\label{entropy-inequality-Id}
    \zeta^\textrm{TIP}_\textrm{isotherm} \, =\,- \frac 1 T \sum_{i=1}^{N-1} ({\bf u}_i-{\bf u}_N) \cdot {\bf d}_i- \frac 1 T \sum_{a=1}^{N_R} R_a \mathcal{A}_a
+ \frac 1 T  \stress^{\rm {irr}} : {\bf D}
%+ {\bf q} \cdot \nabla \frac 1 T
,
\end{equation}
such that we may now repeat precisely the arguments from Section~\ref{section4}
to obtain
\begin{equation}
\label{MS-naiv3}
{\bf d}_i = \sum_{j\neq 1}^N \tau_{ij} ({\bf u}_i -{\bf u}_j),
\end{equation}
where the matrix $[\tau_{ij}]$ has the same properties as the one
in our derivation above. Exploiting Truesdell's argument from
Section \ref{section4} as before, the symmetry of $[\tau_{ij}]$
follows and the binary interaction assumption together with the
positive definiteness again yields $\tau_{ij}=- f_{ij} \rho_i
\rho_j$ with $f_{ij}=f_{ji}>0$ for all $i\neq j$, hence the
Maxwell-Stefan form in the isothermal case.

But note that this derivation does not imply anything on
the dependence of the $f_{ij}$ as functions of the primitive
variables $(T,\rho_1, \ldots ,\rho_N)$. It also does not add
to the understanding of the underlying mechanism as being a
consequence of partial linear momentum exchange. Moreover, note
that this only leads to the standard Maxwell-Stefan equations
in which no information is contained about the diffusion directly
related to momentum transfer during chemical reactions.

In the general, non-isothermal case, the ${\bf d}_i$ from \eqref{di-naiv}
get an additional term $y_i (\rho e +p) T \nabla \frac 1 T$ if the same derivation is used.
This does not coincide with the generalized thermodynamic driving force from \cite{Hirsch}, \cite{Bird}, where the additional term is $h_i T \nabla \frac 1 T$ with the partial enthalpies $h_i$. The latter is precisely the form which we also get below, either in the diffusional
approximation or via the entropy invariant model reduction. This defect of the above derivation
within class-I can only be avoided if the full thermo-diffusive cross-coupling is employed.
\\[1ex]
\noindent
{\bf Derivation of the Maxwell-Stefan equations via diffusional approximation.}
We use a scale-separation argument, but which is only applicable in the non-reactive case.
To this end, we subtract the mass fraction weighted mixture momentum balance
from the closed partial momentum balance \eqref{CIIb} with $C_{ij}=0$ in the absence of
chemical reactions and, first, with $\mathcal{D}_i^T =0$. The resulting equation is
\begin{align}
\ts
& \ts \rho_i (\pt + {\bf v} \cdot \nabla) {\bf u}_i + \rho_i ({\bf u}_i \cdot \nabla ) {\bf v}_i = y_i \nabla p - \rho_i \nabla \mu_i + \rho_i ({\bf b}_i - {\bf b})\nonumber \\
& \ts  + T (h_i - \rho_i \mu_i) \nabla \frac 1 T
+ \div \stress_i^{\rm irr} - y_i \, \div (\sum_k \stress_k^{\rm irr})
- T \sum_j f_{ij}  \rho_i \rho_j ( {\bf u}_i  - {\bf u}_j ),
\end{align}
and our aim is to show that the terms on the left-hand side are negligible for the description
on meso- and macroscopic time scales. To simplify the presentation, we do not consider
the irreversible stress part in the intermediate computations; note that this term
determines the order of the PDE, hence cannot be neglected anyhow.
With this omission, we obtain the dimensionless formulation
\begin{align}
\frac{U}{C} \frac{V}{C}\, y_i
( \pt^* {\bf u}_i^* + {\bf v}^* \cdot \nabla^* {\bf u}_i^* + {\bf u}_i^* \cdot \nabla^* {\bf v}_i^*) = y_i \frac{\nabla^* p^*}{\rho^*}
 - \frac{\rho_0 \mu_0}{p_0} y_i \nabla^* \mu_i^*\hspace{0.5in}\\[1ex]
 - (\frac{h_0}{p_0} \frac{h_i^*}{\rho^*} - \frac{\rho_0 \mu_0}{p_0} y_i \mu_i^*) \nabla^* \ln T^*
+ \frac{\rho_0 BL}{p_0} y_i ({\bf b}_i^* - {\bf b}^*)
- \frac{\rho_0^2 T_0 f_0 L}{p_0} \rho^* T^* \sum_j f_{ij}^*  y_i y_j ( {\bf u}_i^*  - {\bf u}_j^* ),\nonumber
\end{align}
where dimensionless variables are labeled by an asterisk and
the following reference quantities have been employed:
reference diffusion velocity $U$, reference mixture velocity $V$, reference length
$L$, reference mixture time scale $\tau:=L/V$, reference mixture mass density $\rho_0$,
reference pressure $p_0$, reference chemical potential $\mu_0$, reference enthalpy $h_0$,
reference temperature $T_0$ and reference friction coefficient $f_0$.
Furthermore, $C:=\sqrt{p_0 / \rho_0}$ is of the same order of magnitude as the sound speed.
In many applications $C$ is much larger than $U$, usually even significantly larger than $V$.

To proceed, we need to estimate the size of the thermo-mechanical interaction terms, i.e.\ the size
of the ''friction'' coefficients $f_{ij}$, to obtain a reasonable reference value $f_0$.
For this purpose we first establish a relation between the $f_{ij}$ and the so-called Maxwell-Stefan diffusivities $\D_{ij}$,
since quantitative information for the latter is available in the literature.
At this point, consider a binary system under {\it isobaric} (i.e., constant pressure) and isothermal conditions.
Neglecting in the above momentum balance the terms with a factor $U\, V/C^2$ yields
\[
T f_{12} (y_2 \, {\bf j}_1 - y_1 \, {\bf j}_2) = T f_{12} \, {\bf j}_1 = - y_1 \nabla \mu_1.
\]
Specializing to a dilute system with $y_1 \ll 1$, $y_2 \approx 1$, the chemical potential is given as
\[
\ts
\mu_1 = \mu_1^0 + \frac{RT}{M_1} \ln x_1,\vspace{-0.05in}
\]
hence\vspace{-0.05in}
\[
\ts
T f_{12} \, {\bf j}_1 = - \frac{y_1 RT}{M_1 x_1} \nabla x_1
= - \frac{c RT}{\rho} \nabla x_1.
\]
Using $\rho \approx M_2 c \equiv const$ with $c$ the total molar concentration, this finally yields
\[
{\bf j}_1 = - \frac{R}{c\, M_1 M_2 f_{12}} \nabla \rho_1,
\]
which has the form of {\it Fick's law} with the binary diffusivity
\[
\D_{12}:= \frac{R}{c\, M_1 M_2 f_{12}}.
\]
Due to this relationship we further introduce a reference molar concentration $c_0$
and a reference diffusivity $D$.
Letting $M:=\rho_0 / c_0$, $\mu_0 := RT_0/M$ and $\D_{ij}:= \frac{R}{c\,M_i M_j f_{ij}}$, the final form of the dimensionless partial momentum balance is
\begin{align}
\frac{U}{C} \frac{V}{C} y_i
( \pt^* {\bf u}_i^* + {\bf v}^* \cdot \nabla^* {\bf u}_i^* + {\bf u}_i^* \cdot \nabla^* {\bf v}_i^*) =   y_i \frac{\nabla^* p^*}{\rho^*}
 - \frac{c_0 RT_0}{p_0} y_i \nabla^* \mu_i^*
 + \frac{\rho_0 BL}{p_0} y_i ({\bf b}_i^* - {\bf b}^*)\nonumber\\[1ex]
 - (\frac{h_0}{p_0} \frac{h_i^*}{\rho^*} - \frac{c_0 RT_0}{p_0} y_i \mu_i^*) \nabla^* \ln T^*
- \frac{c_0 RT_0}{p_0} \frac{UL}{D} \rho^* T^* \sum_j \frac{y_i y_j}{M_i^* M_j^* \D_{ij}^*}
 ( {\bf u}_i^*  - {\bf u}_j^* ).
\end{align}
Note that the above definition of $\D_{ij}$ is consistent with the relation between the friction factors $f_{ij}$ and the Maxwell-Stefan diffusivities $\D_{ij}$ which was used in Section~\ref{sec:literature-cII} for comparison with the engineering literature.
Now, for multicomponent fluid mixtures not too far from standard conditions, the estimate
\[
\ts
\frac{U}{C} \frac{V}{C}=10^{-8}...10^{-4}
\]
is valid. The other dimensionless coefficients are of order one or above, for instance
\[
\ts
\frac{c_0 RT_0}{p_0}=10^0...10^3 \mbox{ (gas to liquid)},\qquad
\frac{UL}{D} =10^3..10^5.
\]
Consequently, the desired simplification mentioned above yields an excellent approximation under such conditions.
The resulting reduced form of the partial momentum balances reads as
%class-II model reads as
%is equivalent to \eqref{MS1} and reads
%
%
\begin{align}
\label{MS-eqs-nonreactive}
-\sum_{j\neq i} \frac{  y_j {\bf j}_i  - y_i {\bf j}_j }{c\, M_i M_j \D_{ij}} =
\frac{y_i}{ R T} \nabla \mu_i  -\frac{y_i}{ \rho R T} \nabla p
- \frac{y_i}{ R T} ({\bf b}_i - {\bf b})\nonumber \\
+  \frac{ \rho_i \mu_i - h_i}{\rho R} \nabla \frac 1 T
+ \frac{1}{\rho RT} \big(\div \stress_i^{\rm irr} - y_i \, \div (\sum_k \stress_k^{\rm irr})\big),
\end{align}
where we now also put back the irreversible stress parts. If full thermo-diffusion is added and
put into the form mentioned in Section~\ref{sec:literature-cII}, we obtain
\begin{align}
\label{MS-eqs-nonreactive2}
-\sum_{j\neq i} \frac{  y_j {\bf j}_i  - y_i {\bf j}_j }{c\, M_i M_j \D_{ij}}
-\sum_{j\neq i} \frac{\rho y_i \, y_j}{c\, M_i M_j  \D_{ij}}
\Big( \frac{D_i^T}{\rho_i} - \frac{D_j^T}{\rho_j} \Big) \nabla \ln T=
\frac{y_i}{ R T} \nabla \mu_i  -\frac{y_i}{ \rho R T} \nabla p \nonumber \\
- \frac{y_i}{ R T} ({\bf b}_i - {\bf b})
+  \frac{ \rho_i \mu_i - h_i}{\rho R} \nabla \frac 1 T
+ \frac{1}{\rho RT} \big(\div \stress_i^{\rm irr} - y_i \, \div (\sum_k \stress_k^{\rm irr}) \big).
\end{align}
Note that the above diffusional approximation ignores the acceleration of the relative (diffusional) motion.
The cancelation removes sound waves due to diffusional motion, which are related to the so-called phenomenon of {\it second sound}; see, e.g., \cite{M85} for the later phenomenon.
A somewhat similar order-of-magnitude analysis of the terms appearing in the partial momentum
balances has been done in \cite{Whitaker}.
Let us also mention in passing that the appearance of stress induced diffusion has been
investigated in \cite{Annis} via an extended Chapman-Enskog theory.
%\BA CHECK: A. V. Lykov, Yu. A. Mikhailov 1965 \EA

In the Engineering literature on multicomponent diffusion, system \eqref{MS-eqs-nonreactive} is often written using molar based quantities.
With $x_i=c_i/c$, ${\bf J}_i=c_i {\bf u}_i$ and $\mu_i^{m}:=M_i \mu_i$, one obtains
\begin{equation}
-\sum_{j\neq i} \frac{  x_j {\bf J}_i  - x_i {\bf J}_j }{c \, \D_{ij}} =
\frac{c_i}{ R T} \nabla \mu_i^{m}  -\frac{y_i}{R T} \nabla p
+  \frac{h_i- c_i \mu_i^{m}}{R T} \nabla \ln T
- \frac{\rho_i}{ R T} ({\bf b}_i - {\bf b}),
\end{equation}
where we again omitted the irreversible stress part and only consider reduced thermo-diffusion
for brevity.
Observe that the ${\bf J}_i$ do not sum to zero, but satisfy $\sum_i M_i {\bf J}_i =0$.
It is custom to apply the chain rule to separate different contributions from the gradient of the $\mu_i$.
This requires to fix the set of independent variables. With the common choice of $(T,p,x_i, \ldots ,x_{N-1})$,
the notation is
\[
\ts
\nabla \mu_i^m = \nabla_{p,T} \mu_i^m + \frac{\partial \mu_i^m}{\partial p} \nabla p +
\frac{\partial \mu_i^m}{\partial T} \nabla T
\quad \mbox{ with } \quad
\nabla_{p,T} \mu_i^m := \sum_{j=1}^{N-1} \frac{\partial \mu_i}{\partial x_j} \nabla x_j.
\]
We then obtain
\begin{equation}
\label{MS-Eng}
\ts
-R  T  \sum_{j\neq i} \frac{  x_j {\bf J}_i  - x_i {\bf J}_j }{c\, \D_{ij}} =
c_i \nabla_{p,T} \mu_i^{m}  - (\phi_i - y_i) \nabla p
+  (h_i- c_i \mu_i^{m} + c_i T \frac{\partial \mu_i^m}{\partial T}) \nabla \ln T
- \rho_i ({\bf b}_i - {\bf b}),
\end{equation}
where $\phi_i$ denotes the partial molar volume of component $A_i$.
The different contributions on the right-hand side correspond to, from left to right, concentration driven diffusion,
pressure driven diffusion, thermo-diffusion and forced diffusion.
Except for the contribution from the temperature gradient,
the system \eqref{MS-Eng} coincides with the models given in, e.g., \cite{Bird}, \cite{TK-book}. The additional term constitutes a driving force for
thermal diffusion in non-simple mixtures. It is no surprise that this has been missed in the references cited above, since the
standard derivation of the so-called generalized thermodynamic driving forces, going back to \cite{Hirsch} %\BR CHECK \ER,
employs approximate solutions to the multicomponent Boltzmann equations, hence inherently assumes a simple mixture. As explained already above, this contribution to the driving forces vanishes for simple mixtures.\\[1ex]
\noindent
{\bf The reactive extension of the Maxwell-Stefan equations.}
In the case of reacting mixtures, the
scale reduction argument from above is not applicable, because for
reactive flows the term
\[
\rho_i (\pt + {\bf v} \cdot \nabla) {\bf u}_i + \rho_i ({\bf u}_i
\cdot \nabla ) {\bf v}_i,
\]
which has been neglected in the non-reactive case, when summed
over all constituents gives $\sum_i r_i {\bf u}_i$. While this
might be small, it can not be neglected as it is crucial for the thermodynamic consistency.
Instead, the Maxwell-Stefan form is derived by the above entropy invariant model reduction
and was already given in a first form by \eqref{jB} and \eqref{di}.
Brought into the notation of the present section, it reads
\begin{align}
\label{MS-eqs-reactive}
-\sum_{j=1}^N \frac{  y_j {\bf j}_i  - y_i {\bf j}_j }{c\, M_i M_j \D_{ij}}
-\sum_{j=1}^N \gamma_{ij} \Big( \frac{{\bf j}_i}{\rho_i} - \frac{{\bf j}_j}{\rho_j} \Big)
-\sum_{j=1}^N \frac{\rho y_i \, y_j}{c\, M_i M_j  \D_{ij}}
\Big( \frac{D_i^T}{\rho_i} - \frac{D_j^T}{\rho_j} \Big) \nabla \ln T= \nonumber \\
\frac{1}{\rho RT} \Big(
\rho_i \nabla \mu_i  - y_i  \nabla p
- \rho_i ({\bf b}_i - {\bf b})
+  ( h_i- \rho_i \mu_i ) \nabla \ln T
+ \div \stress_i^{\rm irr} - y_i \, \div (\sum_{k=1}^N \stress_k^{\rm irr}) \Big)
\end{align}
with $\gamma_{ij}$ from \eqref{def:gammaij}.
The second term on the left-hand side is new.

Let us close this section with a few remarks on mathematical properties of the resulting
system of strongly coupled diffusion equations.
A closer inspection of the definition of the $\gamma_{ij}$ under the assumption that the
forward reaction rates are modeled such that $R_a^f \to 0$ if $\rho_i \to 0$
for any $i$ for which $A_i$ is a reactand of this reaction, i.e.\ for which $\alpha_i^a >0$,
shows that \eqref{MS-eqs-reactive} has the overall structure
\begin{equation}
\label{MS-structure}
\sum_{j\neq i} a_{ij} y_i y_j ({\bf u}_i - {\bf u}_j) = {\bf F}_i,
\end{equation}
where the matrix $[a_{ij}]$ is positive definite (but, in general, not symmetric).
Given the invertibility, the ${\bf u}_i$ are bounded functions of the ${\bf F}_i$.
As a direct consequence of \eqref{MS-structure}, we then also have the relations
\begin{equation}
\label{MS-structure2}
{\bf j}_i = \rho y_i {\bf u}_i =\frac{\rho}{\sum_{j\neq i} a_{ij} y_j} {\bf F}_i
+ y_i\, \frac{1}{\sum_{j\neq i} a_{ij} y_j} \sum_{j\neq i} a_{ij} {\bf j}_j.
\end{equation}
This shows that the cross-effects vanish in case $y_i\to 0$. Therefore, the solution
will retain non-negative values for all $\rho_i$ if the ${\bf F}_i$ are quasi-positive,
i.e.\ such that ${\bf F}_i\geq 0$ whenever $\rho_i =0$.
An inspection of the corresponding terms in \eqref{MS-eqs-reactive} shows that all of them contain
the factor $y_i$, hence vanish in case $\rho_i=0$, except for $\div \stress_i^{\rm irr}$.
The latter also becomes harmless concerning positivity of solutions if
$\stress_i^{\rm irr}$ contains a factor $y_i$. Recall that a particularly suitable choice
was $\stress_i^{\rm irr}=y_i \sum_k \stress_k^{\rm irr}$ in which case the stress driven
diffusion completely disappears.

For more information on inverting the non-reactive Maxwell-Stefan equations see \cite{DB-MS}.
Concerning criteria for the strong well-posedness of strongly coupled reaction-diffusion
systems and positivity of solutions, the reader is referred in particular to \cite{amann93}.
%
%\begin{align*}
%\ts -\sum_{j\neq i} \frac{  y_j {\bf j}_i  - y_i {\bf j}_j }{D_{ij}} =
%\frac{y_i}{ R T} \nabla \mu_i  -\frac{y_i}{ \rho R T} \nabla p
%+  \frac{ \rho_i \mu_i - h_i}{\rho R} \nabla \frac 1 T
%- \frac{y_i}{R T} ({\bf b}_i - {\bf b}) + \frac{r_i}{ \rho R T} {\bf u}_i
%\end{align*}
%
%Eng.\ literature notation: $\nabla \mu_i = \nabla_{p,T} \mu_i + \partial_p \mu_i \nabla p +
%\partial_T \mu_i \nabla T$.
%\begin{align*}
%\ts   {\bf d}_i =
%\frac{y_i}{ R T}  \nabla_{p,T} \mu_i  + \frac{\phi_i - y_i}{ \rho R T} \nabla p
%+  \frac{h_i - \rho_i s_i T- \rho_i \mu_i }{\rho R T} \nabla \ln T
%- \frac{y_i}{ R T} ({\bf b}_i - {\bf b})
%+ \frac{r_i}{ \rho R T} {\bf u}_i
%end{align*}
%molecular, pressure, thermo- (Sor\'{e}t effect), forced \& \textcolor{red}{reactive} diffusion\\
%
%
%
%
%%%%%%%%%%%%%%%%%%%%%%%%%%%%%%%%
\section{Special free energy densities for fluid mixtures}
\label{sec:free-energies}
 {\bf General context}. We discuss possible
 explicit free energy functions so that the various model schemes of the previous sections
 finally end up with a system of explicit PDEs for the variables. Recall that
class-I models need to know the free energy function
 \begin{equation}\label{SC1b}
    \rho\psi=\rho\tilde\psi(T,\rho_1,\rho_2,...,\rho_N).
\end{equation}
If this function were known we could calculate the specific energy
$e$, the chemical potentials $(\mu_i)_{i\in\{1,2,...,N\}}$ and the
pressure $p$ according to
\begin{equation}\label{SC2}
    e=-T^2\frac{\partial}{\partial T}(\frac{\tilde\psi}{T}),\quad
    \mu_i=\frac{\partial\rho\tilde\psi}{\partial\rho_i},\quad
    p=-\rho\tilde\psi+\sum\limits_{i=1}^N\rho_i\mu_i.
\end{equation}
This is not sufficient for class-II models because here we
additionally need to know constitutive functions for the partial
specific energies $(e_i)_{i\in\{1,2,...,N\}}$ and the partial
pressures $(p_i)_{i\in\{1,2,...,N\}}$, i.e.
\begin{equation}\label{SC3}
    e_i=\tilde e_i(T,\rho_1,\rho_2,...,\rho_N),\quad
    p_i=\tilde p_i(T,\rho_1,\rho_2,...,\rho_N).
\end{equation}
Except in the special case of {\it simple mixtures} (cf.\
\cite{M85}), the functions in \eqref{SC3} cannot be derived from
partial free energy densities in general.\\[1ex]
\noindent
{\bf Simple mixtures}. If the functions in \eqref{SC3} are of the
special form
\begin{equation}\label{SC4}
    e_i=\tilde e_i(T,\rho_i),\quad
    p_i=\tilde p_i(T,\rho_i),
\end{equation}
they can be calculated from partial free energy densities
$\psi_i=\tilde\psi(T,\rho_i)$ and the corresponding mixture is
called {\it simple mixture}. In this case we have
\begin{equation}\label{SC5}
    e_i=-T^2\frac{\partial}{\partial T}(\frac{\tilde\psi_i}{T}),\quad
    \mu_i=\frac{\partial\rho_i\tilde\psi_i}{\partial\rho_i},\quad
    p_i=-\rho_i\tilde\psi_i+\rho_i\mu_i.\vspace{0.1in}
\end{equation}
\noindent
{\bf Mixtures of ideal gases.} A special case of a simple mixture
is a mixture of ideal gases. In the non-degenerate case ideal
gases are characterized by the following constitutive functions
for partial pressures, energies and entropies,
\begin{equation}\label{SC6}
    p_i=\rho_i\frac{k}{m_i}T,\quad
    e_i=z_i\frac{k}{m_i}(T-T^\textrm{R})+e_i^\textrm{R},\quad
    s_i=z_i\frac{k}{m_i}\ln \frac{T}{T^\textrm{R}} -\frac{k}{m_i}
    \ln \frac{\rho_i}{\rho_i^\textrm{R}} +s_i^\textrm{R},
\end{equation}
where $z_i=3/2,5/2~\textrm{or}~3$ for 1-atomic, 2-atomic and more-atomic
constituents, respectively, the index R indicates a reference state and $k$ is the
Boltzmann constant, while $m_i$ denotes the atomic mass of $A_i$.
Usually, the specific entropy is written with $p_i$ as a variable
instead of $\rho_i$. Using $p=\sum_{i=1}^N p_i$, we may write
\begin{equation}\label{SC7}
    s_i(T,p_i)=(z_i+1)\frac{k}{m_i}\ln \frac{T}{T^\textrm{R}} -\frac{k}{m_i}
    \ln \frac{p_i}{p_i^\textrm{R}} +s_i^\textrm{R}=
    s_i(T,p)-\frac{k}{m_i}\ln \frac{p_i}{p}=
    s_i(T,p)-\frac{k}{m_i}\ln x_i.
\end{equation}
With $\psi_i=e_i-Ts_i$, the total free energy density
$\rho\psi=\sum_{i=1}^N \rho_i\psi_i$ can then be represented by
\begin{equation}\label{SC8}
    \rho\psi (T,p,x)
    =\sum\limits_{i=1}^N\rho_i\psi_i(T,p)+n(T,p) \sum\limits_{i=1}^N kT x_i \ln x_i
    .\vspace{0.1in}
\end{equation}
\noindent
{\bf Ideal mixtures.} The last expression has been calculated only for ideal gases.
However, it was observed that there is a large class of mixtures that are appropriately
described by \eqref{SC8}$_2$ with the last term as it stands and with
$\rho_i \psi_i (T,p)$ taken from data tables. Such mixtures are called {\it ideal mixtures}. Note that $\rho_i\psi_i(T,p)$ represents the free energy density of the isolated pure constituent $i$ under the pressure $p$ of the mixture.\\[1ex]
{\bf Elastic mixtures.} We consider a further special fluid
mixture that requires a different treatment. The following may
also be considered as preliminary work to introduce the concept of
incompressibility in the context of fluid mixtures.

Often one meets situations where the various contributions to
the free energy originate from different sources. Let us describe
a typical case. To this end we first transform the variables in
the free energy function \eqref{SC1b} from
$T,\rho_1,\rho_2,...,\rho_N$ to $T,\rho,x_1,...,x_{N-1}$. We write
\begin{equation}\label{SC9a}
    \rho\tilde\psi(T,\rho_1,\rho_2,...,\rho_N)=\rho\hat\psi(T,\rho,x_1,...,x_{N-1})
    \quad\textrm{to calculate}\quad p=\rho^2\frac{\partial\hat\psi}{\partial\rho}.
\end{equation}
Abbreviating $x'=(x_1,x_2,...,x_{N-1})$, we now introduce a
constitutive function $p=\hat p(T,\rho,x')$ for the total pressure
of the mixture which is a response of variations of the total
particle density $n$ measuring the atomic distances. The available
experimental data concern two kinds of variations of $n$: (i)
variations of $n$ at fixed composition and fixed temperature, that
are called elastic deformations and (ii) non-elastic variations at
fixed pressure due to changes of $T$ and $x'$. To describe the
elastic deformation we start from the state $(T,\rho_*,x')$ and
end up in the state $(T,\rho,x')$ and define the mass density
$\rho_*$ by
\begin{equation}\label{SC9}
\rho_*=\hat\rho_*(T,x')\quad\textrm{so that}\quad
    \hat p(T,\rho_*,x')=p^\textrm{R}\quad\textrm{for given}\quad
p=\hat p(T,\rho,x').
\end{equation}
A mixture that is characterized by \eqref{SC9} is called {\it
elastic mixture} (cf.\ \cite{DD08}).

In elastic mixtures we decompose the free energy into non-elastic
and elastic contributions and define
\begin{equation}\label{SC10}
    \psi^\textrm{th}=\hat\psi(T,\rho_*,x')\quad\textrm{and}\quad
\psi^\textrm{el}=\hat\psi-\psi^\textrm{th},
\end{equation}
implying corresponding decompositions of the chemical potentials,
\begin{equation}\label{SC11}
    \mu_i^\textrm{th}=\hat\mu_i(T,\rho_*,x')\quad\textrm{and}\quad
\mu_i^\textrm{el}=\hat\mu_i-\mu_i^\textrm{th}.
\end{equation}
Thus we have achieved an additive decomposition of the free energy
and the chemical potentials into elastic and non-elastic
contributions. The latter are denoted here as thermal
contribution.

In order to complete the determination of the free energy, we
finally assume that, in addition to \eqref{SC9}, also the thermal
parts of the chemical potentials are given:
\begin{equation}\label{SC12}
    \mu_i^\textrm{th}=\hat \mu_i^\textrm{th}(T,x').
\end{equation}
Then the procedure to determine the free energy function is as follows:
\begin{enumerate}
    \item[(i)] Calculate
    \begin{equation}\label{SC13}
    \psi^\textrm{th}=-\frac{p^\textrm{R}}{\rho_*}+
    \sum\limits_{i=1}^N \frac{\rho_{*i}}{\rho_*} \mu_i^\textrm{th}
    \quad\textrm{with}\quad \frac{\rho_{*i}}{\rho_*}=\frac{m_i x_i}{\sum_{j=1}^N m_j x_j}.
\end{equation}
    \item[(ii)] Integrate
    \begin{equation}\label{SC14}
    \frac{\partial\psi^\textrm{el}}{\partial\rho}=\frac{p(T,\rho,x')}{\rho^2}
    \quad\textrm{with}\quad \psi^\textrm{el}(T,\rho_*,x')=0.
\end{equation}
\item[(iii)] Form the total free energy density
\begin{equation}\label{SC14a}
    \rho\psi=\rho\psi^\textrm{el}+\rho\psi^\textrm{th}.
\end{equation}
\item[(iv)] Calculate $\mu_i$ from $\rho\psi$
\begin{equation}\label{SC14b}
    \mu_i=\frac{\partial\rho\psi}{\partial\rho_i}(T,\rho_1,\rho_2,...,\rho_N).
\end{equation}
\item[(v)] Calculate
\begin{equation}\label{SC14c}
    \mu_i^\textrm{el}=\mu_i-\mu_i^\textrm{th}.
\end{equation}
\end{enumerate}
\section{Incompressibility in the context of fluid mixtures}
\label{sec:incomp}
{\bf Definitions of incompressibility}.  The notion of
incompressibility in a single fluid concerns the constitutive law
for the pressure, which we write here as $\rho=\hat\rho(T,p)$. A
single fluid is called incompressible if we have
 $\partial_p\hat\rho=0$. There are reasons to prefer
another definition: (i) incompressibility is a statement on atomic
distances, which may be characterized by the mass density only in
a single fluid. In a mixture, however, the particle densities are
related to the atomic distances but not the total mass density;
(ii) the inequality \eqref{conc8a}$_3$, which is a consequence of
the concavity postulate, reveals that the given definition of
incompressibility implies $\partial_T\hat\rho=0$, i.e.\ the mass
density must be a constant if we only consider the limiting case
$\partial_p\hat\rho=0$ (cf.\ \cite{M85}, \cite{GoMuRu11}).
Therefore we prefer to start with the compressible case and
propose an elastic constitutive law for the pressure containing a
bulk modulus $K$. After exploitation of the 2$^\textrm{nd}$ law of
thermodynamics, we let $K\rightarrow\infty$. Two examples serve to
illustrate the
procedure.\\[1ex]
\noindent
{\bf Example 1}. We consider the constitutive law
\begin{equation}\label{SC15}
    p=p^\textrm{R}+K(\frac{n}{n^\textrm{R}}f(T,x)-1).
\end{equation}
Here $n^\textrm{R}$ denotes some reference value of the total
particle density of the mixture, $K$ is the bulk modulus and the
function $f(T,x)$ characterizes a volumetric expansion due to a
change of temperature and variations of the mole fractions. The
volume expansion is measured at the reference pressure
$p^\textrm{R}$ where we have $n/n^\textrm{R}=f(T,x)$. To calculate
the elastic part of the free energy according to \eqref{SC14}, we
write \eqref{SC15} in the form
\begin{equation}\label{SC16}
    p=p^\textrm{R}+K(\frac{f(T,x)}{M(x)}\frac{\rho}{n^\textrm{R}}-1)=
p^\textrm{R}+K(\frac{\rho}{\rho_*(T,x)}-1)
    \quad\textrm{with}\quad M(x)=\sum\limits_{i=1}^N M_i x_i.
\end{equation}
Recall that the elastic transition concerns
$\rho=M(x)n\leftrightarrows\rho_*=n^\textrm{R}M(x)/f(T,x)$.
Integration of \eqref{SC14} leads to the elastic part of the free
energy density,
\begin{equation}\label{SC17}
    \rho\hat\psi^\textrm{el}(T,\rho,x')=
    (p^\textrm{R}-K)(\frac{\rho}{\rho_*(T,x)}-1)+
    K\frac{\rho}{\rho_*(T,x)}\ln(\frac{\rho}{\rho_*(T,x)}).
\end{equation}
The exploitation of the above scheme to determine
$\mu_i^\textrm{el}$ gives in the limiting case
$K\rightarrow\infty$:
\begin{equation}\label{SC18}
    n=\frac{n^\textrm{R}}{f(T,x)},\quad
    \rho\psi^\textrm{el}(T,p,x')=0\quad
    \mu_i^\textrm{el}=\frac{1}{m_in^\textrm{R}}
    (f+\frac{\partial f}{\partial x_i}-\sum\limits_{j=1}^N\frac{\partial f}{\partial x_j}x_j)(p-p^\textrm{R}).
\end{equation}
%It is important to note that $\psi^\textrm{el}$ and
%$\mu^\textrm{el}$ must be calculated first for the compressible
%case. Hereafter the incompressible limit is carried out. \BR REMARK: WAS PASSIERT SONST? \ER

In the incompressible limit the particle density is no longer an
independent variable because, due to \eqref{SC18}$_1$, it is now
determined by the other variables. However, the number of independent
variables has not changed. In \eqref{SC18}$_3$ the pressure still
appears in the chemical potentials and must now be considered as a
variable that substitutes the particle density as a variable.

The incompressible limit implies a constraint on $\div \V$ that
follows from the algebraic relation
$\rho=M(x)n=n^\textrm{R}M(x)/f(T,x)$. Differentiation with respect
to time and application of the total mass balance yields
\begin{equation}\label{SC19}
    \div \V =-\frac{\partial_t(M(x)/f(T,x))+\V\cdot\nabla(M(x)/f(T,x))}{M(x)/f(T,x)},
\end{equation}
i.e.\ the evolution of $\div \V$ is already given by the evolution
of temperature and mole fractions.\\[1ex]
\noindent {\bf Example 2}. Here we discuss an alternative
constitutive law for the pressure leading to a different
constraint on $\div \V$ in the incompressible limit. We consider
a solution and indicate the solvent by the index $N$ and the
dissolved substances by $i\in\{1,2,...,N-1\}$. For simplicity, in
this example we ignore volumetric changes, i.e. we set $f=1$. We
propose
\begin{equation}\label{SC20}
    p=p_N+\sum\limits_{i=1}^{N-1}p_i\quad\textrm{with}\quad
    p_N=p_N^\textrm{R}+K(\frac{n_N}{n_N^\textrm{R}}-1),\quad
    p_i=\rho_i\frac{k}{m_i}T.
\end{equation}
Thus we describe the solvent by an elastic liquid and the solutes
as a mixture of ideal gases.

The determination of the free energy density and the chemical
potentials runs along the same path as in Example 1. The density
$\rho_*$ for which we have $p=p^\textrm{R}$ is given by
\begin{equation}\label{SC21}
    \rho_*=\frac{p^\textrm{R}-p_N^\textrm{R}+K}{kT+
    (\frac{K}{n_N^\textrm{R}}-kT)x_N} \, M(x).
\end{equation}
The elastic part of the free energy density now reads
\begin{equation}\label{SC22}
    \rho\psi^\textrm{el}=(p_N^\textrm{R}-K)(\frac{\rho}{\rho_*}-1)+
    \left(\frac{kT}{M(x)}+
    (\frac{K}{M(x)n_N^\textrm{R}}-\frac{kT}{M(x)})x_N\right)
    \rho \ln(\frac{\rho}{\rho_*}).
\end{equation}
The further results are exclusively given in the incompressible
limit $K\rightarrow\infty$:
\begin{equation}\label{SC23}
    n_N=n_N^\textrm{R},\quad
    \rho\psi^\textrm{el}(T,p,x')=0,\quad
    (\mu_i^\textrm{el})_{i\in\{1,2,...,N-1\}}=0,\quad
    \mu_N^\textrm{el}=\frac{p-p^\textrm{R}}{m_N n_N^\textrm{N}}.
\end{equation}
In order to compare the results of Example 2 with those of Example
1 we set $f=1$ in \eqref{SC18}. In this case we observe that
\eqref{SC18}$_{1,2}$ agrees with \eqref{SC23}$_{1,2}$, and the
difference of the chemical potentials are also to be expected,
because in Example 1 incompressibility concerns all atomic
distances, while in in Example 2 incompressibility is exclusively
related to the distances between the particles of the solvent.

Accordingly, the resultant constraint refers to $\div \V_N$ and
not to $\div \V$ as in Example 1, because the partial mass
balance of constituent $N$ now reduces to
\begin{equation}\label{SC24}
    \div \V_N=\frac{1}{m_N n_N^\textrm{R}}r_N.
\end{equation}
Thus if there is no mass production of the solvent, which is the
usual case, we have the simple constraint $\div \V_N=0$.\\[1ex]
\noindent {\bf Incompressibility as a limiting case}. The
classical literature frequently considers an incompressible limit
where thermal expansion is still possible. This limiting case
leads to two apparent inconsistencies, (cf.\ \cite{M85},
\cite{GoMuRu11}). One concerns the inequality \eqref{conc8a}$_3$,
\begin{equation}\label{conc8b}
    (\frac{\partial \rho}{\partial T})^2<\frac{c_p\rho^2}{T}\frac{\partial \rho}{\partial
    p}~,
\end{equation}
and the other is related to the coupled systems of partial
differential equations for the variables at hand. Here, we must show that $\div {\bf v} =0$
is compatible with the variation of $\rho$, which seems to contradict the mass balance.
In the following we show that an incompressible limit with thermal expansion is possible.

It is sufficient to discuss the problem in the context of a single
fluid. The variables are then $p$, $\V$ and $T$, the viscous
stress is given by the Navier-Stokes stress and the heat flux is
generated by Fouriers law. The mass density and the internal
energy are calculated from the constitutive laws $\rho=\rho(T,p)$
and $e=e(T,p)$, respectively. In this case the relevant equations
of balance for mass, momentum and internal energy can be written
as
\begin{eqnarray}
\frac{\partial \rho}{\partial T} \dot T+
\frac{\partial \rho}{\partial p} \dot p +\rho \, \div \V&=&0,\label{SC25a}\\
\rho \dot \V+\nabla p-\div \stress^{\rm irr} &=&\rho {\bf g},\label{SC25b}\\
\rho c_p\dot T+\frac{T}{\rho}\frac{\partial \rho}{\partial T}\dot p+\div{\boldsymbol
q}&=& \stress^{\rm irr} : {\bf D},\label{SC25c}
\end{eqnarray}
where $\dot\psi=\partial_t\psi+\V\cdot\nabla\psi$ indicates the
material time derivative. Below, we let $g=||{\bf g}||$.

These equations embody four different phenomena whose importance
may be characterized by four dimensionless numbers which are
called Mach number, Reynolds number, Froude number and
Fourier number, respectively. They are defined via
\begin{equation}\label{SC26}
    \textrm{Ma}^2=\frac{v_0^2\rho_\textrm{R}}{p_\textrm{R}},\quad
    \textrm{Re}=\frac{\rho_\textrm{R}L_0 v_0}{\eta_\textrm{R}},\quad
    \textrm{Fr}^2=\frac{gt_0}{v_0},\quad
    \textrm{Fo}=\frac{\kappa_\textrm{R} t_0}{\rho_\textrm{R}c_p^\textrm{R} L_0^2}.
\end{equation}
After rewriting the equations \eqref{SC25a}-\eqref{SC25c} in a
non-dimensional form, these numbers naturally appear. Let us
introduce $t=t_0 t'$, $x=L_0 x'$, $p=p_\textrm{R}p'$,
$\V=v_0 \V'$ with $v_0:=L_0 / t_0$, $T=T_\textrm{R}T'$,
$\stress^{\rm irr}=\sigma_0 \stress'^{\rm \, irr}$ with
$\sigma_0=\eta_\textrm{R} v_0/L_0$ and $c_p=c_p^\textrm{R} c'_p$,
where the quantities with a prime have no physical dimension. We
set $c_p^\textrm{R}=p_\textrm{R}/\rho_\textrm{R} T_\textrm{R}$ and,
dropping primes, obtain
\begin{eqnarray}
\frac{\partial \rho}{\partial T}~ \dot T+
\frac{\partial \rho}{\partial p}~ \dot p +\rho~\div \V&=&0,\label{SC27a}\\
 \rho \dot \V+\frac{1}{\textrm{Ma}^2}~\nabla p-
\frac{1}{\textrm{Re}}~\div \stress^{\rm irr} &=&
\frac{1}{\textrm{Fr}^2}\rho \, {\bf e}_g,\label{SC27b}\\
 \rho c_p~\dot T+\frac{T}{\rho}\frac{\partial \rho}{\partial
T}~\dot p+\textrm{Fo}~\div{\boldsymbol
q}&=&\frac{\textrm{Ma}^2}{\textrm{Re}}\stress^{\rm irr}:{\bf D},\label{SC27c}
\end{eqnarray}
where ${\bf e}_g$ denotes the unit vector in direction of the
gravity field $\bf g$. Our discussion starts with a simplified
version of the explicit constitutive law \eqref{SC16} which we
write in the form
\begin{equation}\label{SC28}
    \rho=1-\beta T_\textrm{R}(T-1)+\frac{ p_\textrm{R}}{K}(p-1),
\end{equation}
using the dimensionless variables.
We consider water and choose $T_\textrm{R}=293$ K,
$\rho_\textrm{R}=998$ kg/m$^3$ and $p_\textrm{R}= 10^5$ Pa. In the
neigh\-borhood of this state we have a thermal expansion coefficient
of $\beta=2.07\cdot 10^{-4}$ 1/K, a bulk modulus of $K=2.18\cdot 10^9$
Pa and a specific heat of $c_p=4.18\cdot 10^3$ J/kg/K. Moreover, we
have $\eta_\textrm{R}=10^{-3} \mbox{Pa} \, \mbox{s}$ and $\kappa_\textrm{R}=0.6$ W/K/m. Next, we
introduce a small parameter $\varepsilon\ll 1$ and two parameter
$\beta_0$, $\alpha_0$ of order 1. In our example:
$\varepsilon=10^{-4}$, $\beta_0=6.07$, $\alpha_0=0.46$. Then
\eqref{SC28} reads
\begin{equation}\label{SC28b}
    \rho=1-\beta_0 (T-1)\sqrt{\varepsilon}+\alpha_0(p-1)\varepsilon.
\end{equation}
We conclude that the variation of $\rho$ due to compressibility is
about two orders of magnitude smaller than variation due to thermal
expansion. In the limiting case $\varepsilon\rightarrow\infty$ the
smallness parameter drops out from the inequality \eqref{conc8b}
and the remaining inequality reads as
\begin{equation}
\label{inequ-dimless}
\beta_0^2 < c'_p \, \alpha_0.
\end{equation}
Note that $c'_p=12247$ for the example above, hence the inequality \eqref{inequ-dimless} is satisfied.

Next we study the behavior of the equations
\eqref{SC27a}-\eqref{SC27c} in the incompressible limit. We assume
finite values of Re, Fo and consider the low Mach number limit
Ma$~=\sqrt{\varepsilon}$ with a fixed ratio $\gamma=\textrm{Fr}^2/
\textrm{Ma}$. In this regime we will obtain a consistent incompressible
limit $\varepsilon\rightarrow\infty$.

To this end we formally expand the variables
\begin{equation}\label{Sc29}
    p=p_0+p_1 \textrm{Ma}+p_2 \textrm{Ma}^2+...,\quad\V=\V_0+\V_1 \textrm{Ma}+...,\quad
    T=T_0+T_1 \textrm{Ma}+...
\end{equation}
and correspondingly the constitutive functions. In the highest
order, the equations \eqref{SC27a}-\eqref{SC27c} then reduce to
\begin{equation}\label{SC30}
    \div\V_0=0,\quad
    \dot \V_0+\nabla p_2-
\frac{1}{\textrm{Re}}~\div \stress^{\rm irr}_0=-\beta_0(T_0-1),\quad
c_p\dot T_0+\textrm{Fo}~\div{\boldsymbol q_0}=0.
\end{equation}
Furthermore, we obtain $\nabla p_0=0$ and $\nabla
p_1=\frac{\textrm{Ma}}{\textrm{Fr}^2}\, {\bf e}_g$, but these
equations are not needed to solve the coupled system \eqref{SC30}.
Observe that the higher order pressure $p_2$ becomes the Lagrange
multiplier that guarantees the constraint \eqref{SC30}$_1$.  Thus
we have established a thermodynamically consistent limit
describing incompressible behavior with thermal expansion.
Some of the considerations about consistency of the incompressible limit
above are closely related to those in \cite{GoMuRu11}.
%
%
%
%
%%%%%%%%%%%%%%%%%%%%%%%%%%%%%%%%
%
%
%
%
%%%%%%%%%%%%%%%%%%%%%%%%%%%%%%%%
\begin{center}
{\bf Acknowledgement}
\end{center}
The first author (D.B.) gratefully acknowledges financial support by the Deutsche
Forschungsgemeinschaft within the scope of the German Excellence Initiative.
He also wishes to thank Jan Pr\"uss (Halle-Wittenberg) for helpful discussions
on the Maxwell-Stefan equations for reactive mixtures.
The second author (W.D.) is grateful to Clemens Guhlke for numerous
discussions and most valuable comments.
%%%%%%%%%%%%%%%%%%%%%%%%%%%%%%%%
%
}


\begin{thebibliography}{99}
\bibitem{Al09} W.\ Alt: The entropy principle for interfaces. Fluids and Solids.
Advances in Mathematical Sciences and Applications {\bf 19} (2), 585-663 (2009).
\bibitem{amann93} H.\ Amann: Nonhomogeneous linear and quasilinear elliptic and parabolic
boundary value problems, pp.\ 9--126 in
{\it Function spaces, differential operators and nonlinear analysis},
Teubner-Texte Math.\ Vol.\ 133, Teubner, Stuttgart 1993.
\bibitem{Annis} B.K.\ Annis: Stress Induced Diffusion in Monatomic Gases and Gas Suspensions.
Physics of Fluids {\bf 14}, 269-277 (1971).
%The Maxwell-Stefan approach to mass transfer. Chem.\ Eng.\ Sci.\ {\bf 52}, 861-911 (1997).
\bibitem{Bird} R.B.\ Bird, W.E.\ Stewart, E.N.\ Lightfoot:
{\it Transport Phenomena} (2$^{\rm nd}$ edition). Wiley, New York 2007.
\bibitem{Bowen} R.M.\ Bowen:
{\it Theory of Mixtures}, pp.\ 1-127 in Continuum Physics, Vol.\ 3
(A.C.\ Eringen, ed.), Academic Press, New York, 1976.
\bibitem{Bo-JDE} D.\ Bothe: Instantaneous limits of reversible chemical reactions
in presence of macroscopic convection, J.\ Differ.\ Equations {\bf 193}: 27-48 (2003).
\bibitem{DB-MS} D.\ Bothe: On the Maxwell-Stefan equations to multicomponent diffusion,
pp.\ 81-93 in {\it Progress in Nonlinear Differential Equations and their Applications Vol. 60} (P.\ Guidotti, Chr.\ Walker et al., eds),  Springer, Basel 2011.
\bibitem{Burgers} J.M.\ Burgers: The Boltzmann Equation for Flows with Chemical Reactions,
Planetary and Space Science {\bf 3}, 4-11 (1961).
\bibitem{Datta} R.\ Datta, S.\ A.\ Vilekar:
The continuum mechanical theory of multicomponent diffusion in fluid mixtures,
Chem.\ Eng.\ Sci.\ {\bf 65}, 5976--5989 (2010).
\bibitem{DD08} W.\ Dreyer, F.\ Duderstadt: On the modelling of semi-insulating GaAs
 including surface tension and bulk stresses, Proc. R. Soc. Lond. Ser.
  A Math. Phys. Eng. Sci., {\bf 464}: 2693-2720 (2008).
\bibitem{DG13} W.\ Dreyer, C.\ Guhlke: Sharp limit of the viscous Cahn-Hilliard equation and
thermodynamic consistency, WIAS preprint 1771, submitted to SIAM J.\ Appl.\ Math.\ (2013).
\bibitem{DGM13} W.\ Dreyer, C. Guhlke, R. Müller: Rational modeling of electrochemical
 double-layers and derivation of Butler-Volmer equations, WIAS preprint 1860.
\bibitem{Eckart1} C.\ Eckart: The thermodynamics of irreversible processes I. The simple fluid,
Phys.\ Rev.\ {\bf 58}, 267-269, (1940).
\bibitem{Eckart2} C.\ Eckart: The thermodynamics of irreversible processes II. Fluid mixtures,
Phys.\ Rev.\ {\bf 58}, 269-275 (1940).
\bibitem{FK} J.H.\ Ferziger, H.G.\ Kaper: {\it  Mathematical theory of transport
processes in gases}. North-Holland, 1972.
\bibitem{Fick} A.\ Fick: \"Uber Diffusion, Annalen der Physik {\bf
94}, 59-86 (1855).
\bibitem{Fluegge} C.\ Truesdell, R.A. Toupin: The classical field theories. In
{\it  Encyclopedia of Physik Vol.\ III/1. Principals of classical mechanics and field theories} (S.\ Fl\"{u}gge, ed.). Springer, Berlin $\cdot$ G\"{o}ttingen $\cdot$ Heidelberg 1960.
\bibitem{Gallagher} I.\ Gallagher, L.\ Saint-Raymond, B.\ Texier:
From Newton to Boltzmann: hard spheres and short-range potentials.
ArXiv:1208.5753v2 (2013).
\bibitem{dGM} S.R.\ de Groot, P.\ Mazur: {\it  Non-Equilibrium Thermodynamics}. Dover Publications, 1984.
\bibitem{Giovan} V.\ Giovangigli, {\it Multicomponent Flow Modeling},
Birkh\"auser, Boston 1999.
\bibitem{Gouin-mixture} H.\ Gouin, Variational theory of mixtures
in continuum mechanics,
European Journal of Mechanics - B/Fluids {\bf 9}, 469-491 (1990).
\bibitem{GoMuRu11} H.\ Gouin, A.\ Muracchini, T.\ Ruggeri: On the
Müller paradox for thermal incompressible media, Continuum Mech.\
Thermodyn.\ {\bf 24}, 505-513, (2011). %4-6
\bibitem{GENERIC} M.\ Grmela, H.C.\ \"{O}ttinger, Dynamics and thermodynamics of complex fluids.
I. Development of a general formalism Phys. Rev. E {\bf 56}, 6620-6633 (1997).
\bibitem{Diss-Guhlke} C.\ Guhlke: {\it Theorie elektrochemischer Grenzflächen}.
Ph.D.\ dissertation, TU Berlin 2014.
\bibitem{Gurtin71} M.E.\ Gurtin:
On the thermodynamics of chemically reacting fluid mixtures.
Arch.\ Rat.\ Mech.\ Anal.\  {\bf 43}, 198-212 (1971).
\bibitem{HecklM} M.\ Heckl, I.\ M\"uller:
Frame dependence, entropy, entropy flux, and wave speeds in mixtures of gases,
Acta Mech.\ {\bf 50}, 71-95 (1983).
\bibitem{Hirsch} J.O.\ Hirschfelder, C.F.\ Curtiss, R.B.\ Bird:
{\it Molecular Theory of Gases and Liquids} (2$^{\rm nd}$ corrected printing). Wiley, New York 1964.
\bibitem{Hutter-book} K.\ Hutter, K.\ J\"ohnk, {\it Continuum Methods of Physical Modeling},
Springer, Heidelberg 2004.
\bibitem{Kerk} P.J.A.M.\ Kerkhof, M.A.M.\ Geboers:
Analysis and extension of the theory of multicomponent fluid diffusion. Chem.\ Eng.\ Sci.\ {\bf 60}, 3129-3167 (2005).
\bibitem{Kerk-AicheJ} P.J.A.M.\ Kerkhof, M.A.M.\ Geboers:
Toward a Unified Theory of Isotropic Molecular Transport Phenomena. AIChE J.\ {\bf 51}, 79-121 (2005).
Corrections: AIChE J.\ {\bf 57}, 1099 (2011).
\bibitem{KB-book} S.\ Kjelstrup, D.\ Bedeaux, {\it Non-Equilibrium Thermodynamics
of Heterogeneous Systems}, in ''Series on Advances in Statistical Mechanics -- Volume 16'',
World Scientific, Singapore 2008.
\bibitem{KW} R.\ Krishna, J.A.\ Wesselingh:
The Maxwell-Stefan approach to mass transfer. Chem.\ Eng.\ Sci.\ {\bf 52}, 861-911 (1997).
\bibitem{Lebon} G.\ Lebon, D.\ Jou, J.\ Casas-V\'azquez: {\it  Understanding Non-equilibrium
Thermodynamics}.
Springer 2008.
\bibitem{Liu72} I-Shih Liu:
Method of Lagrange multipliers for exploitation of the entropy principle.
Arch.\ Rat.\ Mech.\ Anal.\  {\bf 46}, 131-148 (1972).
\bibitem{Liu} I-Shih Liu, {\it Continuum Mechanics},
Springer, 2002.
\bibitem{Mauri} R.\ Mauri, {\it Non-Equilibrium Thermodynamics in Multiphase Flows},
Springer, 2013.
\bibitem{Max} J.C.\ Maxwell:
On the dynamical theory of gases,
Phil.\ Trans.\ R.\ Soc.\ {\bf 157}, 49-88 (1866).
\bibitem{Meixner43} J.\ Meixner:
Zur Thermodynamik der irreversiblen Prozesse in Gasen mit chemisch reagierenden, dissoziierenden and anregbaren Komponenten, Annalen der Physik {\bf 43}, 244-270 (1943).
\bibitem{Meixner73} J.\ Meixner:
Consistency of the Onsager-Casimir reciprocal relations,
Advances in Molecular Relaxation Processes {\bf 5}, 319-331 (1973).
\bibitem{M68} I.\ M\"uller: A thermodynamic theory of mixtures of fluids,
Arch.\ Rat.\ Mech.\ Anal.\ {\bf 28}, 1-39 (1968).
\bibitem{M75} I.\ M\"uller: Thermodynamics of mixtures of fluids,
J.\ de M\'{e}canique {\bf 14}, 267-303 (1975).
\bibitem{M85} I.\ M\"uller: {\it  Thermodynamics}.
Pitman 1985.
\bibitem{IM-history} I.\ M\"uller: {\it  A History of Thermodynamics}.
Springer 2007.
\bibitem{IM-Ruggeri} I.\ M\"uller, T.\ Ruggeri: {\it  Rational Extended Thermodynamics}.
Springer 1993.
%\bibitem{M-Heckl} M.\ Heckl, I.\ M\"uller: Frame Dependence, Entropy, Entropy Flux,
%and Wave Speeds in Mixtures of Gases,
%Aeta Mechanica 50, 71--95 (1983).
\bibitem{IM-Weiss} I.\ M\"uller, W.\ Weiss:
Thermodynamics of irreversible processes - past and present,
European Physical Journal H{\bf 37}, 139-236 (2012).
\bibitem{M+M} I.\ M\"uller, W.\ H.\ M\"uller: Fundamentals of
Thermodynamics and Applications, Springer 2009.
\bibitem{Pekar} M.\ Pekar:
Thermodynamics and foundations of mass-action kinetics,
Progress in Reaction Kinetics and Mechanisms {\bf 30}, 3-113 (2005).
\bibitem{Samohyl} I.\ Samoh\'{y}l: {\it  Thermodynamics of Irreversible Processes in
Fluid MIxtures (Approached by Rational Thermodynamics)}.
Teubner-Texte zur Physik - Band 12, Leipzig 1987.
\bibitem{Slattery} J.C.\ Slattery: {\it  Advanced Transport Phenomena}.
Cambridge University Press, Cambridge 1999.
\bibitem{Snell} F.\ M.\ Snell, R.\ A.\ Spangler:
A Phenomenological Theory of Transport in Multicomponent Systems, J.\ Phys.\ Chem.\
{\bf 71}, 2503--2510 (1967).
\bibitem{Standart} G.L.\ Standart, R.\ Taylor, R.\ Krishna:
The Maxwell-Stefan formulation of irreversible thermodynamics for simultaneous
heat and mass transfer. Chem.\ Engng.\ Commun.\ {\bf 3}, 277-289 (1979).
\bibitem{Stef} J.\ Stefan:
\"Uber das Gleichgewicht und die Bewegung insbesondere die Diffusion von Gasgemengen,
Sitzber.\ Akad.\ Wiss.\ Wien {\bf 63}, 63-124 (1871).
\bibitem{TK-book} R.\ Taylor, R.\ Krishna:
{\it Multicomponent mass transfer}. Wiley, New York 1993.
\bibitem{CT57} C.\ Truesdell: Sulle basi della termomeccanica,
Accad. Lincei Rendiconti {\bf 22}, 33, 158--166 (1957).
\bibitem{CT62} C.\ Truesdell:
Mechanical basis of diffusion, J.\ Chem.\ Phys. {\bf 37}, 2336-2344 (1962).
\bibitem{CT69} C.\ Truesdell:
{\it Rational Thermodynamics}. McGraw-Hill Series in Modern Applies Mathematics, New York 1969.
\bibitem{TN2003} C.\ Truesdell, W.\ Noll:
{\it The Non-Linear Field Theories of Mechanics {\rm (3$^{rd}$ edition)}}.
Springer 2003.
\bibitem{Whitaker} S.\ Whitaker: Transport processes with heterogeneous reaction.
On the Maxwell-Stefan equations to multicomponent diffusion,
pp.\ 1-94 in {\it Concepts and Design of Chemical Reactors} (A.E.\ Cassano, S.\ Whitaker, eds),  Gordon and Breach, NewYork 1986.
\end{thebibliography}
\end{document}